\RequirePackage{fix-cm}
\documentclass[smallextended]{svjour3}       
\smartqed  
\usepackage{graphicx}
\usepackage{color}
\usepackage{multirow}
\usepackage{graphicx}
\usepackage{subfigure}
\usepackage{epsfig}
\usepackage{amsmath, amssymb}
\usepackage{url}
 \journalname{Statistical Papers}
\begin{document}

\title{Statistical analysis and first-passage-time applications of a lognormal diffusion process with multi-sigmoidal logistic mean
}
\titlerunning{Statistical analysis and FPT applications of a lognormal diffusion process}        
\author{Antonio Di Crescenzo  \and
        Paola Paraggio \and
        Patricia Rom\'an-Rom\'an \and
        Francisco Torres-Ruiz
}
\institute{A. Di Crescenzo, Paola Paraggio \at
              Dipartimento di Matematica, Universit\`a degli Studi di Salerno, Via Giovanni Paolo II n.\ 132, I-84084 Fisciano (SA), Italy \\
              \email{adicrescenzo@unisa.it, pparaggio@unisa.it}             \\
             \emph{ORCID:} of A.D.C. 0000-0003-4751-7341, of P.P. 0000-0002-3308-7937
           \and
           P. Rom\'an-Rom\'an, F. Torres-Ruiz \at
           Department of Statistics and Operations Research, Faculty of Sciences, University of Granada, 18071 Granada, Spain; Institute of Mathematics of the University of Granada (IMAG), Calle Ventanilla, 11, 18001, Granada, Spain \\
           \email{proman@ugr.es, fdeasis@ugr.es} \\
           \emph{ORCID:} of P.R.R. 0000-0001-7752-8290, of F.T.R. 0000-0001-6254-2209
}

\date{Received: date / Accepted: date}

\maketitle

\begin{abstract}
We consider a lognormal diffusion process having a multisigmoidal logistic mean, useful to model the evolution of a population which reaches the maximum level of the growth after many stages. Referring to the problem of statistical inference, two procedures to find the maximum likelihood estimates of the unknown parameters are described. One is based on the resolution of the system of the critical points of the likelihood function, and the other is on the maximization of the likelihood function with the simulated annealing algorithm. A simulation study to validate the described strategies for finding the estimates is also presented, with a real application to epidemiological data. Special attention is also devoted to the first-passage-time problem of the considered diffusion process through a fixed boundary.
%
%
\end{abstract}
%
\section{Introduction}\label{sec:Intro}
%
%
Growth curves with sigmoidal behavior are widely used in several applied fields including biology (see, for instance,
Brauer and Castillo (2012) \cite{BrauerCastilo2012}), software reliability (cf.\ Erto {\it et al.} (2020)\ \cite{Ertoetal2020}, Inoue and Yamada (2013) \cite{InoueYamada2013})
and economics (see, for example, Smirnov and Wang (2020) \cite{SmirnovandWang2020}).
During the times different kinds of sigmoidal curves have been introduced such as logistic, Gompertz, Korf, Bertalanffy, etc.\
Many efforts have been made essentially for two main purposes:
(i) unification of classical models (see Chakraborthy (2019) \cite{Chakrabortyetal2019}), and
(ii) generalizations of growth curves (see, for example, Asadi {\it et al.}\ (2020)  \cite{Asadietal2020}, Di Crescenzo and Spina (2016) \cite{DiCrescenzoSpina2016} and Romero {\it et al.}\  (2016) \cite{Romeroetal2016}).
\par
The differential equations which drive the growth of the aforementioned deterministic models are very useful to describe population dynamics. However, in order to make them more realistic, it is necessary to introduce a noise term in the equation.
In this way, the differential equations are replaced by stochastic ones.
Most of the times, the analysis of the resulting stochastic equation is quite complex, and the transition probability density of the resulting diffusion process cannot be determined (for example, see Campillo {\it et al.}\ (2018) \cite{Campilloetal2018}, in which the authors propose, for this reason, a new approach to find the maximum likelihood estimates). Models based on diffusion processes are commonly used in various fields of applications, for example plant dynamics (cf.\ Rup\v{s}ys {\it et al.}\ (2020)  \cite{Rupsysetal2020}, where   a hybrid growth is based on Gompertz and Vasicek models), resources consumption (for instance, Nafidi {\it et al.}\  (2019) \cite{Nafidietal2019} uses the Brennan-Schwartz process to model electricity consumption in Morocco) or particular fish species growth
(cf.\ a stochastic version of the open-ended logistic model considered in Yoshioka {\it et al.}\ (2019)  \cite{Yoshiokaetal2019}).
\par
In a recent paper, Di Crescenzo {\it et al.}\  (2020) \cite{DiCrescenzoetal2020} focuses on the generalization of the classical logistic growth model introducing more than one inflection point. To this end, firstly, two different birth-death processes, one with linear birth and death rates and the other with quadratic rates were considered. Then, a diffusive approximation was performed  leading to a non-homogeneous lognormal diffusion process with mean of multi-sigmoidal logistic type. Attention was also given to the description of its main features of interest in applied contexts. For instance,
the mean of the process is a generalized version of the classical logistic function (see, for instance,
Di Crescenzo and Paraggio (2019) \cite{DiCrescenzoParaggio2019}) with more than one inflection point.
The transition probability density of the process has been obtained explicitly and has been applied to plant dynamics.
\par
Starting from the theoretical results of the previous works, in the present paper we approach the problem of the inference of the stochastic model.
This is done by means of the maximum likelihood method, thanks to the availability in closed form of the likelihood function.
We also address the treatment of some collateral problems that emerge in the development carried out, such as: (i) obtaining initial solutions to solve the system of likelihood equations, and (ii) bounding the parametric space for addressing the estimation by metaheuristic procedures. All development is supported by simulation examples.
Subsequently, in order to provide an example of application to real phenomena, we adopt the proposed model to describe
the behavior of the data on the evolution of COVID-19 in different European countries during the two first waves of infection.

Indeed, some of the main features of the diffusion process, such as the mean, the mode and the quantiles,
may be used for prediction purposes and they are expressed as a function of the parameters of the process.
\par
The problem of parameters estimation has been considered in several papers, for instance in Shimizu and Iwase (1987) \cite{ShimizuIwase1987} and in Tanaka (1987) \cite{Tanaka1987}. See also  the more recent works of Garcia (2019) \cite{Garcia2019},
in which the author converts the maximization of the likelihood function into an equivalent problem regarding the minimization of a square error, and of Ramos-\'Abalos {\it et al.}\ (2020) \cite{Ramosetal2020} where maximum likelihood estimates of the parameters of the powers of the homogeneous Gompertz diffusion process are obtained.
\par
Two different strategies to obtain the maximum likelihood estimates of the parameters are introduced.
The first is based on the solution of the system of the critical points of the likelihood function, and the other stems from a meta-heuristic optimization method (simulated annealing) to maximize the likelihood function.
\par
This is the outline of the content of the paper. In Section 2, the most relevant characteristics of the deterministic and the corresponding stochastic model are recalled. Then, the problem of finding the maximum likelihood estimates of the involved parameters is described in Section 3.
In several contexts of population dynamics, it may be relevant to know how long the population spends below a certain control threshold.
For this reason the  first-passage-time (FPT) problem is also addressed. More precisely, in Section 4, the \textbf{\textsf{R}}-package \textsf{fptdApprox} (see \cite{fptdApprox}) is used to determine the approximated FPT density of the lognormal diffusion process through a constant boundary. With the purpose of validating the described procedures for finding the maximum likelihood estimates, a simulation study is presented in Section 5. Finally, in Section 6 we propose an application of the model to real data concerning the COVID-19 infections in France, Italy, Spain and United Kingdom.
%
\section{The multisigmoidal logistic model and the corresponding diffusion process}\label{sec:Section1}
\begin{figure}[t]
	\centering
	\subfigure[]{\includegraphics[scale=0.42]{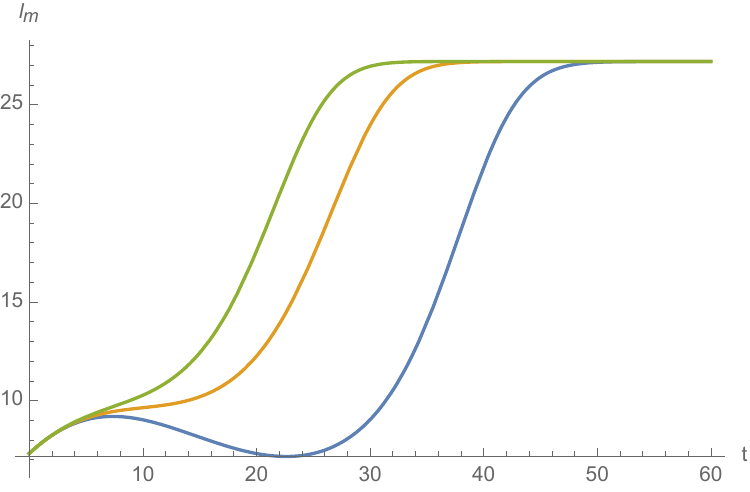}}\quad
	\subfigure[]{\includegraphics[scale=0.42]{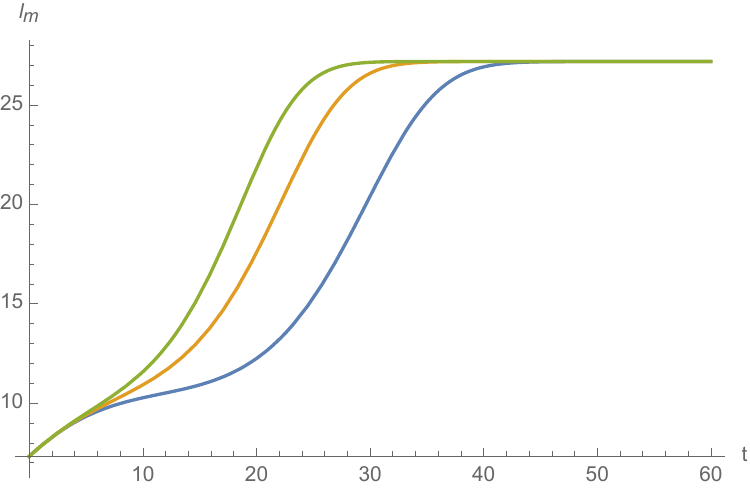}}
	\caption{The multisigmoidal logistic function for some choices of the parameters: $t_0=0$, $l_0=\frac{10}{1+\eta}$, $\eta=e^{-1}$, $\beta_1=0.1$, (a)  $\beta_2=-0.009$ and, from bottom to top, $\beta_3=0.0002, 0.0003, 0.0004$; (b)  $\beta_2=-0.007$ and, from bottom to top, $\beta_3=0.0002, 0.0003, 0.0004$.}
	\label{fig:Figure1}
\end{figure}

%
%
Consider the classical logistic equation
$$
\frac{d}{dt}l(t)= rl(t)\left[1-\frac{\eta}{C}l(t)\right], \qquad  t\ge t_0,
$$
with $r, \eta, C>0$. If the intrinsic growth rate $r$ is replaced by a polynomial $P(t)$, then the solution of this equation, with the initial condition $l(t_0)=l_0$, is given by
$$
l(t)=\frac{l_0 e^{Q(t)-Q(t_0)}}{1-\frac{\eta}{C}l_0\left(1-e^{Q(t)-Q(t_0)}\right)}, \qquad t\ge t_0,
$$
where $Q(t)-Q(t_0)=\int_{t_0}^tP(\tau)d\tau$. With the hypotesis that $Q(t)\to +\infty$ when $t\to \infty$, the carrying capacity of this generalized model is given by $\frac{C}{\eta}$, and thus it is independent from the initial condition $l_0$. In order to obtain a generalized logistic function in which the carrying capacity is dependent on the initial condition, we consider the following equation (cf.\  Di Crescenzo {\it et al.}\  (2020) in \cite{DiCrescenzoetal2020})
\begin{equation}\label{odelm}
	\frac{d}{dt}l_m(t)=h_\theta(t)l_m(t), \qquad t\ge t_0,
\end{equation}
with
\begin{equation}\label{htheta}
	h_\theta(t):=\frac{P_\beta(t)e^{-Q_\beta(t)}}{\eta+e^{-Q_\beta(t)}},
\end{equation}
for $\eta>0$, $\theta=(\eta,\beta^T)^T$ with $\beta^T=(\beta_1,\dots, \beta_p)\in\mathbb R^p$,  where
\begin{equation}\label{Qbeta}
	Q_\beta(t)=\sum_{i=1}^p \beta_it^i,\qquad \beta_p>0,
\end{equation}
and $P_\beta(t)=\displaystyle\frac{d}{dt}Q_\beta (t)$. Under these assumptions, the solution of the ordinary differential equation (\ref{odelm}), with initial condition $l_m(t_0)=l_0$, is the so-called multisigmoidal logistic function  given by
\begin{equation}\label{nuovalm}
	l_m(t)=l_0\frac{\eta + e^{-Q_\beta(t_0)}}{\eta + e^{-Q_\beta(t)}}, \qquad t\ge t_0.
\end{equation}
We point out that the function $l_m$ may exhibit more than one inflection point, and its carrying capacity is
\begin{equation}\label{lim}
	\lim_{t\to\infty}l_m(t)=l_0\frac{\eta+e^{-Q_\beta(t_0)}}{\eta}=\frac{C}{\eta},
\end{equation}
where $C=C(l_0,\eta,\beta, t_0)=l_0\left(\eta+e^{-Q_\beta(t_0)}\right)$ and $Q_\beta$ is defined in Eq.\ \eqref{Qbeta}.
It is easy to note that the function (\ref{nuovalm}) is not monotonous in general, since the monotonicity intervals depend on the coefficients $\beta_1,\dots,\beta_p$ of the polynomial $Q_\beta$, and the carrying capacity is the maximum value attainable by the function $l_m$. See Figure \ref{fig:Figure1} for some plots of the multisigmoidal logistic function.
\par
\begin{figure}[t]
	\centering
	\subfigure[]{\includegraphics[scale=0.32]{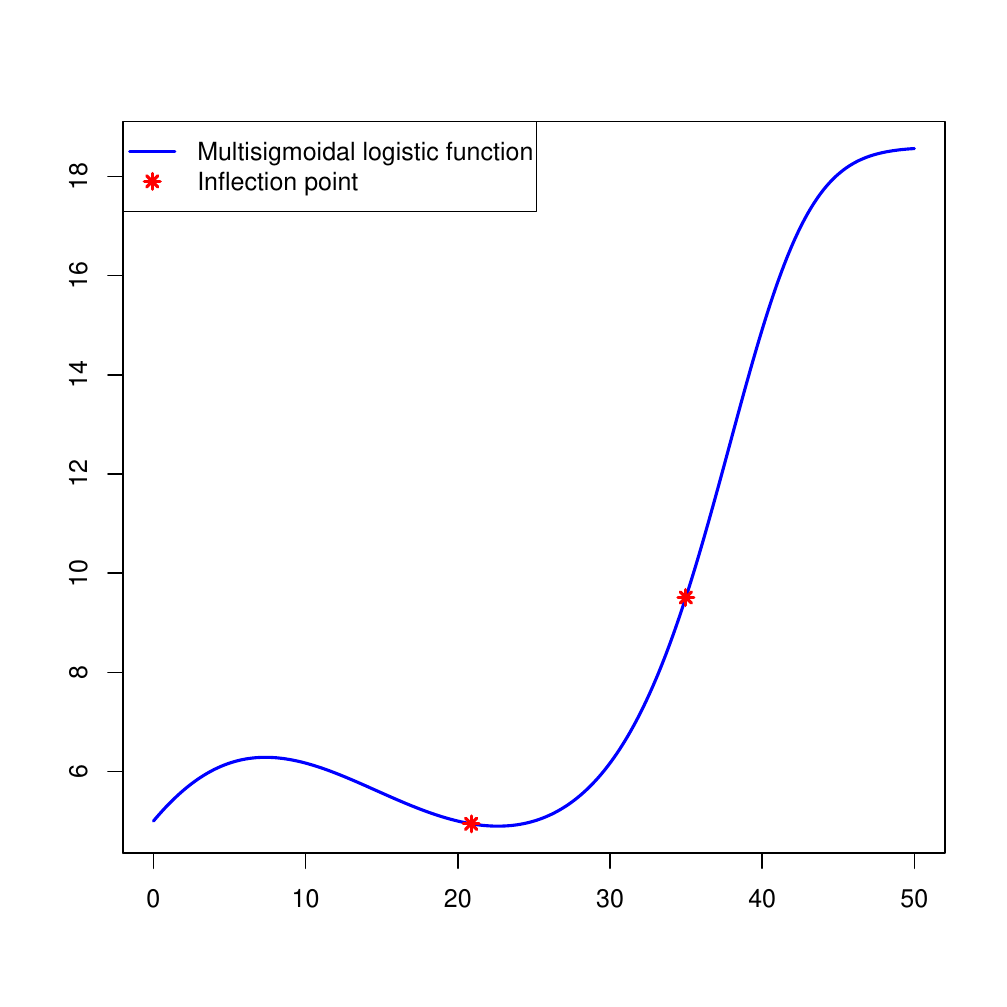}}\quad
	\subfigure[]{\includegraphics[scale=0.32]{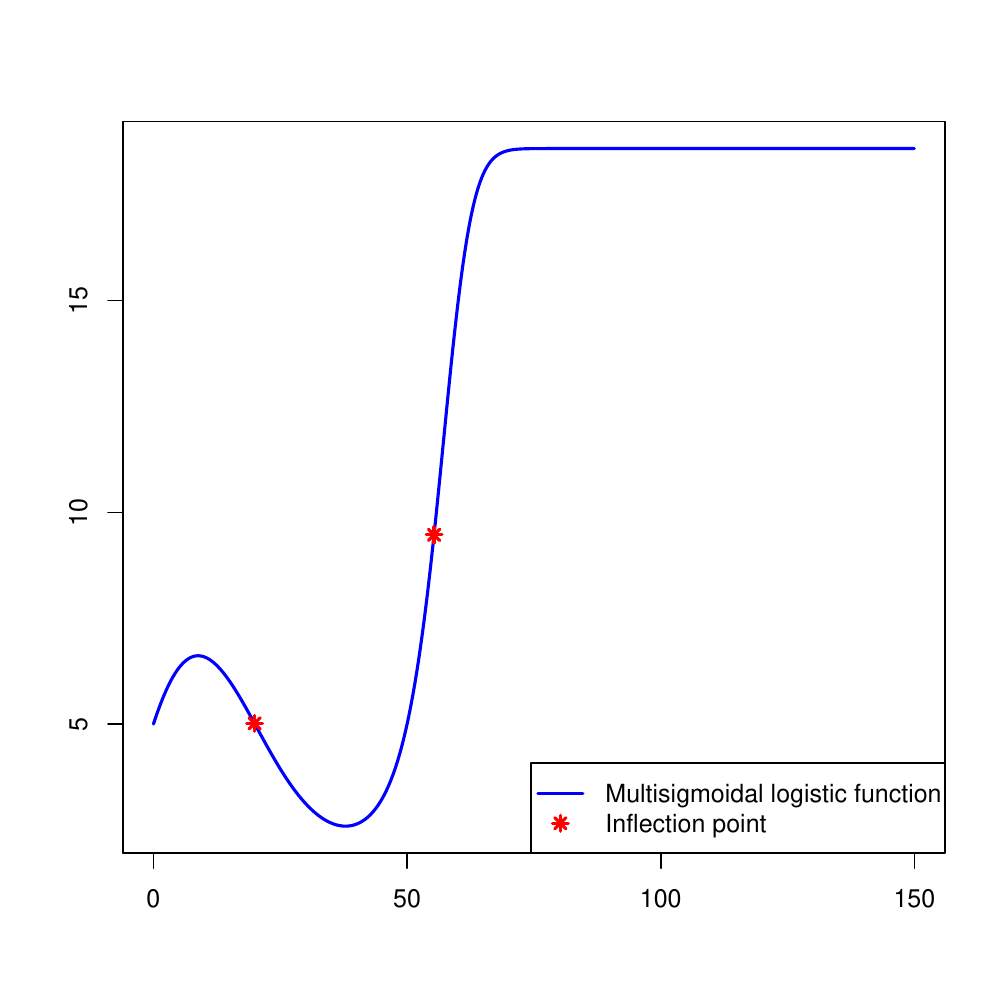}}
	\caption{The multisigmoidal logistic function and the corresponding inflection points for $t_0=0$, $l_0=5$, $\eta=e^{-1}$, $\beta_1=0.1$, (a)  $\beta_2=-0.009$ and $\beta_3=0.0002$; (b)  $\beta_2=-0.007$ and $\beta_3=0.0001$.}
	\label{fig:Figure2}
\end{figure}

	
The investigation of the inflection points in the case of multisigmoidal growth curves are of great interest. Unfortunately, since the expression of function \eqref{nuovalm} is quite complex, these points cannot be obtained explicitly, but it is possible to provide an equation in the unknown $t$ solved by the inflection points, that is
\begin{equation}\label{eqinflec}
	\frac{d}{dt}P_\beta(t)=P_\beta^2(t)\left[\frac{\eta-e^{-Q_\beta(t)}}{\eta+e^{-Q_\beta(t)}}\right].
\end{equation}
%
In Figure \ref{fig:Figure2}, the multisigmoidal logistic function and the corresponding inflection points are shown for some choices of the parameters.
%
\subsection{The corresponding diffusion process}\label{2.2}
%
In Di Crescenzo {\it et al.}\ (2020) \cite{DiCrescenzoetal2020},  a special time-dependent lognormal diffusion process $\left\{X(t);t\in I\right\}$ has been considered, with $I=[t_0,+\infty)$ and infinitesimal moments
\begin{equation}\label{infmom}
	A_1(x,t)=h_\theta(t)x,\qquad A_2(x)=\sigma^2x^2,
\end{equation}
where $h_\theta$ is defined in \eqref{htheta}, $\theta= (\eta,\beta^T )^T$ and $\sigma>0$. The aforementioned process is determined by the following stochastic differential equation, obtained from Eq.\ \eqref{odelm} by adding a multiplicative noise term,
\begin{equation}\label{sde}
	dX(t)=h_\theta(t)X(t)dt+\sigma X(t)dW(t),
	\qquad
	X(t_0)\stackrel{d}{=}X_0,
\end{equation}
where $\stackrel{d}{=}$ means equality in distribution, and
where $W(t)$ denotes a Wiener process independent from the (possibly random) initial state $X_0$, for $t\ge t_0$. We point out that this is not the only way to randomize the growth deterministic equation. Indeed, in the case of random catastrophes, it may be more appropriate to consider as a noise term a Poisson process (see for example Schlomann (2018) \cite{Schlomann2018}). The solution of Eq.\ \eqref{sde} is
\begin{equation}\label{process}
	X(t)=X_0\exp\left[H_\xi(t_0,t)+\sigma\left(W(t)-W(t_0)\right)\right],\qquad t\ge t_0
\end{equation}
with
\begin{equation}\label{H}
	H_\xi(t_0,t)=\int_{t_0}^{t}h_\theta (\tau)d\tau-\frac{\sigma^2}{2}(t-t_0)
	=\log\left[\frac{\eta+e^{-Q_\beta (t_0)}}{\eta+e^{-Q_\beta (t)}}\right]-\frac{\sigma^2}{2}(t-t_0).
\end{equation}
The existence and uniqueness of solution of the linear stochastic differential equation  (\ref{sde})
is ensured by virtue of the continuity of function $h_{\theta}(t)$
(see, for example, Arnold (1974) \cite{Arnold}).
Moreover,  if  either
$X_0$ is degenerated at $x_0$, in the sense that
$\mathbb{P}\left[X(t_0)=x_0\right]=1$,
or $X_0$ follows a lognormal distribution
$\Lambda_1\left(\mu_0,\sigma_0^2\right)$, then the finite dimensional distributions of the process are lognormal. Namely, for any $n\in\mathbb N$ and $t_0\leq t_1<\ldots<t_n$, the vector $\left(X(t_1),\dots, X(t_n)\right)^T$ follows an $n$-dimensional lognormal distribution $\Lambda_n\left(\epsilon, \Sigma\right)$, where the entries of the vector $\epsilon$ are given by
\begin{equation*}
	\epsilon_i=\mu_0+H_\xi(t_0,t_i)=\mu_0+\log\left[\frac{\eta+e^{-Q_\beta (t_0)}}{\eta+e^{-Q_\beta (t_i)}}\right]-\frac{\sigma^2}{2}(t_i-t_0),\qquad i=1,\dots,n,
\end{equation*}
and the components of the matrix $\Sigma=(\sigma_{i,j})$ are given by
\begin{equation*}
	\sigma_{i,j}=\sigma_0^2+\sigma^2\left(\min\left(t_i,t_j\right)-t_0\right),\qquad i,j=1,\dots,n.
\end{equation*}
Further,  the conditional distribution of the process follows a lognormal distribution, i.e.\ for $s<t$
\begin{equation*}
	\left[X(t)|X(s)=z\right]\sim\Lambda_1\left(\log z+\log\left[\frac{\eta+e^{-Q_\beta (s)}}{\eta+e^{-Q_\beta (t)}}\right]-\frac{\sigma^2}{2}(t-s), \sigma^2(t-s) \right).
\end{equation*}
From the above mentioned distributions, some characteristics associated to the process can be obtained
(cf.\ Di Crescenzo {\it et al.}\  (2020) \cite{DiCrescenzoetal2020}). For example,
the mean of $X(t)$ conditional on $X(t_0)=x_0$
is given by
\begin{equation}\label{conmean}
	m(t|t_0)=\mathbb E\left[X(t)|X(t_0)=x_0\right]=x_0\frac{\eta+e^{-Q_\beta(t_0)}}{\eta+e^{-Q_\beta(t)}},
	\qquad t\ge t_0.
\end{equation}
Moreover, if $X(t_0)\stackrel{d}{=}X_0$ then the mean of $X(t)$   is
\begin{equation}\label{unconmean}
	m(t)=\mathbb E\left[X(t)\right]=\mathbb E\left[X_0\right]\frac{\eta+e^{-Q_\beta(t_0)}}{\eta+e^{-Q_\beta(t)}},\qquad t\ge t_0,
\end{equation}
and the $\alpha$-percentiles for $t\ge t_0$ are
\begin{equation}\label{alphapercentile}
	C_\alpha[X(t)]=\frac{\eta+e^{-Q_\beta(t_0)}}{\eta+e^{-Q_\beta(t)}}\exp\left(\mu_0-\frac{\sigma^2}{2}(t-t_0)+z_\alpha\sqrt{\sigma_0^2+\sigma^2(t-t_0)}\right),
\end{equation}
for $0<\alpha<1$, where $z_\alpha$ is the $\alpha$-percentile of the standard normal random variable.
Note that the conditional mean \eqref{conmean} and the mean \eqref{unconmean} are multisigmoidal logistic functions of $t$, in the sense that they solve the multisigmoidal logistic equation \eqref{odelm}.
%
\section{Maximum likelihood estimations}\label{section3}
The stochastic model introduced in Section \ref{2.2} can be employed in several applications,
especially for describing real populations that exhibit a growth pattern with more than one inflection point. Clearly, in order to apply this model to real data,
the unknown parameters need to be estimated. In Section \ref{2.2} we obtained
the distribution of the diffusion process $X(t)$ defined in  \eqref{process}. Now we propose to estimate the parameters by means of the classical maximum likelihood method.
The adoption of this strategy is particularly suggested by the availability in closed form of the transition distribution
 of the process $X(t)$. Hence, we follow the same lines
introduced in Rom\'an-Rom\'an {\em et al.} (2018)  \cite{RomanTorres2018} for general lognormal  diffusion processes.
We consider a discrete sampling of $X(t)$ based on $d$ independent sample paths, with $n_i$ different observation instants for the $i$-th sample path,
i.e.\ $t_{ij}$, $j=1,\dots, n_i$, for $i=1,\dots,d$.
For simplicity, assume that the first observation time is identical for any trajectory, i.e.\ $t_{i1}=t_0$, $i=1,\dots,d$.
Moreover, let the vector {$\mathbb X_i=\left(X(t_{i1}),\dots, X(t_{in_i})\right)^T$} contain the variables of the $i$-th sample path, for $i=1,\dots, d$,
and let $\mathbb X=\left(\mathbb X_1^T|\dots| \mathbb X_d^T\right)^T$.
By supposing that $X(t_0)$ follows a one-dimensional lognormal distribution $\Lambda_1\left(\mu_1,\sigma_1^2\right)$
and by considering the transitions of the process $X(t)$, the probability density function of  $\mathbb X$ has the following expression
\begin{equation} \label{1}
	f_{\mathbb X}(x)=\prod_{i=1}^{d}\exp\left(-\frac{\left(\log x_{i1}-\mu_1\right)^2}{x_{i1}\sigma_1 \sqrt{2\pi}}\right)\cdot\prod_{j=1}^{n_i-1}\frac{\exp\left(-\frac{\left[\log\left(\frac{x_{i,j+1}}{x_{ij}}\right)-m_{\xi}^{i,j+1,j}\right]^2}{2\sigma^2\Delta_i^{j+1,j}}\right)}{x_{ij}\sigma\sqrt{2\pi \Delta_{i}^{j+1,j}}},
\end{equation}
where $x^T=\left(x_{1,1},\dots,x_{1,n_1}|\dots|x_{d,1},\dots,x_{d,n_d}\right)\in \mathbb R_+^{n+d}$ is a vector of dimension $n+d$, with
\begin{equation} \label{eq:n}
	n=\sum_{i=1}^d (n_i-1).
\end{equation}
Recalling (\ref{H}), the parameters in (\ref{1}) are given by
\begin{equation}\label{csi}
	m_{\xi}^{i,j+1,j}:=H_\xi\left(t_{ij},t_{i,j+1}\right)=\log\left[\frac{\eta+e^{-Q\beta(t_{ij})}}{\eta+e^{-Q\beta(t_{i,j+1})} }\right]-\frac{\sigma^2}{2}\Delta_{i}^{j+1,j},
\end{equation}
and
$$\Delta_i^{j+1,j}:=t_{i,j+1}-t_{ij},\qquad j=1,\dots,n_i-1,\quad i=1,\dots,d,$$
with $$\xi=\left(\theta^T,\sigma^2\right)=\left(\eta,\beta^T,\sigma^2\right).$$
In order to obtain a more manageable expression of the density \eqref{1},  the following change of variables may be considered:
\begin{equation*}
	\begin{aligned}
		&V_{0i}=X_{i1}, \qquad i=1,\dots,d\\
		&V_{ij}=\left(\Delta_i^{j+1,j}\right)^{-1/2}\log\frac{X_{i,j+1}}{X_{ij}},\qquad j=1,\dots,n_i-1,\;\; i=1,\dots, d.
	\end{aligned}
\end{equation*}
Hence, the probability density function of the vector $\mathbb V=\left[\mathbb V_0^T|\mathbb V_1^T|\dots|\mathbb V_d^T\right]=\left[\mathbb V_0^T|\mathbb V_{(1)}^T\right]^T$, with $\mathbb V_{(1)}^T=\left(\mathbb V_1^T|\dots|\mathbb V_d^T\right)$, and $\mathbb V_i^T=(V_{i1}, V_{i2}, \ldots, V_{i n_i})$,
is
\begin{equation*}
	f_{\mathbb V}(v)=\frac{\exp\left[-\frac{1}{2\sigma_1^2}
		\left(lv_0-\mu_1\mathbb I_d\right)^T\left(lv_0-\mu_1\mathbb I_d\right)\right]}{\prod_{i=1}^d v_{0i}\left(2\pi\sigma_1^2\right)^{d/2}}
	\cdot \frac{\exp\left[-\frac{1}{2\sigma^2}\left(v_{(1)}-\gamma^\xi\right)^T\left(v_{(1)}-\gamma^\xi\right)\right]}{\left(2\pi \sigma^2\right)^{n/2}}
\end{equation*}
for $v=(v_{01},\dots, v_{0d}, v_{11},\dots, v_{1(n_1-1)},\ldots, v_{d1},\dots, v_{d(n_d-1)})\in \mathbb R^{n+d}$, with $n$ defined in (\ref{eq:n}),
where $lv_0=\left(\log v_{01},\dots,\log v_{0d}\right)^T$, and $\mu_1\in \mathbb R$, $\sigma_1^2\in \mathbb R_+$,
$\mathbb I_d=(1,\dots,1)^T_{d\times 1}$, with
$\gamma^\xi=(\gamma^\xi_{11},\ldots, \gamma^\xi_{1(n_1-1)},\ldots, \gamma^\xi_{d1},\ldots, \gamma^\xi_{d(n_d-1)})^T
\in\mathbb R^{n\times 1}$ and $\gamma^\xi_{ij}=\left(\Delta_i^{j+1,j}\right)^{-1/2}m_\xi^{i,j+1,j}$,
for $j=1,\dots,n_i-1$ and $i=1,\dots,d$.
\par
By setting  $\alpha=\left(\mu_1,\sigma_1^2\right)^T$ and supposing that $\alpha$ and $\xi$ are functionally independent, the log-likelihood function is given by
\begin{equation}\label{loglik}
	L_{\mathbb V}\left(\alpha,\xi\right)
	=\tilde L_{\mathbb V}(\xi)-\frac{(n+d)}{2}\log 2\pi -\frac{d}{2}\log \sigma_1^2-\sum_{i=1}^d \log v_{0i}-\frac{\sum_{i=1}^d\left(\log v_{0i}-\mu_1\right)^2}{2\sigma_1^2},
\end{equation}
with
\begin{equation*}
	\tilde L_{\mathbb V}(\xi)=-\frac{n}{2}\log \sigma^2-\frac{Z_1+\Phi_\xi-2\Gamma_\xi}{2\sigma^2}
\end{equation*}
and
\begin{equation*}
	Z_1=\sum_{i=1}^d\sum_{j=1}^{n_i-1}v_{ij}^2,\qquad\Phi_\xi=\sum_{i=1}^d\sum_{j=1}^{n_i-1}\frac{\left(m_\xi^{i,j+1,j}\right)^2}{\Delta_i^{j+1,j}},\qquad\Gamma_\xi=\sum_{i=1}^d\sum_{j=1}^{n_i-1}\frac{v_{ij}m_\xi^{i,j+1,j}}{\left(\Delta_i^{j+1,j}\right)^{1/2}}.
\end{equation*}
The maximum likelihood estimations (MLEs) of $\alpha=\left(\mu_1,\sigma_1^2\right)^T$	
can be computed easily. Indeed, by differentiating $L_{\mathbb V}$, from (\ref{loglik}) we obtain
%
%
%
\begin{equation}\label{pardistrin}
	\hat\mu_1=\frac{1}{d}\sum_{i=1}^d \log v_{0i},\qquad \hat\sigma_1^2=\frac{1}{d}\sum_{i=1}^d \left(\log v_{0i}-\hat\mu_1\right)^2.
\end{equation}
Further on, in order to find the maximum likelihood estimates of $\xi$, two different approaches are available:\\
(i) solving the nonlinear system $\frac{\partial}{\partial \xi}\tilde L_{\mathbb V}=0,$
\\
(ii) maximizing the objective function $\tilde L_{\mathbb V}$.
\par
Hereafter, in the Sections \ref{section3.1} and \ref{section3.2} we provide a  description of the two strategies,
whereas in Section \ref{sec:simulation} we present an application to a simulation study that involves the given strategies.
\par
The availability  of the probability density function of  $\mathbb X$ in (\ref{1}) allows to obtain
explicitly the log-likelihood function given in (\ref{loglik}).
Consequently, following the maximum likelihood estimation procedure,
in Section \ref{section3.1} we obtain the associated system of equations, the final form being reported
in Eq.\ (\ref{sist3}) below. However, since such system does not have an explicit solution,  its resolution
must be obtained by  adopting numerical methods.

\subsection{Solving the nonlinear system} \label{section3.1}
Recalling that  $\theta^T=(\theta_0,\theta_1,\dots,\theta_p)=(\eta,\beta_1,\dots,\beta_p)$,
the partial derivatives of $\tilde L_{\mathbb V}$ are given by
\begin{equation*}
	\frac{\partial}{\partial \sigma^2}\tilde L_{\mathbb V}
	=-\frac{n}{2\sigma^2}+\frac{Z_1+\Phi_\xi-2\Gamma_\xi}{2\sigma^4}+\frac{Y_\xi}{2\sigma^2}-\frac{Z_2}{2\sigma^2},
\end{equation*}
\begin{equation*}
	\frac{\partial}{\partial \theta}\tilde L_{\mathbb V}=-\frac{1}{2\sigma^2}\left[\frac{\partial}{\partial \theta}\Phi_\xi-2\frac{\partial }{\partial \theta}\Gamma_\xi\right],
\end{equation*}
where
\begin{equation*}
	Y_\xi:=\sum_{i=1}^d\sum_{j=1}^{n_i-1}m_\xi^{i,j+1,j},\qquad Z_2:=\sum_{i=1}^d\sum_{j=1}^{n_i-1}v_{ij}\left(\Delta_i^{j+1,j}\right)^{1/2},
\end{equation*}
\begin{equation*}
	\frac{\partial}{\partial \theta} \Phi_\xi=\sum_{i=1}^{d}\sum_{j=1}^{n_i-1}\frac{2m_\xi^{i,j+1,j}\frac{\partial}{\partial \theta}m_\xi^{i,j+1,j}}{\Delta_i^{j+1,j}},
\end{equation*}
\begin{equation*}
	\frac{\partial}{\partial \theta}\Gamma_\xi=\sum_{i=1}^d\sum_{j=1}^{n_i-1} \frac{v_{ij}}{\left(\Delta_i^{j+1,j}\right)^{1/2}}\cdot \frac{\partial}{\partial \theta}m_\xi^{i,j+1,j},
\end{equation*}
with
$$\frac{\partial}{\partial \theta}m_\xi^{i,j+1,j}=\left(\frac{\partial}{\partial\theta_0},\dots,\frac{\partial}{\partial \theta_p}\right)m_{\xi}^{i,j+1,j}.$$	
Hence, the MLEs are the solutions of the following system of $p+2$ nonlinear equations
\begin{equation}\label{sist}
	\begin{cases}
		&-n\sigma^2+Z_1+\Phi_\xi-2\Gamma_\xi+Y_\xi\sigma^2-Z_2\sigma^2=0,
		\\[2mm]
		&\displaystyle{\sum_{i=1}^d\sum_{j=1}^{n_i-1}\frac{m_\xi^{i,j+1,j}}{\Delta_i^{j+1,j}}\cdot\frac{\partial}{\partial \theta}m_\xi^{i,j+1,j}-\sum_{i=1}^d\sum_{j=1}^{n_i-1}\frac{v_{ij}}{\left(\Delta_i^{j+1,j} \right )^{1/2}}\cdot\frac{\partial}{\partial \theta}m_\xi^{i,j+1,j}=0.}
	\end{cases}
\end{equation}
By defining the following quantities
\begin{equation}\label{lXi}
	\begin{aligned}
		&_{l}\Delta_{\theta}^{i,m}:=\frac{1}{\eta+e^{-Q_\beta(t_{im})}}\left(-t_{im}^l\right),\\
		&\bar{\delta}_{l0}:=1-\delta_{l0}=\begin{cases}
			0,\quad l=0\\
			1,\quad l\neq 0
		\end{cases},\\
		&_lD_\theta^{i,m,n}:={_l\Delta_\theta^{i,m}\left(e^{-Q_\beta(t_{im})}\right)^{\bar{\delta}_{l0}}}-{_l\Delta_\theta^{i,n}\left(e^{-Q_\beta(t_{in})}\right)^{\bar{\delta}_{l0}}},\quad m>n,
	\end{aligned}
\end{equation}
with $l=0,1,\dots,p$, the last $p+1$ equations of the system \eqref{sist} can be written as follows
\begin{equation*}
	-\sum_{i=1}^d\sum_{j=1}^{n_i-1}\frac{m_\xi^{i,j+1,j}}{\Delta_i^{j+1,j}}{_lD_\theta^{i,j+1,j}}+\sum_{i=1}^d\sum_{j=1}^{n_i-1}\frac{v_{ij}}{\left(\Delta_i^{j+1,j}\right)^{1/2}}{_lD_\theta^{i,j+1,j}}=0,\qquad l=0,1,\dots,p.
\end{equation*}
Substituting the expression \eqref{csi} of $m_\xi$ in the previous equations, one has
\begin{equation*}
	Y_l^\theta+\frac{\sigma^2}{2}W_l^\theta+X_l^\theta=0,\qquad l=0,1,\dots,p,
\end{equation*}
%
%
%
where, for any $l=0,1,\dots,p$, one has
\begin{equation*}
	\begin{aligned}
		&W_l^\theta:=\sum_{i=1}^d\sum_{j=1}^{n_i-1} {_lD_\theta^{i,j+1,j}}=\sum_{i=1}^d {_lD_\theta^{i,n_i,1}}, \\
		&Y_l^\theta:=\sum_{i=1}^d\sum_{j=1}^{n_i-1}
		\frac{1}{\Delta_i^{j+1,j}}\log\left[\frac{\eta+e^{-Q_\beta(t_{i,j+1})}}{\eta+e^{-Q_\beta(t_{ij})}}\right]{_lD_\theta^{i,j+1,j}}, \\
		&X_l^\theta:=\sum_{i=1}^d\sum_{j=1}^{n_i-1}\frac{v_{ij}}{\left(\Delta_{i}^{j+1,j}\right)^{1/2}}{_lD_\theta^{i,j+1,j}}.
	\end{aligned}
\end{equation*}
Hence, until now, the expression of the system solved by the MLEs is
\begin{equation}\label{sist2}
	\begin{cases}
		-n\sigma^2+Z_1+\Phi_\xi-2\Gamma_\xi+Y_\xi\sigma^2-Z_2\sigma^2=0\\[2mm]
		Y_l^\theta+\frac{\sigma^2}{2}W_l^\theta+X_l^\theta=0,\qquad\qquad l=0,1,\dots,p.
	\end{cases}
\end{equation}
The first equation of system \eqref{sist2} can be further simplified. Indeed, by setting
\begin{equation}\label{Z3}
	\begin{aligned}
		&\lambda_\theta ^{i,m,n}:=\log \frac{\eta+e^{-Q_\beta(tin)}}{\eta+e^{-Q_\beta(tim)}},\quad m>n\qquad Z_3:=\sum_{i=1}^d \Delta_i^{n_i,1},\\ &A_\theta:=\sum_{i=1}^d\sum_{j=1}^{n_i-1}\frac{\left(\lambda_\theta^{i,j+1,j}\right)^2}{\Delta_i^{j+1,j}},\quad B_\theta:=\sum_{i=1}^d\sum_{j=1}^{n_i-1}\frac{v_{ij}\lambda_\theta^{i,j+1,j}}{\left(\Delta_i^{j+1,j}\right)^{1/2}},\quad C_\theta:=\sum_{i=1}^d\sum_{j=1}^{n_i-1}\lambda_\theta^{i,j+1,j},
	\end{aligned}
\end{equation}
one has
\begin{equation*}
	\Phi_\xi=A_\theta+\frac{\sigma^4}{4}Z_3-\sigma^2C_\theta,
	\qquad \Gamma_\xi=B_\theta-\frac{\sigma^2}{2}Z_2,
	\qquad Y_\xi=C_\theta-\frac{\sigma^2}{2}Z_3.
\end{equation*}
Consequently, the system \eqref{sist2} finally becomes
\begin{equation}\label{sist3}
	\begin{cases}
		\sigma^2\left(n+\frac{\sigma^2}{4}Z_3\right)-Z_1-A_\theta+2B_\theta=0\\[2mm]
		Y_l^\theta+\frac{\sigma^2}{2}W_l^\theta+X_l^\theta=0,\qquad l=0,1,\dots,p.
	\end{cases}
\end{equation}
Note that \eqref{sist3} is a system of $p+2$ equations in the unknowns
contained in $\xi=(\eta,\beta_1,\dots,\beta_p,\sigma^2)$.
\begin{remark}
	For the first equation of the system \eqref{sist3} in the unknown $\sigma^2$, since $Z_3>0$, $n>0$ and
	$$
	Z_1+A_\theta-2B_\theta=\sum_{i=1}^{d}\sum_{j=1}^{n_i}\left(v_{ij}-\frac{\lambda_\theta^{i,j+1,j}}{\left(\Delta_{i}^{j+1,j}\right)^{1/2}}\right)^2\ge 0,
	$$
	the only acceptable solution is
	$$
	\sigma^2={2}\,\frac{-n+\sqrt{n^2+Z_3(Z_1+A_\theta-2B_\theta)}}{Z_3}.
	$$
\end{remark}
Clearly, since in general  system \eqref{sist3} cannot be solved analytically, then a numerical approach is needed.
Specifically, we adopt the well-known Newton-Raphson method to solve \eqref{sist3} (for instance, see Dennis and Schnabel (1996) \cite{{DenisSchnabel1996}}).
For such an iterative method, an initial approximation for the solutions of the system is needed.
It can be obtained by a procedure similar to that used by Rom\'an-Rom\'an {\em et al.}\ (2019)  \cite{Romanetal2019}.
For the initial solution of the vector $\theta= \left(\eta, \beta_1,\dots,\beta_p\right)^T$, by considering the  multisigmoidal logistic function, i.e.
\begin{equation*}
	l_m(t)=\frac{C}{\eta+e^{-Q_\beta(t)}},\qquad t\ge t_0,
\end{equation*}
it can be supposed, without loss of generality, that $t_0=0$ (see Remark 2.1 of Di Crescenzo {\em et al.}\ (2020) \cite{DiCrescenzoetal2020}), so that
\begin{equation*}
	Q_\beta(t)+\log\eta=-\log\left(\frac{C/\eta}{l_m(t)}-1\right).
\end{equation*}
Then, considering the sampling $\mathbb X=\left(\mathbb X_1^T| \dots| \mathbb X_d^T \right)^T$ defined in Section \ref{section3}, consisting of $d$ independent sample paths of the process $X(t)$,
for simplicity  we suppose that any sample path of the process has the  same number of observations,
i.e.\ $n_i=N$ for any $i=1,\dots, d$. However, the following remarks hold even in more general cases.
Moreover, let $m_j$ be the values of the mean of the sample paths at the time $t_j$, for $j=1,\dots, N$, that is
\begin{equation}
	m_j=\frac{1}{d}\sum_{i=1}^{d} x_{ij}, \qquad j=1,2,\dots,N,
	\label{eq:smean}
\end{equation}
where $x_{ij}$ is the value of the $i$-th sample path at the time $t_j$.
\par
In general, the carrying capacity $C/\eta$ is unknown.
We suppose that the observations are available over a large time interval, such that the evolution of the population is terminated over
such an interval. Hence, the carrying capacity $C/\eta$ can be approximated with the last value of the sample mean $m_N$.
This approximation can be adopted also in the other cases, since it is used just to construct an initial solution
for the parameters of the Newton-Raphson method for the estimate of $\theta$.
Thus, we can consider a polynomial regression for the pairs
$$
\left(t_j, -\log\left(\frac{m_N}{m_j}-1\right)\right), \qquad
j=1,2,\dots, N-1.
$$
The coefficients $(\hat \beta_1,\dots,\hat \beta_p,\log \hat \eta)$  of the approximating polynomial will be the initial values
for the parameters $(\beta,\log\eta)$. Thus, the initial solution for $\eta$ is given by $\hat \eta$.
\par
Finally, in order to construct the initial solution of $\sigma^2$, let us now recall that
for a lognormal distribution $Y\sim \Lambda_1(\alpha,\delta)$, one has $\log Y\sim \mathcal N(\alpha,\delta)$, so that
the quantity $2\log\frac{m}{m^g}$ gives an approximation for $\delta$,
where $m$ and $m^g$  are respectively the arithmetic sample mean and the geometric sample mean
of a random sample $(y_1,\ldots, y_n)$ from $Y$.
Hence, one has
\begin{equation*}
	\alpha\approx \frac{1}{n}\sum_{i=1}^n \log y_i,
	\qquad
	e^\alpha\approx e^{\frac{1}{n}\sum_{i=1}^n \log y_i}=\left(\prod_{i=1}^n y_i\right)^{1/n}=m_g.
\end{equation*}
Since   $\mathbb E[Y]=e^{\alpha+\delta/2}$ is estimated by the sample mean $m$, we have
\begin{equation*}
	m\approx e^{\alpha+\delta/2}\approx m_g \cdot e^{\delta/2},
	\qquad
	\delta\approx 2\log \frac{m}{m_g}.
\end{equation*}
As a consequence, in our setting an estimate for $\sigma_0^2+\sigma^2t_j$ is given by
$$
\sigma_j^2=2\log\frac{m_j}{m_j^g},
\qquad j=1,\dots,N,
$$
where $m_j$ and $m_j^g$ denote respectively the arithmetic and the geometric sample mean
of the observations performed at the time $t_j$.
Hence, an initial approximation for $\sigma^2$ can be obtained by performing a simple linear regression of  $\sigma_j^2-\sigma_0^2$ against $t_j$.
\par
In conclusion,
in order to obtain the maximum likelihood estimates of the parameters
contained in $\xi=(\eta,\beta_1,\dots,\beta_p,\sigma^2)$,
the steps of the proposed strategy to solve the system  \eqref{sist3} are:
\begin{itemize}
	\item[(i)]
	finding an initial solution for the parameters $\eta$ and $\beta$ with a polynomial regression of $-\log\left(\frac{m_N}{m_j}-1\right)$ against $t_j$, for any $j=1,\dots, N-1$;
	\item[(ii)]
	finding an initial solution for $\sigma^2$ with a simple linear regression of $\sigma_j^2-\sigma_0^2$ against $t_j$,
	with $\sigma^2_j=2\log\frac{m_j}{m_j^g}$, for any $j=1,\dots,N$ {and where $\sigma_0^2$ can be obtained by means of the second of Eqs.\ \eqref{pardistrin}};
	\item[(iii)]
	using the Newton-Raphson method to solve the system \eqref{sist3}, with the initial solutions determined at steps (i) and (ii).
\end{itemize}
The adoption of the above strategy requires to start from good initial solutions for the unknown parameters.
Unfortunately, even in this case it is not always possible to guarantee the convergence of this method.
For this reason, recently various procedures have been proposed aimed at addressing
the maximization of the likelihood function, by viewing this as a direct optimization problem.
Indeed, there is a wide range of stochastic metaheuristic methods, which can be classified into two large families:
those based on trajectories and those based on swarms.
Hereafter, in Section \ref{section3.2} we employ one of the most widely used, the Simulated Annealing.
This method requires necessarily to bound the parametric space,
and this matter is the object of Section \ref{sect:pspace}.

%
\subsection{Maximizing the log-likelihood function}\label{section3.2}
Let us now illustrate a strategy based on Simulated Annealing (S.A.) and
finalized to obtain the MLEs for the parameters of the process (\ref{process}).
We first provide a brief description of this method in Section \ref{sect:SA}.
Then, in Section \ref{sect:pspace} we describe a suitable criterion to restrict
the parametric space, this being essential to apply  the S.A.\ method in the remainder of the paper.

\subsubsection{Brief notes on Simulated Annealing}\label{sect:SA}
%
The aim of this section is to determine the MLEs by using the S.A.\ algorithm. The aforementioned method, introduced by Kirkpatrick {\em et al.}\ (1983) in \cite{Kirkpatricketal1983}, is a meta-heuristic optimization algorithm used for problems like finding $\displaystyle \arg\min_{\theta\in \Theta} f(\theta)$. It is considered more suitable with respect to other numerical algorithms since it needs less restrictive conditions regarding the regularity of the domain $\Theta$ and the analytical properties of the objective function $f$.
The algorithm works such that in every step a random point is chosen in the solution space. If the new solution is better than the previous one, then the latter is replaced. Otherwise, if the new solution is worse than the previous, then the latter may be replaced with a probability rate $\rho=\min\{\exp(-\Delta f/T),1\}$ which depends on the increase of the objective function $\Delta f=f(\xi)-f(\theta_0)$ and on a suitable scale factor $T$, that is named `temperature' in agreement with the metallurgical process of annealing that inspired this algorithm. We recall that the S.A.\ is successful because it avoids local minima. In recent years it has been widely used in the context of estimation in diffusion processes (see, for example Luz Sant'Ana
{\em et al.} (2018) \cite{Luzetal2018} and  Rom\'an-Rom\'an and Torres-Ruiz (2015) \cite{RomanTorres2015}).
\par
In this context, the algorithm works in the following way. It begins with an initial choice $\theta_0$ for the parameters of interest, then $\xi$ is generated from an uniform distribution
in a neighborhood $\nu(\theta_0)$ of $\theta_0$.
Then, a new value $\theta_1$ of $\theta$ is obtained in such a way
\begin{equation*}
	\theta_1=\begin{cases}
		\xi,&\qquad\text{ with probability } \rho\\
		\theta_0,&\qquad\text { with probability } 1-\rho.
	\end{cases}
\end{equation*}
Consequently, if $f(\xi)\le f(\theta_0)$, then $\rho=1$ and therefore $\theta_0$ is replaced by $\xi$. Otherwise, if $f(\xi)>f(\theta_0)$, then $\xi$ may be accepted anyway with probability $\rho\in (0,1)$. The temperature $T$ is defined in such a way that at the beginning the probability of accepting $\xi$ is high, and during the execution of the algorithm the function $T$ decreases.  The initial temperature $T_0$ must be sufficiently large so that the algorithm accept the solutions which let the objective function increases with a large probability $p_0$. In literature, the choices of the initial parameters are usually $p_0=0.9$ and $T_0=-\Delta f^+/\log p_0$, where $\Delta f^+$ denotes the average increase of the objective function in an application test where all the solutions which cause an increase are accepted. The cooling process which defines the temperature $T$ is usually chosen of geometric type, i.e.\  $T_i=\gamma T_{i-1}$ for $i=1,2,\dots$. Usually the constant $\gamma $ is chosen among $0.8$ and $0.99$ in order to have a slow cooling procedure. In our case, we set  $\gamma=0.95$.
In any iteration of the algorithm, a chain of $L$ new solutions is obtained, for $L=50$.
As required, the algorithm stops when at least one of the following rules is satisfied: (i) the last $L$ obtained values are equal, (ii) the maximum number of iterations ($1000$, in our case) is attained,  (iii) the final temperature $T_F=10^{-7}$ is reached.
%
\subsubsection{Bounding the parametric space}\label{sect:pspace}
S.A.\  needs a restriction of the solution space $\Theta$, namely the set which contains the parameters
$\xi=(\eta,\beta^T,\sigma^2)$. Until now, this space is continuous and unbounded, since
\begin{equation*}
	\Theta=\left\{(\eta,\beta^T,\sigma^2): \eta>0, \beta_1,\dots,\beta_{p-1}\in \mathbb R, \beta_p>0, \sigma^2>0\right\}.
\end{equation*}
We consider $0<\sigma<0.1$ so that the simulated sample paths are less variable around the sample mean, and
thus the multisigmoidal logistic profile is advisable. For the parameters $\beta=\left(\beta_1,\dots,\beta_p\right)^T$, we  find the confidence intervals by using the data of the polynomial regression performed previously to find the initial solutions. More in detail, it is known that the carrying capacity of the multisigmoidal logistic model with $t_0=0$ is $l_0 \left(1+\frac{1}{\eta}\right)$ (see Eq.\ \eqref{lim}). The carrying capacity can be approximated with the last value of the sample mean, whereas the initial value $l_0$ with the first value of the sample mean (\ref{eq:smean}), so that one has
\begin{equation}\label{star}
	m_N\approx m_1\left(1+\frac{1}{\eta}\right).
\end{equation}
From Eq.\ \eqref{star}, it easily follows
$\eta\approx \left(\frac{m_N}{m_1}-1\right)^{-1}$
and thus an approximation of $\eta$ is
$$
\hat \eta=\left(\frac{m_N}{m_1}-1\right)^{-1}.
$$
Considering Eqs.\  \eqref{nuovalm} and \eqref{lim}, for $t_0=0$ one has
\begin{equation*}
	l_m(t)= \frac{C}{\eta+e^{-Q_\beta(t)}},
\end{equation*}
so that
\begin{equation*}
	Q_\beta(t)=-\log\left(\frac{C}{l_m(t)}-\eta\right).
\end{equation*}
Hence, by replacing $\eta$ with its estimate $\hat \eta$, we can use the resulting confidence intervals of the parameters of the polynomial regression as intervals of variation for the parameters $\beta$ of the diffusion process. We adopt a confidence level equal to $0.999$, to attain a high probability that the true parameters $\beta$ belong to the computed intervals.
\par
In order to approximate the range of variation of $\eta$, from Eq.\ \eqref{star} we have that the last value of the $i$-th sample path satisfies
\begin{equation*}
	x_{i,n_i}\approx x_{i,1}+\frac{x_{1,i}}{\eta}, \qquad i=1,2,\dots,d,
\end{equation*}
where $x_{i,j}$ with $i=1,2,\dots,d$ and $j=1,2,\dots,n_i$ are the sample data.
Hence, for the range of variation of $\eta$ one has  $\eta\in(a,b)$, where
\begin{equation}\label{intereta}
	a:=\min_{1\leq i\leq d}\left(\frac{x_{i,n_i}}{x_{i,1}}-1\right)^{-1},
	\qquad
	b:=\max_{1\leq i\leq d}\left(\frac{x_{i,n_i}}{x_{i,1}}-1\right)^{-1}.
\end{equation}
\par
In conclusion, the following bounded intervals are employed:
\\
$\bullet$ \ for $\beta_1,\dots,\beta_p$ we consider the confidence intervals of the coefficients of the polynomial regression of $-\log\left[\left(\frac{m_N}{m_j}-1\right) \hat \eta\right]$ against $t_j$, for $j=1,\dots,N$, where $\hat\eta=\displaystyle \left(\frac{m_N}{m_1}-1\right)^{-1}$,
\\
$\bullet$ \ for $\eta$ we consider the interval $I_\eta=\left(a,b\right)$, with $a$ and $b$ defined in \eqref{intereta},
\\
$\bullet$ \ for $\sigma^2$ we consider the interval $I_{\sigma^2}=\left(0,0.01\right)$.
%
\subsection{Asymptotic distribution of the MLEs}\label{distribMLE}
On the ground of the results given in Section 5 of Rom\'an-Rom\'an {\em et al.}\ (2018) \cite{RomanTorres2018}, in this section we aim to determine the asymptotic distribution of the MLEs (i) of the parameters $\mu_1,\sigma_1^2$ of the initial distribution,  and (ii) of the parameters	$\xi=(\eta,\beta^T,\sigma^2)$ of the process.
\par
(i) \ The {exact} distribution of $\widehat\mu_1$ is normal $\mathcal{N}\left(\mu_1,\frac{\sigma_1^2}{d}\right)$, whereas the {exact} distribution of $d\,\frac{\widehat\sigma_1^2}{\sigma_1^2}$ is chi-square $\chi_{d-1}^2$,
cf.\ Rom\'an-Rom\'an {\em et al.}\ (2018) \cite{RomanTorres2018}.
\par
(ii) \ The asymptotic distribution of $\widehat\xi$ is a $(p+2)$-dimensional normal distribution with mean $\xi$ and covariance matrix $I(\xi)^{-1}$, i.e.\ $\mathcal N_{p+2}\left(\xi, I(\xi)^{-1}\right)$, where $I(\xi)$ denotes Fisher's information matrix of $\xi$. For the diffusion process $X(t)$ with a multisigmoidal logistic mean, $I(\xi)\in\mathbb{R}^{(p+2)\times(p+2)}$ can be expressed as
\begin{equation}\label{infmatrix}
	I(\xi)=\frac{1}{\sigma^2}\begin{pmatrix}
		\Xi_\xi & -\frac{1}{2}\left(\frac{\partial}{\partial \theta}\gamma_{\xi} \right )\\[1.5ex]
		-\frac{1}{2}\left(\frac{\partial}{\partial \theta}\gamma_{\xi}\right )^T& \frac{n}{2\sigma^2}-\frac{Z_3}{4},
	\end{pmatrix},
\end{equation}
where $\Xi_\xi\in\mathbb R^{(p+1)\times(p+1)}$ is given by
\begin{equation*}
	\Xi_\xi=\sum_{i=1}^{d}\sum_{j=1}^{n_i-1}(\Delta_i^{j+1,j})^{-1}\left(\frac{\partial}{\partial \theta}m_\xi^{i,j+1,j}\right)\left(\frac{\partial}{\partial \theta}m_\xi^{i,j+1,j}\right)^T,
\end{equation*}
with
\begin{equation*}
	\left(\frac{\partial}{\partial \theta}m_\xi^{i,j+1,j}\right)^T=\left(-{_0D_\theta^{i,j+1,j}},\dots,-{_pD_\theta^{i,j+1,j}}\right),
\end{equation*}
and ${_lD_\theta}^{i,j+1,j}$ is defined in the third of Eqs.\ \eqref{lXi}. Moreover,  $\frac{\partial}{\partial \theta}\gamma_{\xi}\in\mathbb{R}^{(p+1)\times 1}$ is defined as
\begin{equation*}
	\frac{\partial}{\partial \theta}\gamma_{\xi}=\sum_{i=1}^d\sum_{j=1}^{n_i-1}\frac{\partial}{\partial \theta}m_\xi^{i,j+1,j}.
\end{equation*}
Finally, $Z_3$ is given in the second of Eqs.\ \eqref{Z3}. We point out that the matrix \eqref{infmatrix} will be used in Section \ref{sec:simulation} to determine the asymptotic variances for the estimates of the parameters and an approximation of the confidence intervals. Indeed, by applying the delta method (cf.\ Oehlert (1992) \cite{Oehlert1992}), any $q$-parametric function $g(\hat \xi)$ with $q\le p+2$ asymptotically has a $q$-dimensional normal distribution, i.e.\ (cf.\ Rom\'an-Rom\'an {\em et al.}\ (2018) \cite{RomanTorres2018})
\begin{equation*}
	\mathcal N_{q}\left(g(\xi), \nabla g(\xi)^T I(\xi)^{-1}\nabla g(\xi)\right),
\end{equation*}
where $\nabla g(\xi)$ is the vector of partial derivatives of $g(\xi)$ with respect to $\xi$.
\par
In the following section we address a relevant problem for the applications, namely
the FPT problem of the diffusion process $X(t)$ through a continuous boundary.
Subsequently, in Section \ref{sec:simulation} we adopt a simulation-based approach as the  basis of
both computational methods described so far, namely the Newton-Raphson method and the S.A.\ method.
The estimates of the parameters obtained through these methods are then used to perform
inference on the FPT density.
%
\section{First-passage-time problem}\label{sec:FPT}
%
The FPT problem of a stochastic process $X(t)$ through a boundary $S(t)$ is a problem of great interest in many fields of application, such as medicine, biology  or mathematical finance, since the threshold $S(t)$ may represent a critical value of the modeled population size. Considering a  stochastic process $\left\{X(t);t_0\le t\le T\right\}$, the FPT of the process $X(t)$ through the continuous boundary $S(t)$, given $X(t_0)=x_0$, is defined as the following random variable
\begin{equation*}
	T=\begin{cases}
		\displaystyle \inf_{t\ge t_0}\left\{X(t)>S(t)|X(t_0)=x_0\right\}, \quad x_0<S(t_0)\\
		\displaystyle\inf_{t\ge t_0}\left\{X(t)<S(t)|X(t_0)=x_0\right\}, \quad x_0>S(t_0).
	\end{cases}
\end{equation*}
Finding the expression of the distribution of the variable $T$ is hard in general. However, in literature there are several studies for particular types of processes, for example diffusion processes. It has been shown that if $S(t)$ is a continuous and differentiable function, then the density of $T$, denoted by $g\left(S(t),t|x_0,t_0\right)$, solves a II-kind Volterra equation (cf.\ Eq.\ (2.4) of Buonocore {\em et al.}\ (1987) \cite{Buonocoreetal1987}). The aforementioned Volterra equation has an explicit solution only for certain special boundaries (see for example Sections 2.3 and 4.3 of Giorno and Nobile (2019) \cite{GiornoNobile2019} in which the FPT density through special boudaries has been obtained for the restricted Gompertz-type diffusion processes). In certain instances, it is appropriate to adopt numerical procedures in order to approximate its solution. To this aim Buonocore {\em et al.}\ (1987)  \cite{Buonocoreetal1987} proposed a simple but efficient algorithm,  based on the composite trapezoidal formula.
More in detail, Theorem 4 of  Buonocore {\em et al.}\ (1987) \cite{Buonocoreetal1987} proves the convergence of the approximated FPT density to the theoretical one.
However, the application of the proposed numerical procedure requires (i) the choice of a suitable step $h$ of integration which ensures a good approximation of the real solution, (ii) the choice of an initial time instant $t_0$ and (iii) the choice of the final time instant $T=t_0+Nh$.
Rom\'an-Rom\'an {\em et al.}\ (2008) in \cite{Romanetal2008} studied the problems related to the practical application of the numerical procedure. The first problem is linked with a suitable choice of $h$. Indeed, taking into account the result of Theorem 4 of Buonocore \textit{et al.}\ (1987) \cite{Buonocoreetal1987}, it is easy to note that the convergence is ensured when $h\to 0^+$. Consequently, the value of $h$ should be small enough, but sufficiently far from $0$. Indeed, if $h$ is  excessively small, then the computational cost may increase in vain because, with a larger integration step,  a similar approximation may be obtained with a smaller number of iterations. On the other hand, if $h$ is excessively large, the approximation may be unsatisfactory. These problems depend on the localization of the FPT $T$, and may be solved if the range of variation of $T$  is known. For this reason, Rom\'an-Rom\'an {\em et al.}\ (2008) in \cite{Romanetal2008} introduced a function, called `FPT location' (FPTL), finalized to obtain, from a heuristic point of view, the range of variation of $T$. Specifically, the FPTL function is defined as follows
\begin{equation*}
	\begin{aligned}
		FPTL(t)&=\begin{cases}
			\mathbb P\left[X(t)>S(t)|X(t_0)=x_0\right],\quad x_0<S(t_0)\\
			\mathbb P\left[X(t)<S(t)|X(t_0)=x_0\right],\quad x_0>S(t_0)
		\end{cases}\\
		&=\begin{cases}
			1-F\left(S(t),t|x_0,t_0\right),\quad x_0<S(t_0)\\
			F\left(S(t),t|x_0,t_0\right),\quad x_0>S(t_0),
		\end{cases}
	\end{aligned}
\end{equation*}
where $F(x,t|x_0,t_0)$ is the transition distribution of the process $X(t)$. Referring to the diffusion process
with infinitesimal moments given by Eq.\ \eqref{infmom},
and by considering a fixed and constant boundary $S>x_0$, given $X(t_0)=x_0$, the FPTL function for the process $X(t)$ is given by
\begin{equation*}\label{fptl}
	FPTL(t)=1-\Phi\left(C(t)\right),
\end{equation*}
where $\Phi$ is the standard normal distribution and
\begin{equation*}
	C(t)=\frac{1}{\sigma\sqrt{t-t_0}}\left\{\log S-\log x_0 -\log\left[\frac{\eta+e^{-Q_\beta(t_0)}}{\eta+e^{-Q_\beta(t)}}\right]+\frac{\sigma^2}{2}(t-t_0)\right\}.
\end{equation*}
%
%
%
The information provided by the FPTL function is relevant for an efficient application of the algorithm proposed by Buonocore {\it et al.}\ (1987) in \cite{Buonocoreetal1987}. Indeed, thanks to the FPTL function, an adaptive step of integration can be obtained. In this way, the execution time of the algorithm 
is reduced.
%
%
%
\begin{example}{\rm
		Let $X(t)$ be a diffusion process with infinitesimal moments   \eqref{infmom}, with $p=3$, $Q_\beta(t)=0.1t-0.009t^2+0.0002t^3$, $\eta=e^{-1}$, $\sigma=0.01$, $t_0=0$, and $X_0=5$ a.s. See Figure \ref{fig:Figure3} for the plot of 100 simulated sample paths of the process. Let us study the FPT density through the fixed boundary $S=3\, x_0=15$, by using the information provided by the FPTL function and the \textbf{\textsf{R}} package \textsf{fptdApprox} (for references, see Rom\'an-Rom\'an {\it{et al.\ }} (2012) \cite{Romanetal2012}, (2014) \cite{Romanetal2014} and \cite{fptdApprox}).
		\begin{figure}[t]
			\centering
			\includegraphics[scale=0.3]{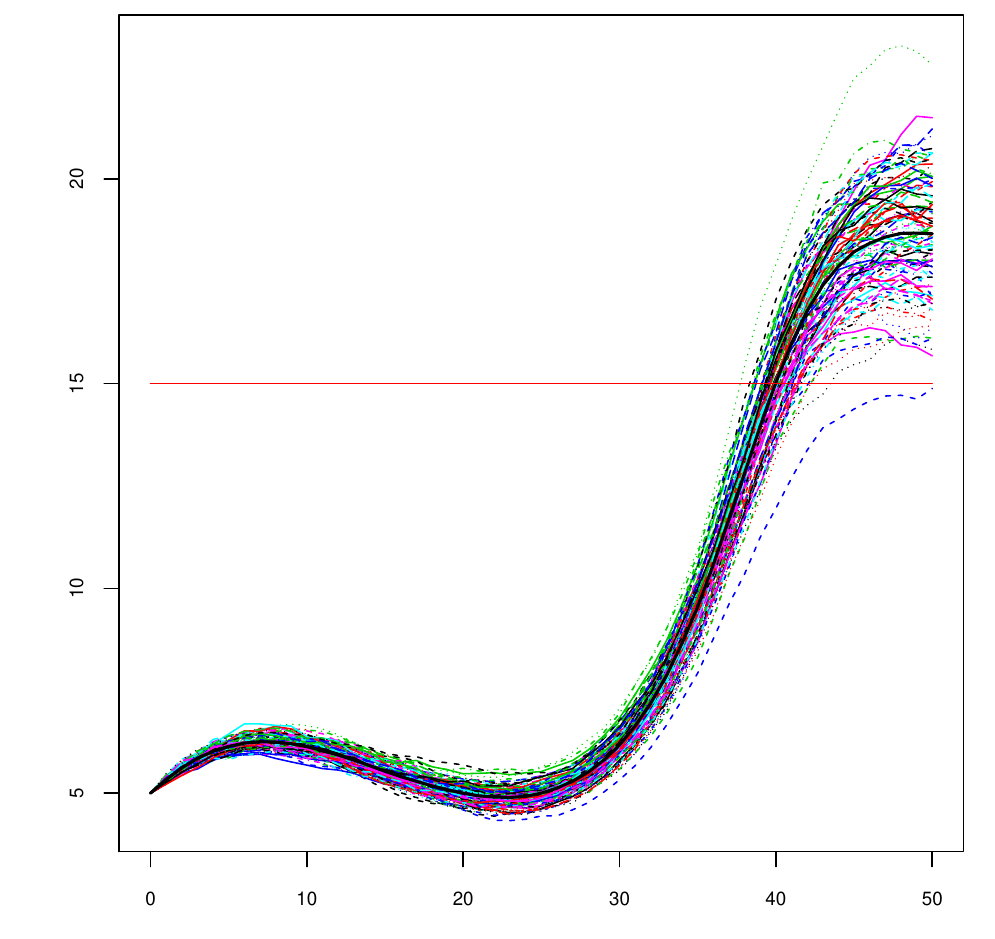}
			\caption{100 simulated sample paths of the process $X(t)$ with $p=3$, $Q_\beta(t)=0.1t-0.009t^2+0.0002t^3$, $\eta=e^{-1}$, $\sigma=0.01$, $t_0=0$ and $x_0=5$. The black line represents the sample mean of the process, while the red line represents the  boundary $S=15$.}
			\label{fig:Figure3}
		\end{figure}
		Figure \ref{fig:Figure4}-(a) shows the FPTL function (obtained by means the function \textsf{FPTL} of the package \textsf{fptdApprox}), whereas
		\begin{figure}[t]
			\centering
			\subfigure[]{\includegraphics[scale=0.35]{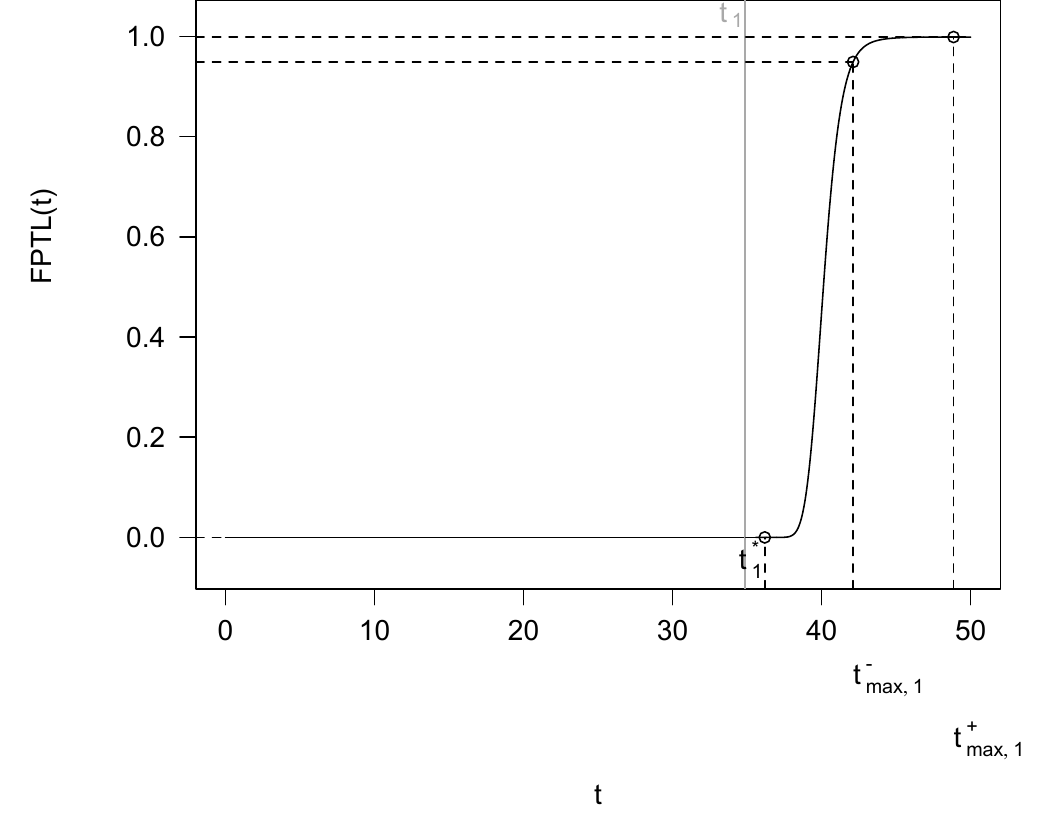}}
			\subfigure[]{\includegraphics[scale=0.35]{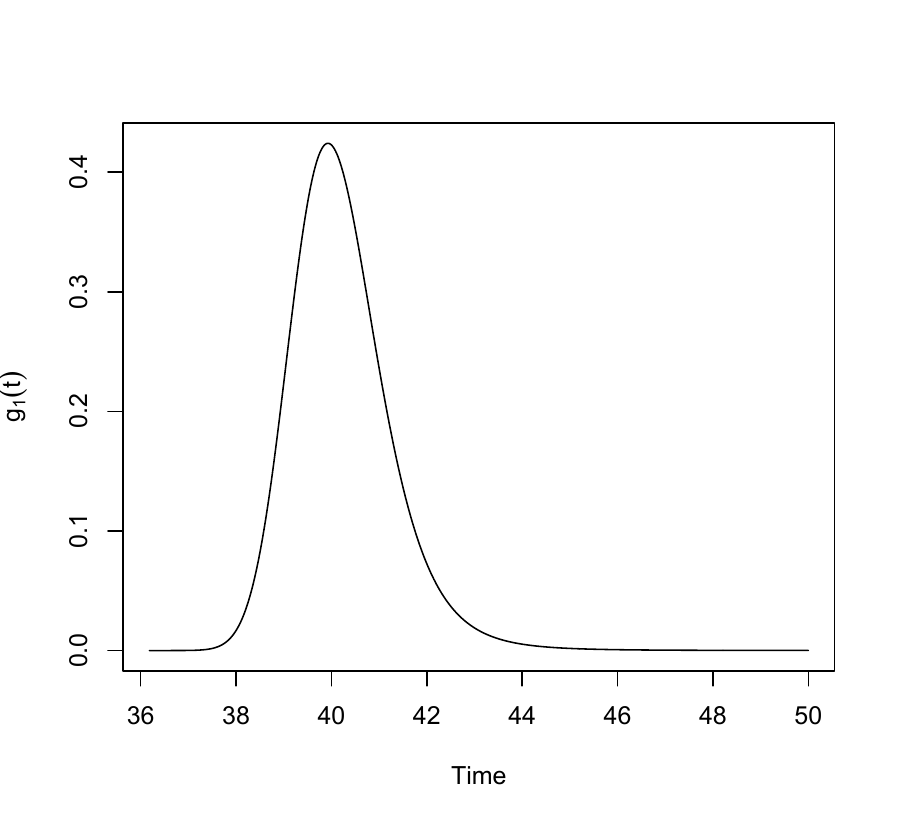}}
			\caption{(a) The FPTL function and (b) the approximated FPT density of the process $X(t)$ through the constant boundary $S=15$, for the same assumptions of Figure \ref{fig:Figure3}.}
			\label{fig:Figure4}
		\end{figure}
		%
		the approximated FPT density (obtained using
		the package \textsf{fptdApprox}) is plotted in Figure \ref{fig:Figure3}-(b).
		Other useful quantities related to the FPT density are given in Table \ref{Tab:table4}.
		\begin{table}[t]
			\caption{The mean, the standard deviation, the mode, the first, the fifth and the ninth decile of the FPT of the process $X(t)$ through the boundary $S=15$.}
			\label{Tab:table4}
			\centering
			\begin{tabular}{l|l|l|l|l|l}
				mean 			& st. dev.                	& mode                	& $1^{st}$ decile                   & $5^{th}$ decile  	& $9^{th}$ decile \\ \hline
				$40.18765$ 		&$1.568392$ 							&$39.92321$ 			&$39.02346$ 						&$40.11264$			& $41.58065$
			\end{tabular}
	\end{table} }
\end{example}
%
\section{Simulation}\label{sec:simulation}
\begin{figure}[t]
	\centering
	\hspace*{-1cm}
	\subfigure[]{\includegraphics[scale=0.35]{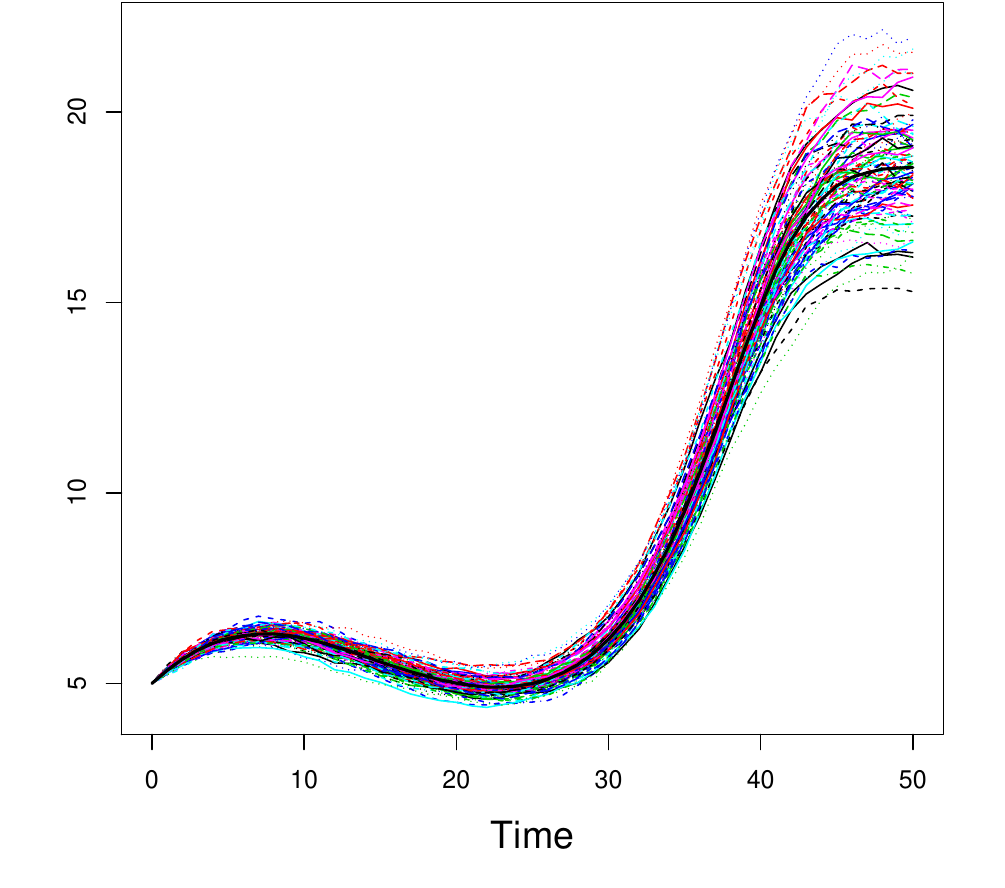}}\quad
	\subfigure[]{\includegraphics[scale=0.35]{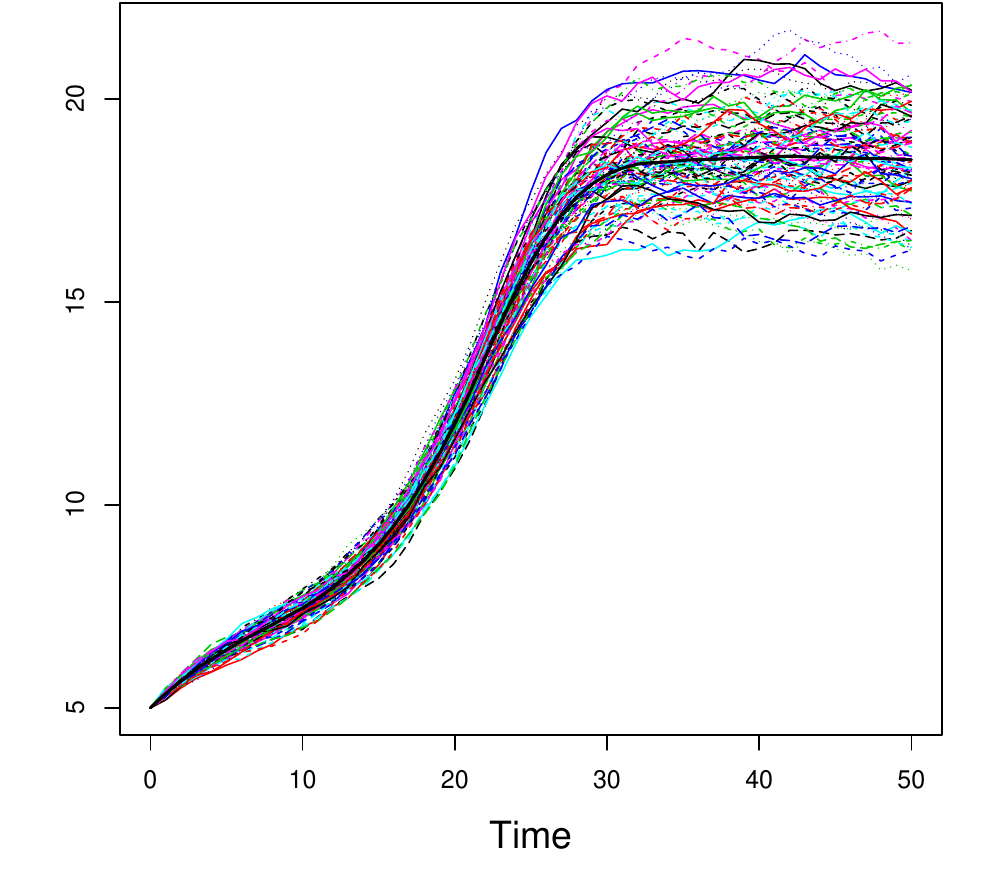}}
	\caption{100 simulated sample paths of the diffusion process for $\sigma=0.01$, $\eta=e^{-1}$ and (a) $\beta_1=0.1$, $\beta_2=-0.009$, $\beta_3=0.0002$ and (b) $\beta_1=0.1$, $\beta_2=-0.007$, $\beta_3=0.0003$ (simulation study).}
	\label{fig:Figure5}
\end{figure}

%
In Section \ref{section3}, two procedures have been introduced  to obtain the MLEs of the parameters involved in the diffusion process \eqref{process}. The former procedure is based on the numerical resolution of a system of nonlinear equations, whereas the latter is based on the application of S.A.\ algorithm. In this section, a simulation study is developed to verify the validity of the two aforementioned procedures.
We consider the diffusion process $X(t)$ with infinitesimal moments \eqref{infmom}, for $p=3$, and  $\beta_1\in\left\{0.1,0.5\right\}$, $\beta_2\in\left\{-0.009,-0.007\right\}$, $\beta_3\in\left\{0.0002, 0.0004\right\}$, $\eta\in\left\{e^{-1},e^{-3}\right\}$ and $\sigma\in\left\{0.01,0.05\right\}$. These choices of the parameters are performed
arbitrarily, to obtain different patterns of the growth curve. For example, the choice $\beta_1=0.1$, $\beta_2=-0.009$, $\beta_3=0.0002$, $\eta=e^{-1}$ refers to the case of a non monotonous multisigmoidal logistic function, whereas the choice $\beta_1=0.1$, $\beta_2=-0.007$, $\beta_3=0.0003$ and $\eta=e^{-1}$ to the case of an increasing multisigmoidal logistic curve (see Figure \ref{fig:Figure5}). To estimate the parameters in $\xi$, we consider the $32$ combinations of the values of the parameters listed in Table \ref{Tab:table1}, with $x_0=5$ in every case. For each case, we simulate $200$ sample paths of $X(t)$, by generating $501$ simulated points at equidistant times for  $0\leq t\leq 50$.
%
\begin{table}[t]
	\caption{The values of the parameters (simulation study).}
	\label{Tab:table1}
	\tiny
	\begin{tabular}{llllll}
		case no. & $\beta_1$                 & $\beta_2 $                  & $\beta_3$                   & $\eta$                      & $\sigma$ \\ \hline
		1         & $0.1$ 					& $-0.009$ 						& $0.0002$					 & $e^{-1}$ 				& $0.01$ \\ 
		2         & $''$                    & $''$                          & $''$                          & $''$                          & $0.05$ \\ 
		3         & $''$                        & $''$                          & $''$                          & $e^{-3}$ & $0.01$ \\ 
		4         & $''$                        & $''$                          & $''$                         &  $''$                         & $0.05$ \\ 
		5         & $''$                        & $''$                          & $0.0003$ 						&$e^{-1}$ & $0.01$ \\ 
		6         & $''$                        & $''$                          & $''$                          &  $''$                   & $0.05$ \\ 
		7         & $''$                        & $''$                          & $''$                          & $e^{-3}$ & $0.01$ \\ 
		8         & $''$                       	& $''$                        &  $''$                         &  $''$                         & $0.05$ \\ 
		9         & $''$                        & $-0.007$  						& $0.0002$ 						& $e^{-1}$ & $0.01$ \\ 
		10        & $''$                        & $''$                          & $''$                          &     $''$                      & $0.05$ \\ 
		11        & $''$                        & $''$                          & $''$                          & $e^{-3}$ & $0.01$ \\ 
		12        & $''$                        & $''$                          & $''$                          &         $''$                  & $0.05$ \\ 
		13        & $''$                        & $''$                          & $0.0003$ 						&$e^{-1}$ & $0.01$ \\ 
		14        & $''$                        & $''$                          & $''$                          &        $''$                   & $0.05$ \\ 
		15        & $''$                        & $''$                          & $''$                          &$e^{-3}$ & $0.01$ \\ 
		16        & $''$                        & $''$                          & $''$                          &            $''$               & $0.05$ \\ \hline
	\end{tabular}
	\begin{tabular}{llllll}
		case no. & $\beta_1$                 & $\beta_2 $                  & $\beta_3$                   & $\eta$                      & $\sigma$ \\ \hline
		17         & $0.5$ 					& $-0.009$ 						& $0.0002$					 & $e^{-1}$ 				& $0.01$ \\ 
		18         & $''$                    & $''$                          & $''$                          & $''$                          & $0.05$ \\ 
		19       & $''$                        & $''$                          & $''$                          & $e^{-3}$ & $0.01$ \\ 
		20         & $''$                        & $''$                          & $''$                         &  $''$                         & $0.05$ \\ 
		21         & $''$                        & $''$                          & $0.0003$ 						&$e^{-1}$ & $0.01$ \\ 
		22         & $''$                        & $''$                          & $''$                          &  $''$                   & $0.05$ \\ 
		23         & $''$                        & $''$                          & $''$                          & $e^{-3}$ & $0.01$ \\ 
		24         & $''$                       	& $''$                        &  $''$                         &  $''$                         & $0.05$ \\ 
		25         & $''$                        & $-0.007$  						& $0.0002$ 						& $e^{-1}$ & $0.01$ \\ 
		26        & $''$                        & $''$                          & $''$                          &     $''$                      & $0.05$ \\ 
		27        & $''$                        & $''$                          & $''$                          & $e^{-3}$ & $0.01$ \\ 
		28        & $''$                        & $''$                          & $''$                          &         $''$                  & $0.05$ \\ 
		29        & $''$                        & $''$                          & $0.0003$ 						&$e^{-1}$ & $0.01$ \\ 
		30        & $''$                        & $''$                          & $''$                          &        $''$                   & $0.05$ \\ 
		31        & $''$                        & $''$                          & $''$                          &$e^{-3}$ & $0.01$ \\ 
		32        & $''$                        & $''$                          & $''$                          &            $''$               & $0.05$ \\ \hline
	\end{tabular}
\end{table}
\par
The remainder of this section is organized as follows: (a) since the degree of the polynomial $Q_\beta$ is unknown a priori, we propose the use of the strategy described in Rom\'an-Rom\'an {\em et al.}\ (2019) \cite{Romanetal2019}, by increasing the degree until the goodness of fit is optimal; (b) considering the degree obtained at the step (a), we use the two procedures described in Sections \ref{section3.1} and \ref{section3.2} to find the MLEs of the parameters.
\par
The choice of the best degree of the polynomial $Q_\beta$ is performed under the goodness of fit criteria based on the four following measures:
\begin{itemize}
	\item[(i)] the absolute relative error ($RAE$) between the sample mean and the estimated mean, i.e.
	\begin{equation*}
		RAE_p=\frac{1}{N}\sum_{i=1}^N\frac{\left|m_i-\hat{\mathbb E}(X^{(p)}(t_i))\right|}{m_i},\qquad p=2,3,\dots,
	\end{equation*}
	where $\hat{\mathbb E}(X^{(p)}(t_i))$ denotes the mean of the estimated process considering a polynomial $Q_\beta$ of degree $p$;
	\item[(ii)] the Akaike information criterion ($AIC$), which is defined as
	\begin{equation*}
		AIC_p=2(p+2)-2L_{\mathbb V}(\hat\alpha, \hat\xi), \qquad p=2,3\dots,
	\end{equation*}
	\item[(iii)] the Bayesian information criterion ($BIC$), which is given by
	\begin{equation*}
		BIC_p=(p+2)\log(n)-2L_{\mathbb V}(\hat\alpha, \hat\xi), \qquad p=2,3\dots,
	\end{equation*}
	where $n$ represents the number of observations,
	\item[(iv)] the resistor-average distance  ($D_{RA}$) between the sample distribution $f_C$ and the $p$-th estimated distribution $f_{S_p}$, for $p=2,\dots,6$, which is defined as the following harmonic mean  (cf.\ Johnson and Sinanovic (2001) \cite{JohnsonSinanovic2001}):
	\begin{equation*}
		D_{RA}(f_C||f_{S_p})(t)=\frac{D_{KL}(f_C||f_{S_p}) (t)\cdot D_{KL}(f_{S_p}||f_{C}) (t)}{D_{KL}(f_C||f_{S_p}) (t) + D_{KL}(f_{S_p}||f_{C}) (t)}, \qquad t\ge t_0,
	\end{equation*}
	where $D_{KL}$ denotes the Kullback-Leibler divergence. Assuming that the sample distribution is  lognormal  with parameters
	$$
	\mu_C(t)\approx\widehat{\mu}_t= \log(m_g(t)),
	\qquad
	\sigma_C^2(t)\approx\widehat{\sigma}_t^2=2\log\frac{m(t)}{m_g(t)},
	$$
	and that the estimated distribution is lognormal with parameters
	$$
	\mu(t)\approx\widehat{\mu}_0+H_{\widehat\xi}(t_0,t),
	\qquad
	\sigma(t)\approx\widehat{\sigma}_0^2+\widehat{\sigma}^2(t-t_0),
	$$
	the Kullback-Leibler divergence between the sample distribution $f_C$ and the $p$-th estimated distribution $f_{S_p}$ for $p=2,\dots,6$ is given by, for any $t\ge t_0$
	\begin{equation*}
		D_{KL}(f_C||f_{S_p})(t)=\frac{1}{2}\left[\log\left(\frac{\widehat{\sigma}_0^2+\widehat\sigma^2(t-t_0)}{\widehat{\sigma}_t^2}\right)+\frac{\widehat{\sigma}_t^2+\left(\widehat\mu_t-\widehat{\mu}_0-H_{\widehat\xi}(t_0,t)\right)^2}{\widehat{\sigma}_0^2+\widehat\sigma^2(t-t_0)}-1\right],
	\end{equation*}
	with $H_{\widehat \xi}(t_0,t)$ defined in \eqref{H}. Clearly, if the theoretical distribution of the process is known, one can alternatively compute the resistor-average distance between the theoretical and the estimated distribution.
	We consider the expected distance and the median of the distance as reference values for the resistor-average distance.	%
\end{itemize}
\par
\begin{figure}[t]
	\label{fig:Figure6}
	\subfigure[]{\includegraphics[scale=0.32]{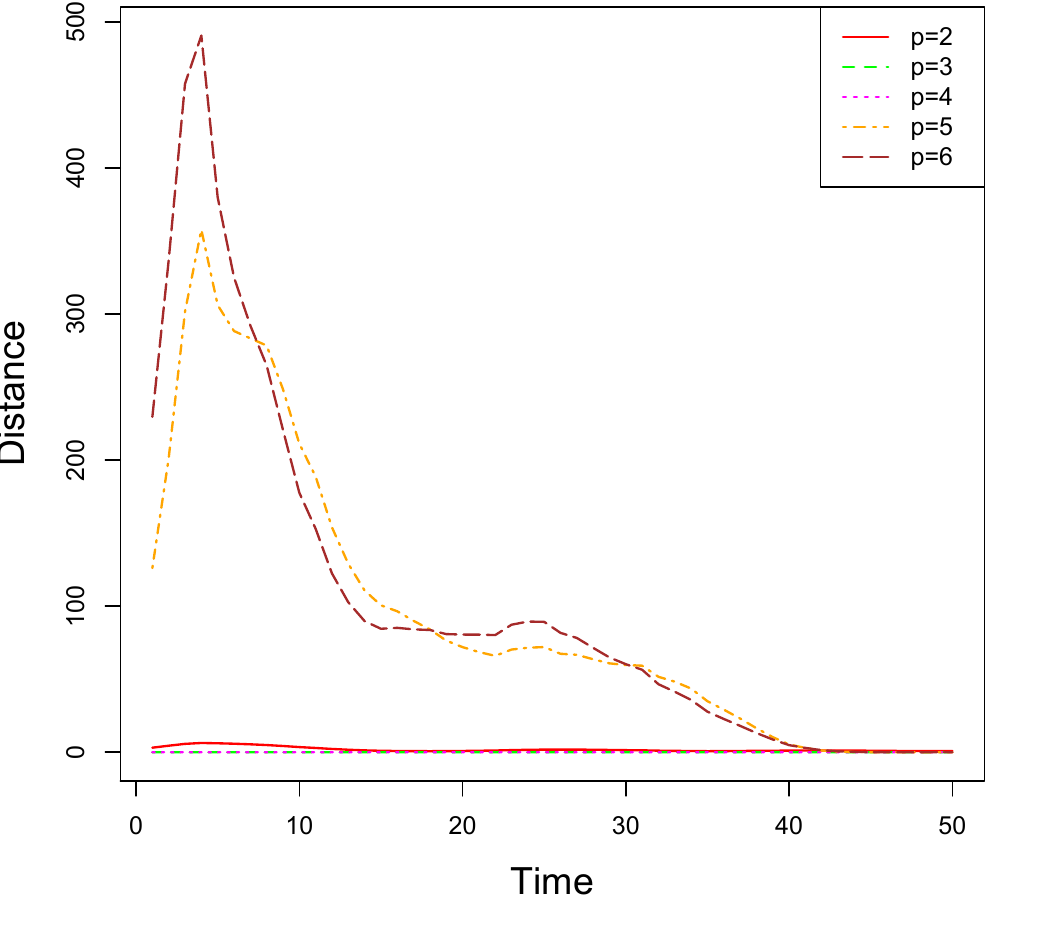}}\quad
	\subfigure[]{\includegraphics[scale=0.32]{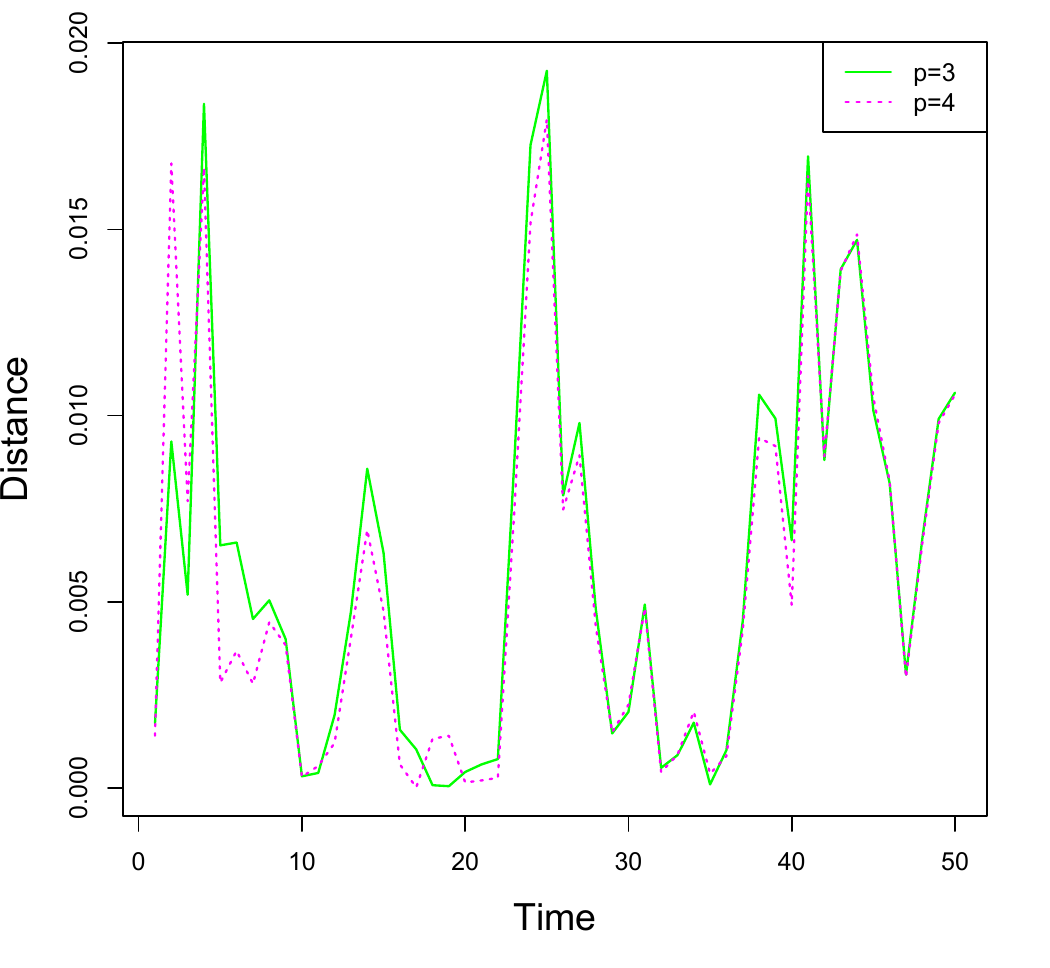}}\\
	\subfigure[]{\includegraphics[scale=0.32]{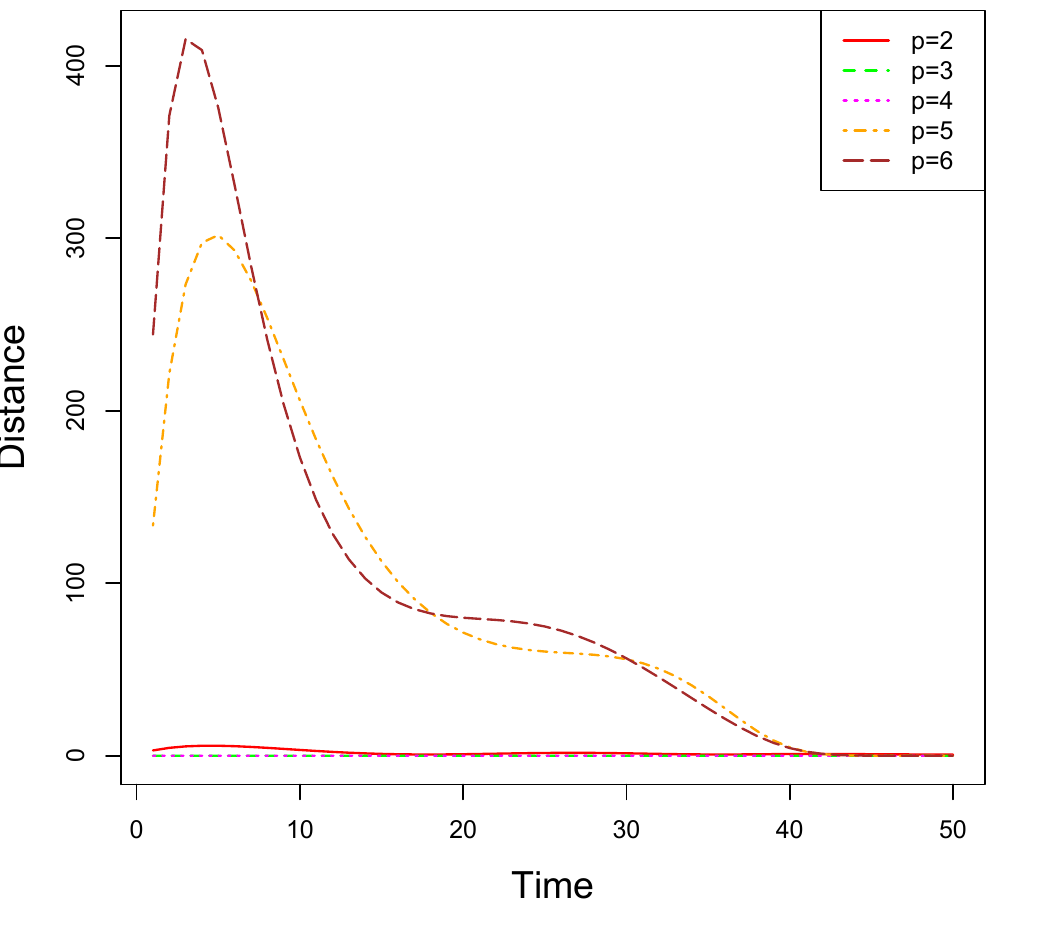}}\quad
	\subfigure[]{\includegraphics[scale=0.32]{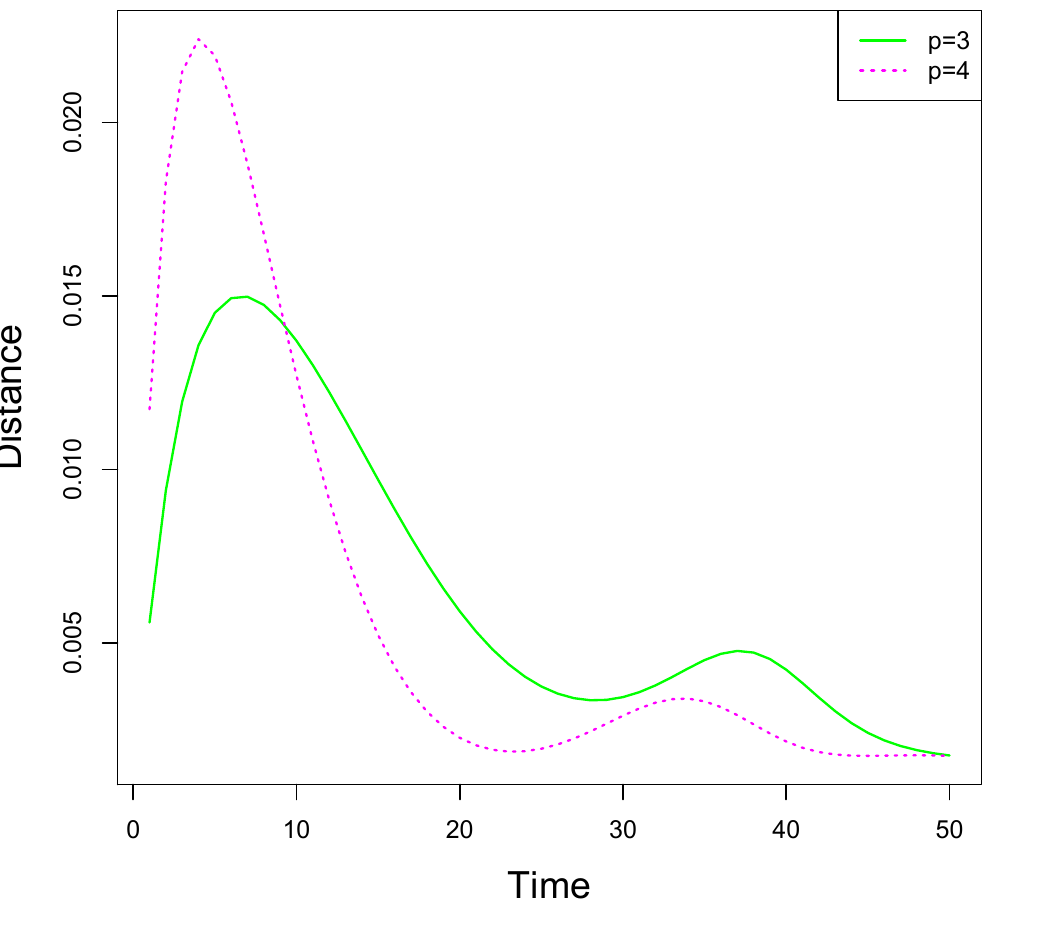}}
	\caption{The resistor average distance between (a)-(b) the sample and the estimated distribution, and (c)-(d) the theoretical and estimated distribution for the case 1 of Table \ref{Tab:table1}, for different degrees of the polynomial (simulation study).}
\end{figure}

In cases (ii) and (iii), the stochastic model is characterized by $p+2$ parameters. Moreover, $L_\mathbb{V}(\alpha,\xi)$ is defined in \eqref{loglik}, and $\hat\alpha$ and $\hat\xi$ are the MLEs of the parameters $\alpha$ and $\xi$. The best fit is  attained for the smallest value of the considered goodness measures. Table \ref{tab:Tabella15} shows the estimated parameters for the case no.\ $1$ of Table \ref{Tab:table1}, which is obtained by solving the system \eqref{sist3} for different degrees of the polynomial $Q_\beta$. Furthermore, the results about the goodness of measures are given in Table \ref{tab:Tabella12} and in Figure \ref{fig:Figure6}. It can be noticed that the estimated parameters for $p=3$ and $p=4$ are almost identical, and that $\beta_4$ is very close to zero in the case $p=4$. Hence, the results concerning the measures of goodness obtained in these two cases are quite similar.
This conclusion is also confirmed by the analysis of the RAE measures  (in Table \ref{tab:Tabella12}),
that are often used to measure the fit error of the model in terms of the fit of the mean function. The analysis is performed
in terms of the scale of judgment of the model accuracy based on the Mean Absolute Percentage Error (MAPE),
cf.\ Klimberg {\em et al.}\ (2010) \cite{Klimbergetal2010} and Lewis (1982) \cite{Lewis1982}.
Indeed, the judgment suggested by the MAPE shows that $p=3$ and $p=4$ are referred as highly accurate,
whereas $p=2$ is evaluated as good forecast, with both $p=5$ and $p=6$ considered as reasonable forecast.
Consequently, the choice $p=3$ is taken as the best, since it involves the lowest number of parameters.
\begin{table}[t]
	\caption{The estimated parameters obtained by solving system \eqref{sist3} for different degrees of the polynomial $Q_\beta$ (simulation study).}
	\label{tab:Tabella15}
	\centering
	\tiny
	\begin{tabular}{l|l|l|l|l|l}
		& value     										&$\beta_1$ 						&$\beta_2$ 						&$\beta_3$ 							&$\beta_4$ \\ \hline
		\multirow{2}{*}{$p=2$} & initial    	&$-0.19227458$               &$0.006134021$            &--                         				&--                         \\
		& estimated 									&$-0.02167632$               &$0.001417851$             &--                                      &--                         \\ \hline
		\multirow{2}{*}{$p=3$} & initial   		&$-0.02332665$               &$-0.005246427$           &$0.0001724780$                &--                         \\
		& estimated 									&$0.10200076$                &$-0.009115730$           &$0.0002019617$                &--                         \\ \hline
		\multirow{2}{*}{$p=4$} & initial        &$-0.16488586$               &$0.011922169$             &$-0.0004348227$                &$6.54664e$-$03$                         \\
		& estimated 									&$0.10313862$               &$-0.009310721$          &$0.0002108884$                 &$-1.240487e$-$07$                         \\ \hline
		\multirow{2}{*}{$p=5$} & initial   		&$-0.23209031$                 &$0.024605250$           &$-0.0012041705$                 &$2.521331e$-$05$                         \\
		& estimated 									&$-0.23209031$                 &$0.024605250$           &$-0.0012041705$                 &$2.521331e$-$05$                         \\ \hline
		\multirow{2}{*}{$p=6$} & initial   		&$-0.36570406$                 &$0.060641889$            &$-0.0044843917$               &$1.579124e$-$04$                         \\
		& estimated 									&$-0.36570406$                 &$0.060641889$            &$-0.0044843917$               &$1.579124e$-$04$
	\end{tabular}
	\begin{tabular}{l|l|l|l|l|l}
		& value     										&$\beta_5$ 						&$\beta_6$ 						&$\eta$ 							&$\sigma^2$ \\ \hline
		\multirow{2}{*}{$p=2$} & initial    	&--               					  &--           						&$0.3688831$                 &$0.0092356136$                         \\
		& estimated 									&--               					  &--             						&$0.2342132$                  &$0.0013732663$                         \\ \hline		
		\multirow{2}{*}{$p=3$} & initial   		&--               					  &--           						&$0.3688831$                &$0.0092356136$                          \\
		& estimated 									&--             					&--           							&$0.3649943$                &$0.000103298$                         \\ \hline
		\multirow{2}{*}{$p=4$} & initial        &--           						&--          							 &$0.3688831$                &$0.0092356136$                             \\
		& estimated 									&--       							&--         							 &$0.3649325$                 &$0.0001032229$                         \\ \hline
		\multirow{2}{*}{$p=5$} & initial   		&$-1.572548e$-$07$                 &--       							&$0.3688831$                 &$0.0092356136$                      \\
		& estimated 									&$-1.572548e$-$07$                &--          							&$0.3688831$                &$0.0000859656$                         \\ \hline
		\multirow{2}{*}{$p=6$} & initial   		&$-2.617632e$-$06$          &$1.706174e$-$08$            &$0.3688831$               &$0.0092356136$                           \\
		& estimated 									&$-2.617632e$-$06$              &$1.706174e$-$08$            &$0.3688831$                 &$0.0000859656$
	\end{tabular}
\end{table}
\begin{table}[t]
	\caption{The goodness measures for different degree. For the resistor-average distance $D_{RA}$, the estimated and the theoretical distributions are considered (simulation study).}
	\label{tab:Tabella12}
	\tiny
	\centering
	\begin{tabular}{rrrrrrr}
		degree & $RAE$                 & $BIC$                  & $AIC$			&median of $D_{RA}$ 		&mean of $D_{RA}$		                  \\ \hline
		$2$ 	&$0.163198755$	&$-6986.46$			&$-7006.983$ 	&$1.076721237$		&$1.874847040$				\\
		$3$		&$0.001667815$	&$-10261.07$		&$-10286.721$	&$0.001035228$		&$0.001668560$		\\
		$4$		&$0.001390662$	&$-10254.13$	&$-10284.916$		&$0.001149273$		&$0.001916131$		\\
		$5$		&$0.413978453$	&$34449.77$		&$34449.77$		&$54.690728902$			&$84.305882377$		\\
		$6$		&$0.416847669$	&$47306.94$		&$47306.94$		&$69.316641926$			&$94.836470468$		\\ \hline
	\end{tabular}
\end{table}
\par
The same result can be obtained for the other parameters choices, but it is omitted for brevity.
Here, we limit  to mention that the AIC and its Bayesian version, the BIC,
provide a global measure of the adjustment to the model in terms of the likelihood that the model itself
gives to the observed sample, so that these measures also allow for model selection criteria.
The AIC and the BIC are seen often as complementary measures to the use
of the Resistor Average Distance between the sample and the theoretical distributions of the model.
However, there is no criterion that indicates that one measure is better than another, and thus in general
the use of several alternative measures is recommended, as usual in practical applications.
In our analysis, the coincidence of the conclusions suggested by these measures supports the final decision.
Hence, from now on, a polynomial of degree $p=3$ will be considered.
\par
\begin{figure}[t]
	\centering
	\hspace*{-1cm}
	\subfigure[]{\includegraphics[scale=0.32]{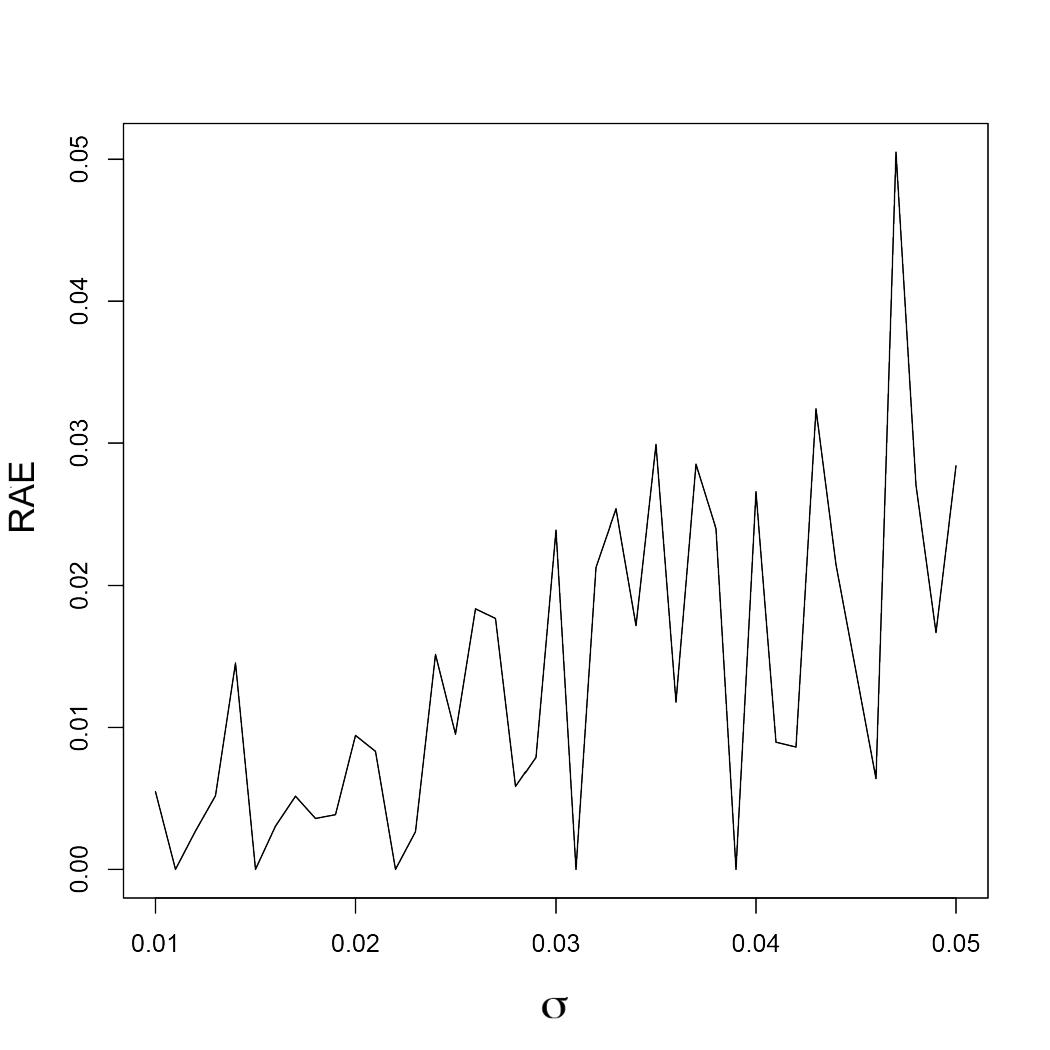}}\quad
	\subfigure[]{\includegraphics[scale=0.32]{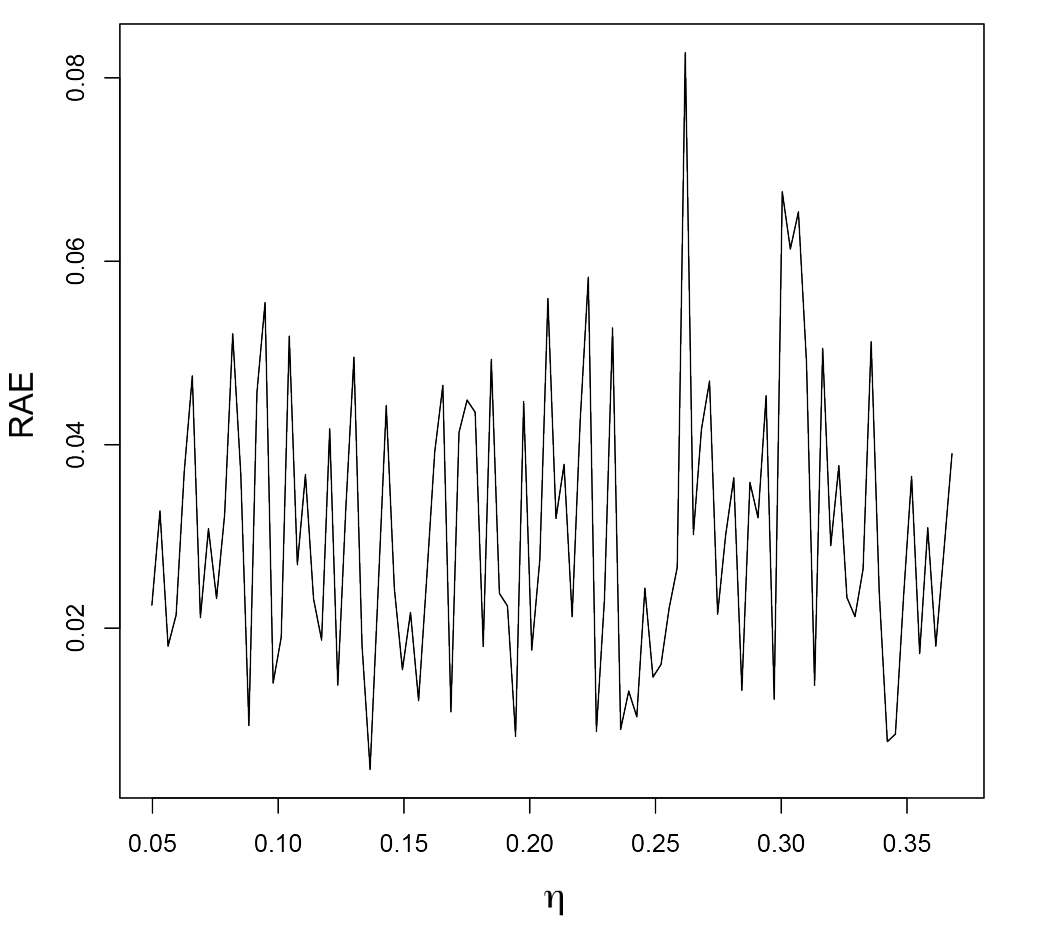}}\quad
	\subfigure[]{\includegraphics[scale=0.32]{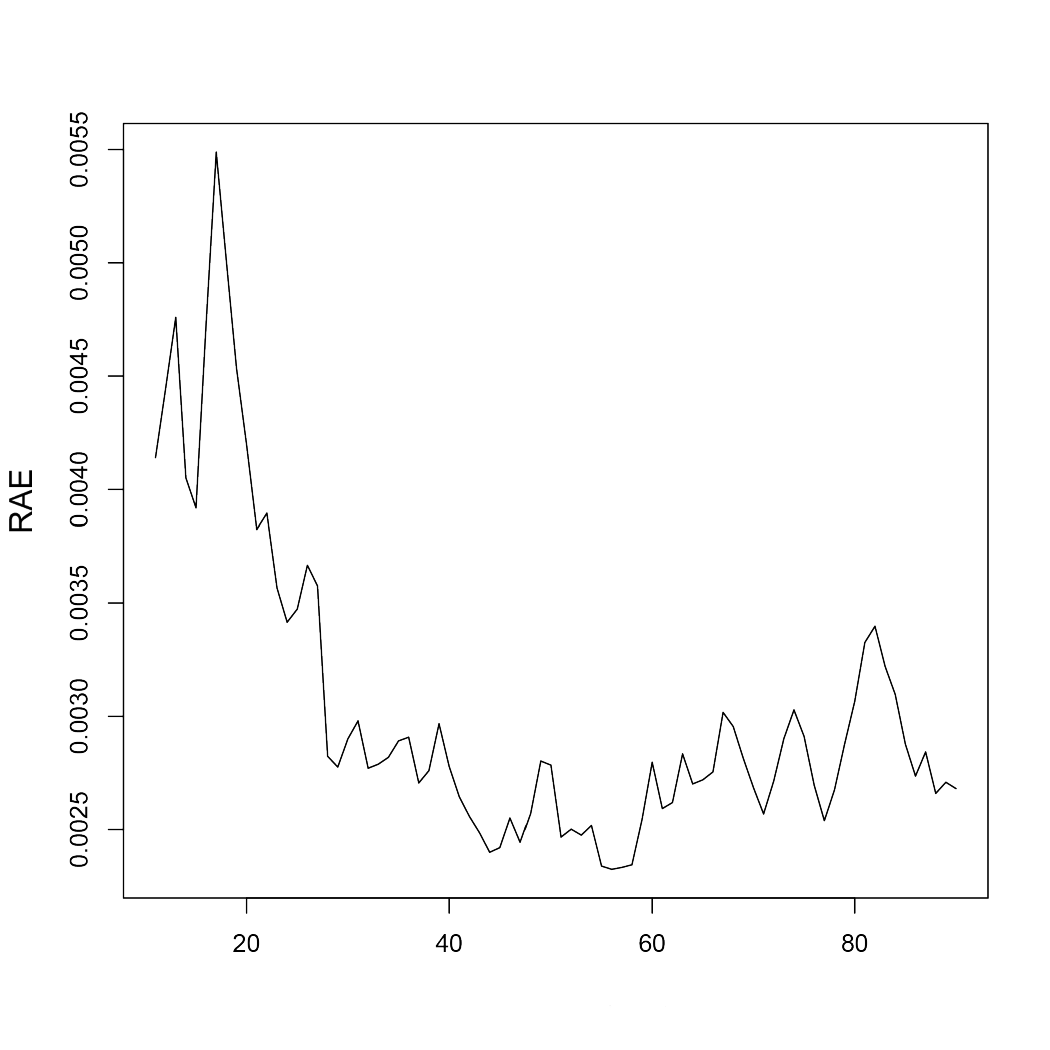}}
	\caption{The $RAE$ for (a) $\eta=e^{-1}$ and $\sigma\in[0.01,0.05]$, (b) $\sigma=0.05$ and $\eta\in\left[e^{-3},e^{-1}\right]$ and (c) $\sigma=0.01$, $\eta=e^{-1}$ and with respect of the number of replications. In all the cases $Q_\beta(t)=0.1t-0.009t^2+0.0002t^3$ (simulation study).}
	\label{fig:Figure7}
\end{figure}
\begin{figure}[t]
	\centering
	\subfigure[]{\includegraphics[scale=0.35]{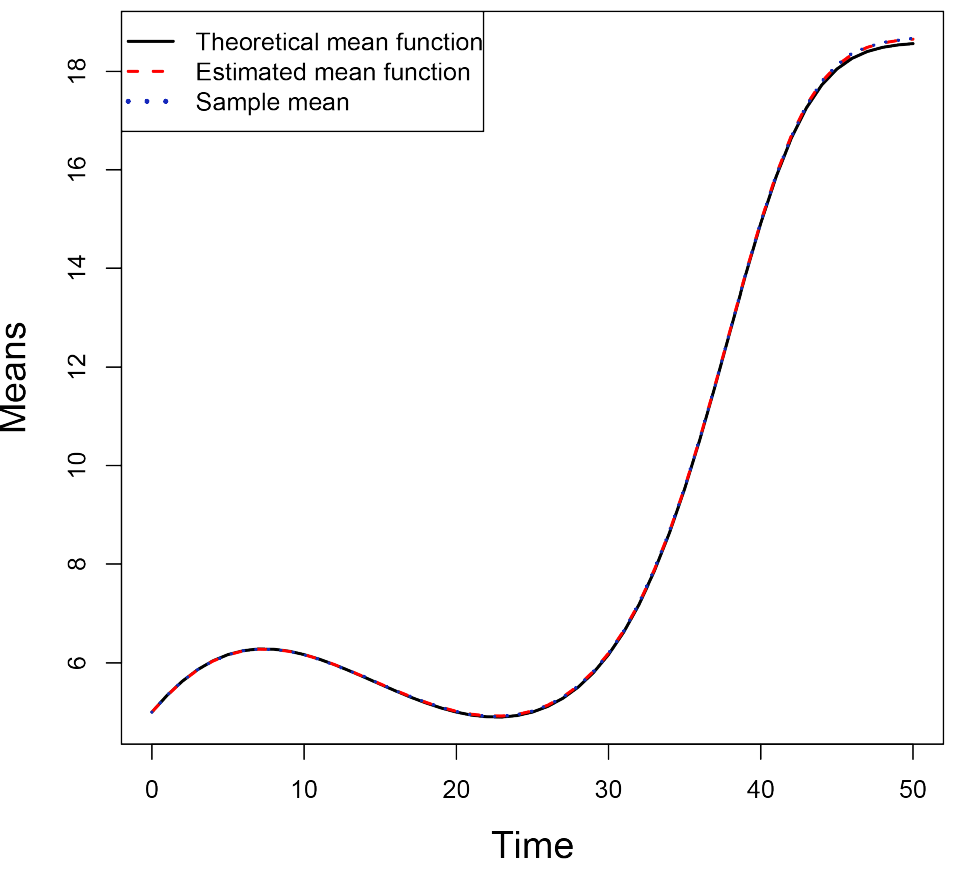}}
	\subfigure[]{\includegraphics[scale=0.35]{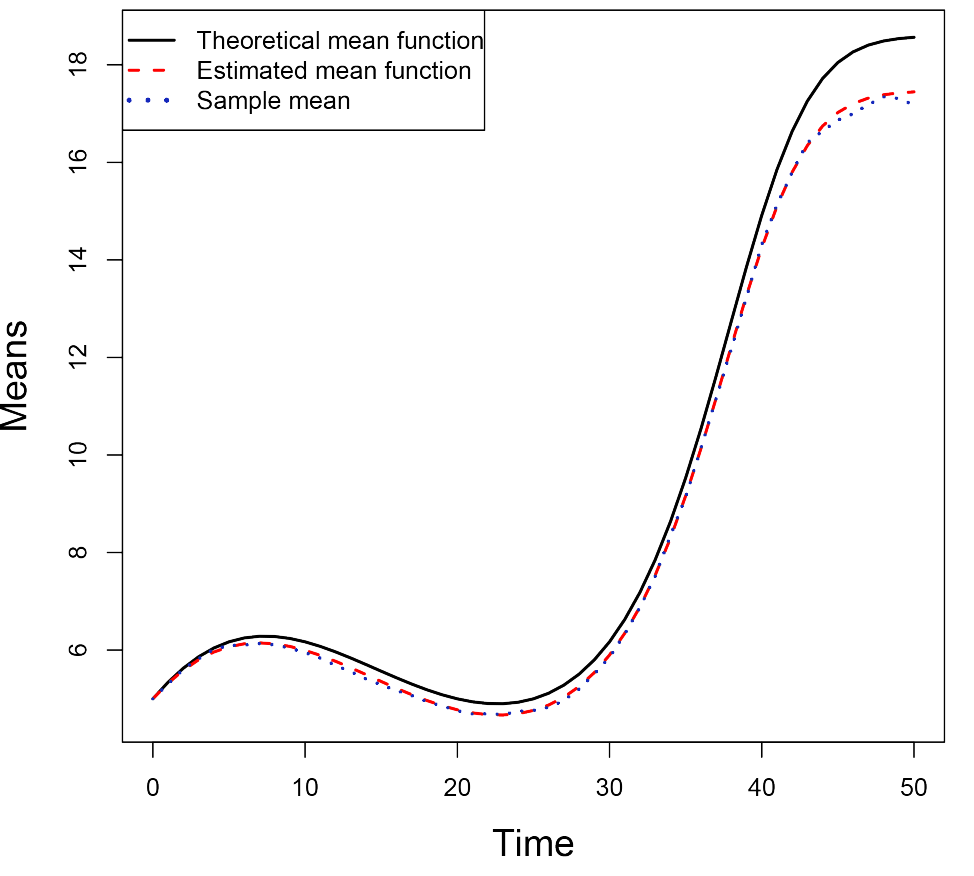}}
	\subfigure[]{\includegraphics[scale=0.35]{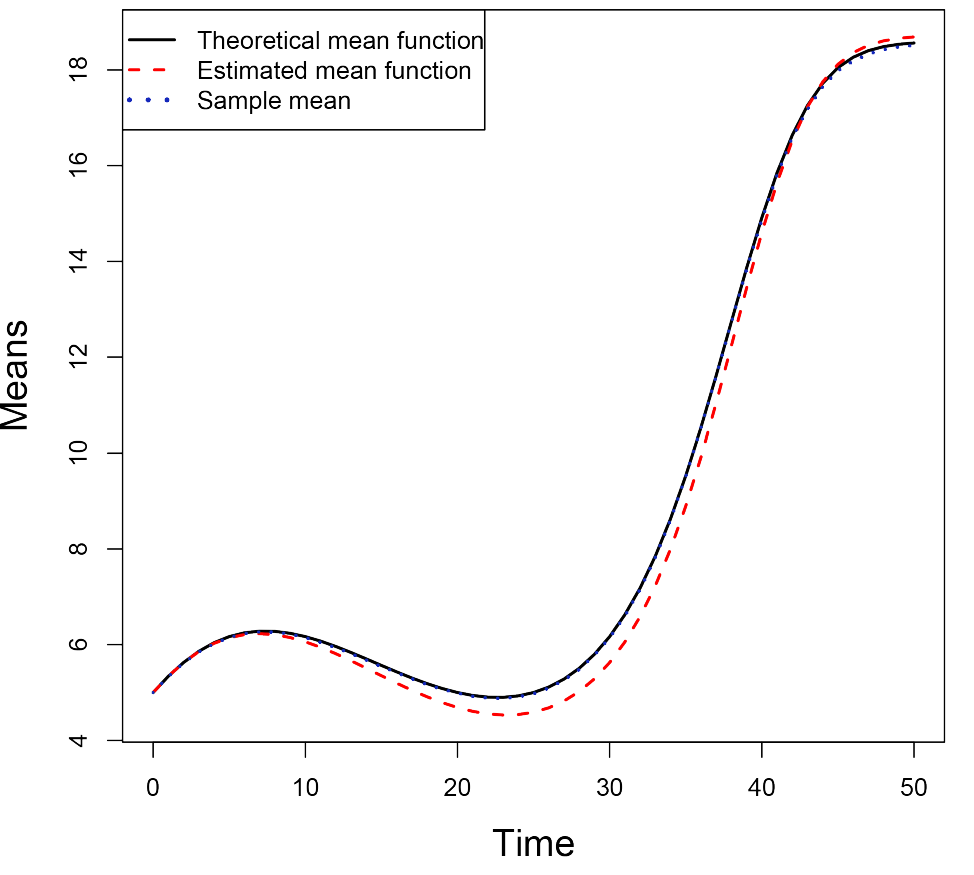}}
	\subfigure[]{\includegraphics[scale=0.35]{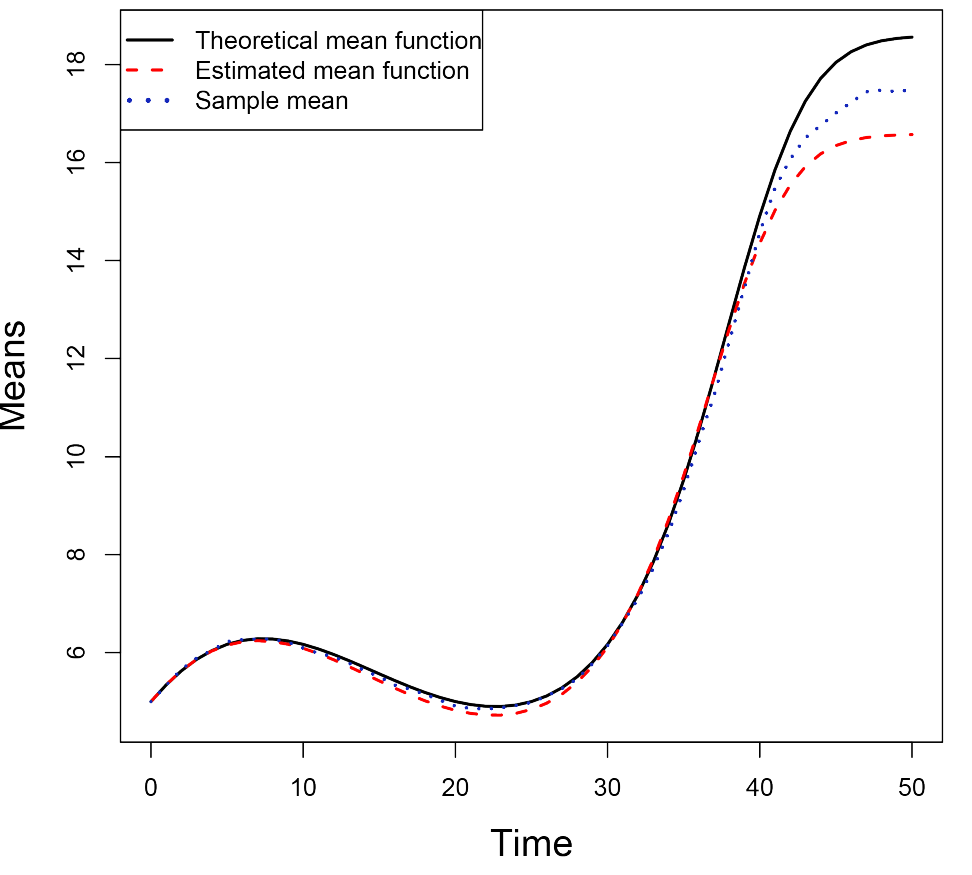}}
	\caption{The theoretical, sample and estimated means of the process $X(t)$ for the parameters of the cases number $1$ and $2$ of Table \ref{Tab:table1} (from left to right). The results are obtained via Newton-Raphson method in (a) and (b) and via S.A. in (c) and (d). (Simulation study).}
	\label{fig:Figure8}
\end{figure}

Table \ref{Tab:table2} shows the estimated values of the parameters obtained by solving the nonlinear system \eqref{sist3} by means of the Newton-Raphson method. These values provide good parameters estimates, especially when $\sigma$ is small. The last column of the Table \ref{Tab:table2} contains the $RAE$. In this case, it is defined  as
\begin{equation}\label{RAEdef}
	RAE_3=\frac{1}{N}\sum_{i=1}^N\frac{\left|m_i-\hat{\mathbb E}(X^{(3)}(t_i))\right|}{m_i},
\end{equation}
where $m_i$ are the values of the sample mean and $\hat{\mathbb E}(X^{(3)}(t_i))$ are the values of the estimated mean at the time $t_i$ considering a polynomial of degree $p=3$.
For a comparison between $\sigma$ or $\eta$ and the $RAE$, see Figure \ref{fig:Figure7} (a)-(b): it can be noticed that the value of the $RAE$ shows an increasing trend with respect to the parameter $\sigma$, whereas it shows a constant trend with respect to $\eta$. In Figure \ref{fig:Figure8} (a)-(b) the theoretical, sample and estimated sample means for the parameters choices number $1$ and $2$ of the Table \ref{Tab:table1} are shown. Clearly, the best estimation is obtained when $\sigma$ is small.
\begin{table}[t]
	\caption{The estimated values of the parameters obtained by solving the system (simulation study).}
	\label{Tab:table2}
	\tiny
	\centering
	\begin{tabular}{llllllll}
		case no. & $\hat\beta_1$                 & $\hat\beta_2 $                  & $\hat\beta_3$                   & $\hat\eta$                      & $\hat\sigma$  &$RAE$\\ \hline
		1         & $0.1008708$  				  &$-0.009083141$  					&$0.0002016053$  				  &$0.3748345$   					& $0.009948509$ &$0.009948509$\\ 2         & $0.09647959$                   &$-0.009128214$                   &$0.0002035713$                   &$0.4174093 $                 &$0.04861361$                           &$0.00968444$ \\ 
		3         & $0.1006515$                 &$-0.009032199  $                   &$0.0002004328$                   &$0.04941139 $  				&$0.01006651$ &$0.001394514$ \\ 
		4         & $0.09813563$                   &$-0.008927315$                    &$0.0001989717$                   &$0.05311648$                   & $0.05045509$ &$0.004833166$\\ 
		5         & $0.09917466$                   &$-0.008943618$                   &$0.0002994268$ 					&$0.3684543$					 & $0.009948515$ &$0.0007958334$\\ 
		6         & $0.1014043$                    &$-0.008998974 $                  &$0.0002981807$                    &$0.3892023$                         & $0.05016611$ &$0.004735055$\\ 
		7         & $0.1000035$                    &$-0.008977959$                  & $0.000299122$                   &$0.0499255$ 				& $0.01019311$ &$0.001671998$ \\ 
		8         & $0.0998637 $                     &$-0.009123996$                 &$0.0003010265$             &$0.05329735$                          & $0.04990423$  &$0.006741047$ \\ 
		9         & $0.0996273$                    &$-0.006921284 $ 				&$0.000198009 $ 				&$0.3691476$ 			&$0.01007512$ &$0.001185446$ \\ 
		10        & $0.09453137$                     &$-0.006522038$                          &$0.0001887209$                          &$0.3961056$                           & $0.05014104$  &$0.008249555$ \\ 
		11        & $0.0994893$                     &$-0.006961563$                           &$0.000199408$                           & $0.05006629$ & $0.01005502$ &$0.001227369$\\ 
		12        &  $0.09991228$                   &$-0.007300655$                         &$0.0002073627$                          &$0.05440441$                           & $0.04977635$ &$0.008537779$ \\ 
		13        & $ 0.09871006$                        &$-0.006934799$                       &$0.0002995149$               & $0.3732344 $ & $0.009929474$  &$0.001391576$ \\ 
		14        & $0.1089433$                       &$-0.00769021$                        &$0.0003170773$              & $0.3827415$ & $0.05020807$  &$0.007503426$ \\ 
		15        & $0.09915895$                        &$-0.006970525$                           &$0.0003002155$                           & $0.05005078$ & $0.009887142$ &$0.001247189$\\ 
		16        & $0.099621$                        &$-0.007308948$                          &$0.0003089962$                           &$ 0.05292541$                           & $0.05039523$  &$0.01051539$ \\ \hline
	\end{tabular}
	\begin{tabular}{llllllll}
		case no. & $\hat\beta_1$                 & $\hat\beta_2 $                  & $\hat\beta_3$                   & $\hat\eta$                      & $\hat\sigma$  &$RAE$\\ \hline
		17         & $0.5027283$ & $-0.009975521$ &$0.0002718814$  & $0.3675535$ & $0.01005231$ &$0.001725124$ \\ 
		18         & $0.4994427$ & $-0.0113279$ &$0.0008357429$  & $0.3763182$   & $0.04958661$  &$0.01836523$\\ 
		19        & $0.5009121$  & $-0.009399375$ &$0.0002318097$  & $0.04989883$ & $0.01014064$ & $0.001231152$ \\ 
		20        & $0.4889807$  &$-0.007392059$  &$0.0001686769$  &$0.05133072$   & $0.05002574$ & $0.02180217$ \\ 
		21        & $0.4966796$  &$-0.005064475$  &$9.070535e$-$05$  &$0.3645112$ 	& $0.008760448$ &$ 0.00882512$ \\ 
		22        & $0.5169986$  &$-0.008982219$  &$0.0007673324$  &$0.3842942$   & $0.05027921$ &$0.01926412$ \\ 
		23        & $0.5007486$  &$-0.00918498$  &$0.0003122469$  &$0.05015442$   & $0.009886917$ &$0.001718233$ \\ 
		24        & $0.5004829$     &$-0.009757746$  &$0.0004111398$  &$0.05113107$   & $0.0501877$  &$0.03498688$ \\ 
		25        & $0.4992278$  &$-0.007530001$  &$0.0002736762$  &$0.367121$  & $0.01017701$ &$0.003303379$ \\ 
		26        & $0.522767$  &$-0.01616745$  &$0.001661617$  &$0.3978223$  & $0.0504873$ &$0.02065117$\\ 
		27        & $0.4947951$  &$-0.004930811$  &$-3.751225e$-$05$   &$0.04872577$  & $0.0100036$ &$0.01431122$ \\ 
		28        & $0.4973394$  &$-0.004424543$  &$-5.398909e$-$05$  &$0.0509033$   & $0.04989332$ &$0.008392656$ \\ 
		29        & $0.5007914$  &$-0.005683407$  &$0.0001988298$  & $0.3686014$ & $0.01002725$ &$0.001054943$ \\ 
		30        & $0.6589188$  &$-0.01472611$  &$7.767629e$-$05$  & $0.3884501$  & $0.05398134$  &$0.02949292$ \\ 
		31        & $0.5005393$   &$-0.007046342$  &$0.0003130846$  & $0.04990925$ & $0.009948212$  &$0.002032919$ \\ 
		32        & $0.4944903$  &$-0.006618593$  &$0.0004977852$  & $0.0531368$ & $0.05019473$ &$0.01046167$\\ \hline
	\end{tabular}
\end{table}
\par
Further on, the estimated values obtained via S.A.\ are given in Table \ref{Tab:table7}, whose last column contains the value of the $RAE$ defined in Eq.\ \eqref{RAEdef}.
Since S.A.\ is a heuristic algorithm,  the $MLEs$ have been computed as the average of the results obtained by 10 uses of the procedure.
Figures \ref{fig:Figure8} (c)-(d) provide the theoretical, the sample and the estimated (via S.A.) means for the cases no.\ $1$ and no.\ $2$ of the Table \ref{Tab:table1}. In addition, in Figure \ref{fig:Figure7} (c), the trend of the $RAE$ is plotted as a function of the number of replications: clearly, the goodness of the results improves as the number of replications increases.
\begin{table}[t]
	\caption{The estimated values of the parameters obtained via S.A. (simulation study).}
	\label{Tab:table7}
	\tiny
	\centering
	\begin{tabular}{llllllll}
		case no. & $\hat\beta_1$                 & $\hat\beta_2 $                  & $\hat\beta_3$                   & $\hat\eta$                      & $\hat\sigma^2$  &$RAE$\\ \hline
		1         & $0.101561$  				  &$-0.009082552$  					&$0.0002018522$  				  &$0.3625206$   					& $9.989558e$-$05$ &$0.01145807$\\
		2         & $0.09954966$                   &$-0.009144748$                   &$0.0002033654$                &$0.4041588 $                           &$0.002549043$  &$0.04471631$\\ 
		3         & $0.1131058$                   &$-0.00993448$                    &$0.0002154949$                   &$0.05040806$                   & $0.0001547924$ &$0.02141325$\\ 
		4        & $0.1091467$                   &$-0.009814184$                   &$0.0002144671$ 					&$0.05347671$					 & $0.002515041$ &$0.0371569$\\ 
		5         & $0.1180621$                   &$-0.01109313$                   &$0.0003614234 $ 					&$0.3760972$					 & $0.0001248723 $ &$0.01352716$\\ 
		6         & $0.1278859$                   &$-0.01283086$                   &$0.0004142981$ 					&$0.4122873$					 & $0.00251747$ &$0.01633801 $\\ 
		7         & $0.1060342$                    &$-0.009544155$                  &$0.00031209$                    &$0.0501014$                         & $0.0001160831$ &$0.007163973$\\ 
		8         & $0.1233499$                     &$-0.01139476$                 &$0.000355587$             &$0.05196034$                          & $0.002510716$  &$0.02342116$ \\ 
		9         & $0.05999024$                    &$-0.003213622 $ 				&$0.0001156288$ 				&$0.3619285$ 			&$0.00019005$ &$0.02598964$ \\ 
		10        & $0.1281187$                     &$-0.01053716$                          &$0.000303384$                          &$0.4643666$                           & $0.002569706$  &$0.04634721$ \\ 
		11        & $0.07833233$                     &$-0.005157051$                           &$0.0001608136$                           & $0.05258961$ & $0.0001810726$ &$0.04867231$\\ 
		12        &  $0.118225$                   &$-0.008582805$                         &$0.0002342187$                          &$0.05132632$                           & $0.002638776$ &$0.035553561$ \\ 
		13        & $ 0.1177451$                        &$-0.00947922$                       &$0.0003863382$               & $0.3979672$ & $0.0002070029$  &$0.02261541$ \\ 
		14        & $0.1155341$                       &$-0.009715339$                        &$0.0004058845$              & $0.4262232$ & $0.002535962$  &$0.01010711$ \\ 
		15        & $0.09977635$                        &$-0.007003481$                           &$0.0003003812$                           & $0.04986716$ & $0.0001068239$ &$0.001787891$\\ 
		16        & $0.1240548$                        &$-0.009570475$                          &$0.0003701909$                           &$ 0.05082801$                           & $0.002543112$  &$0.02769778$ \\ \hline
	\end{tabular}
	\begin{tabular}{llllllll}
		case no. & $\hat\beta_1$                 & $\hat\beta_2 $                  & $\hat\beta_3$                   & $\hat\eta$                      & $\hat\sigma^2$  &$RAE$\\ \hline
		17         & $0.4996174$  				  &$-0.007431545$  					&$1.808038e$-$05$  				  &$0.3697584$   					& $9.898729e$-$05$ &$0.003974179$\\
		18         & $0.5116561$                   &$-0.01727471$                   &$0.0002780168$                &$0.3700344$                           &$0.002477853$  &$0.01019564$\\ 
		19         & $0.4908564 $                   &$-0.007392949$                    &$5.594539e$-$05$                   &$0.04968137$                   & $0.0001481544$ &$0.008720396$\\ 
		20        & $0.5013114$                   &$-0.00809688$                   &$0.0001788676$ 					&$0.0530825$					 & $0.00273835$ &$0.009876328$\\ 
		21         & $0.6249235$                   &$-0.02166049$                   &$0.000398154$ 					&$0.4112333 $					 & $0.000224436 $ &$0.06819835$\\ 
		22         & $0.5686069$                   &$-0.02374947$                   &$0.000553332$ 					&$0.3847588$					 & $0.002566421$ &$0.01345391$\\ 
		23         & $0.4865293$                    &$-0.006591622$                  &$5.45598e$-$05$                    &$0.05038445$                         & $0.0001598932$ &$0.01251157$\\
		24         & $0.4833729$                     &$-0.00633273$                 &$0.000290948$             &$0.05788054$                          & $0.002589166$  &$0.03791299$ \\ 
		25         & $0.4994063 $                    &$-0.008634032$ 				&$9.26605e$-$05$ 				&$0.3633383$ 			&$0.0001204934$ &$0.003431125$ \\ 
		26        & $0.4987191$                     &$-0.004809727$                          &$6.356344e$-$05$                          &$0.3925613$                           & $0.002584155$  &$ 0.02507006$ \\ 
		27        & $0.5200495$                     &$-0.01434754$                           &$0.001020145$                           & $0.05181291$  &$0.0001540365$ &$0.02603051$\\ 
		28        &  $0.4965627$                   &$-0.004536852$                         &$0.0001049973$                          &$0.05176822$                           &$0.002566167$ &$0.01328192$ \\ 
		29        & $0.4859707$                        &$-0.005580455$                       &$9.861833e$-$05$               & $0.3638371$ & $0.0001267092$  &$0.006797028$ \\ 
		30        & $0.5420568$                       &$-0.0276007$                        &$0.003931064$              & $0.3872892$  & $0.002639974$  &$0.01427894$ \\ 
		31        & $0.4917976$                        &$-0.004818751$                           &$0.0001038265$                           & $0.04923269$ & $0.0001633201$ &$0.01411135$\\ 
		32        & $0.5002401$                        &$-0.008964678$                          &$0.0006969377$                           &$0.05423012$                           & $0.002528847$  &$0.02389477$ \\ \hline
	\end{tabular}
\end{table}
Moreover, Table \ref{tab:Tabella16} contains the estimated values of the parameters (obtained by solving the system \eqref{sist3}), as well as their real values and the asymptotic estimation error. Finally, Table \ref{tab:Tabella16} provides various confidence intervals obtained by applying the delta method and using the distribution given in Section \ref{distribMLE} for the case no.\ 1 of Table \ref{Tab:table1}.
\begin{table}[t]
	\caption{The estimated values of the parameters (obtained by solving the system \eqref{sist3}), their real values, their asymptotic estimation error and their $95\%$, $90\%$ and $75\%$ confidence intervals (simulation study).}
	\label{tab:Tabella16}
	\tiny
	\centering
	\begin{tabular}{rrrr}
		Parametric function & $\eta$                 & $\beta_1$                  & $\beta_2$		 \\ \hline
		Estimated value 	 &$0.3748345$	&$0.1008708 $	&$-0.0090831$ \\
		Real value				&$0.3678794$	&$0.1000000$   &$-0.0090000$	\\
		Standard error		&$1.218018e$-$03$	&$5.355428e$-$05$	&$2.683426e$-$07$ \\
		$95\%$ confidence interval		&$(0.3064315,0.4432375)$	&$(0.0865276,0.1152140)$	&$(-0.0100984,-0.0099352)$ \\
		$90\%$ confidence interval		&$(0.3174289,0.4322401)$	&$(0.0888336,0.1129080)$	&$(-0.0099352,-0.0082311)$ \\
		$75\%$ confidence interval   &$(0.3346872,0.4149818)$		&$(0.0925245, 0.1092891)$  &$(-0.0096790, -0.0084872)$ \\ \hline
	\end{tabular}
	\begin{tabular}{rrr}
		Parametric function 		&$\beta_3$  &$\sigma^2$                   \\ \hline
		Estimated value 	 &$0.0002016$     &$0.0000990$\\
		Real value				&$0.0002000$  &$0.0001000$ \\
		Standard error		 &$9.218516e$-$11$  &$3.918248e$-$12$\\
		$95\%$ confidence interval		 &$(0.0001828,0.0002204)$ &$(0.0000951,0.0001028)$\\
		$90\%$ confidence interval		&$(0.0001858,0.0002174)$ &$(0.0000957,0.0001022)$\\
		$75\%$ confidence interval    &$(0.0001906, 0.0002126)$  &$(0.0000967, 0.0001012)$\\ \hline
	\end{tabular}
\end{table}
%
\subsection{Approximation of FPT density}\label{sec5.1}
\begin{figure}
	\centering
	\subfigure[]{\includegraphics[scale=0.32]{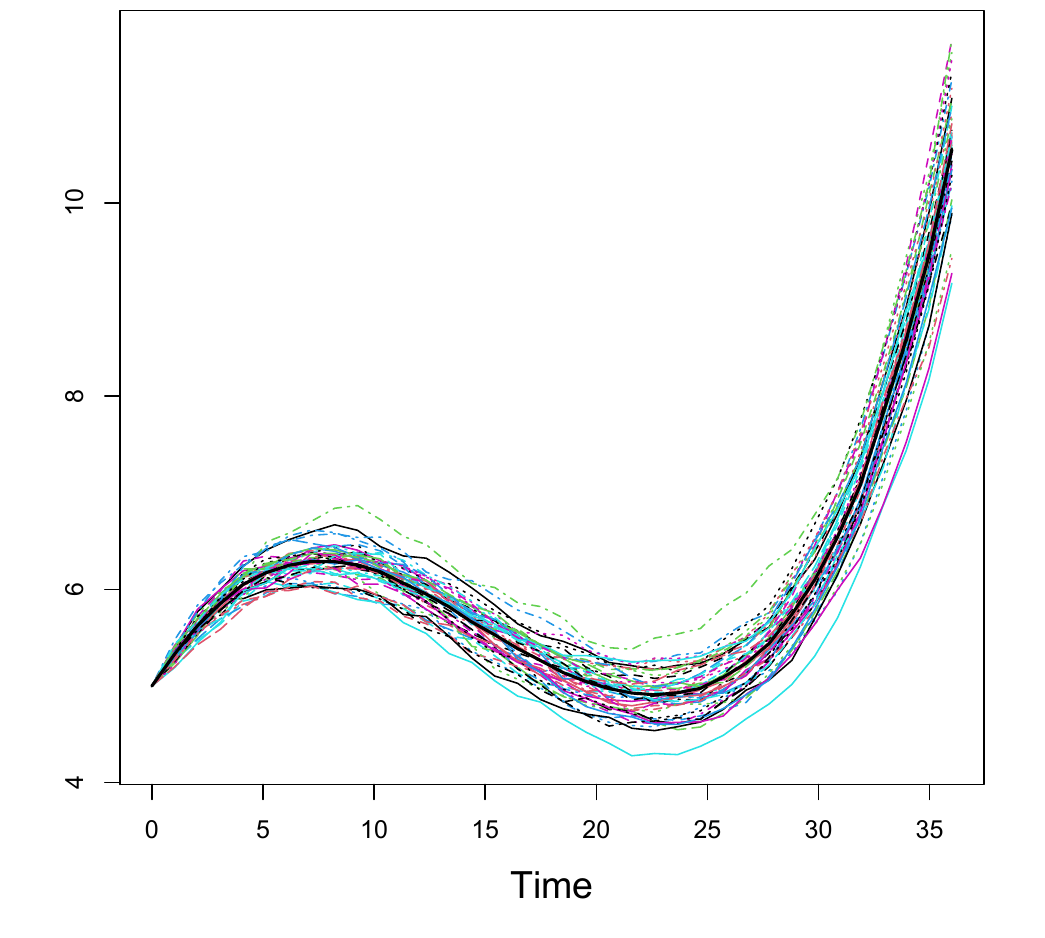}}\quad
	\subfigure[]{\includegraphics[scale=0.32]{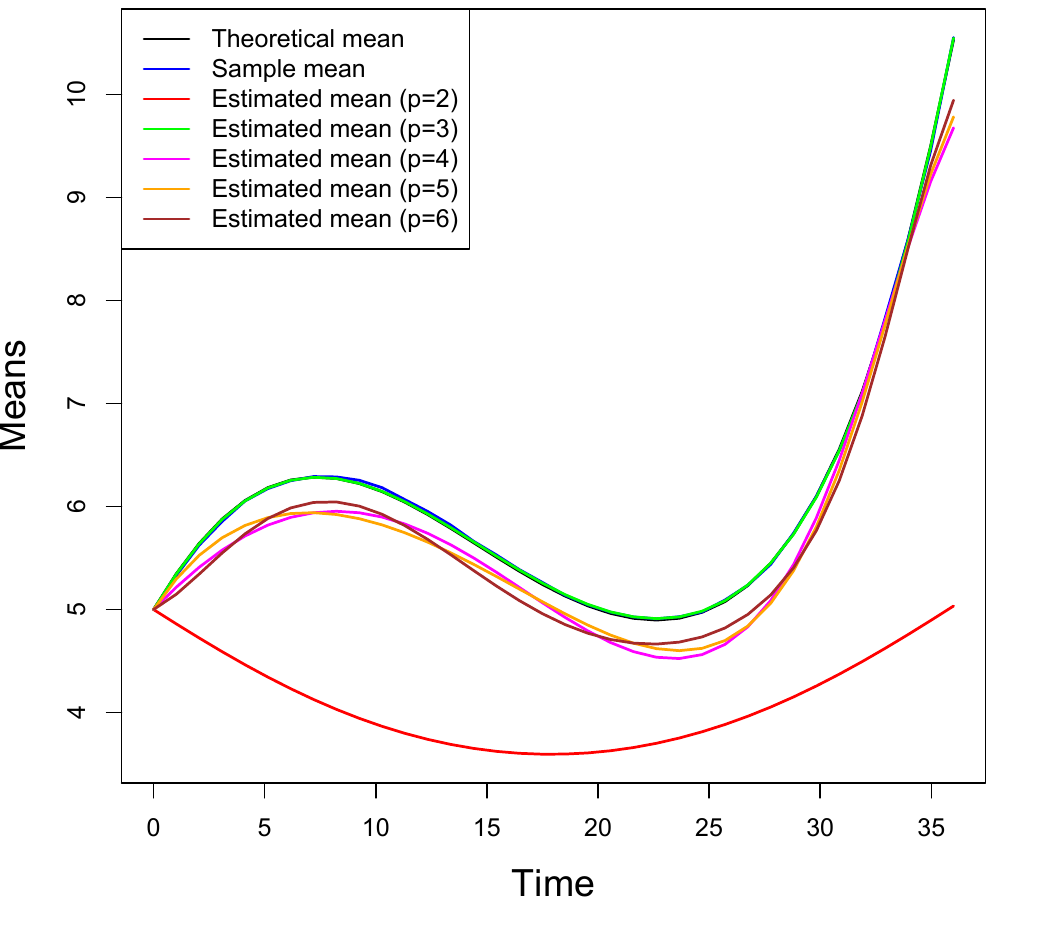}}
	\caption{(a) $50$ simulated sample paths of the diffusion process $X(t)$ for $\sigma=0.01$ , $\eta=e^{-1}$ and $Q_\beta(t)=0.1t-0.009t^2+0.0002t^3$. (b) Theoretical, sample and estimated means of the process $X(t)$ for the FPT density approximation (simulation study - FPT problem).}
	\label{fig:Figure9}
\end{figure}

\begin{figure}[t]
	\centering
	\subfigure[]{\includegraphics[scale=0.32]{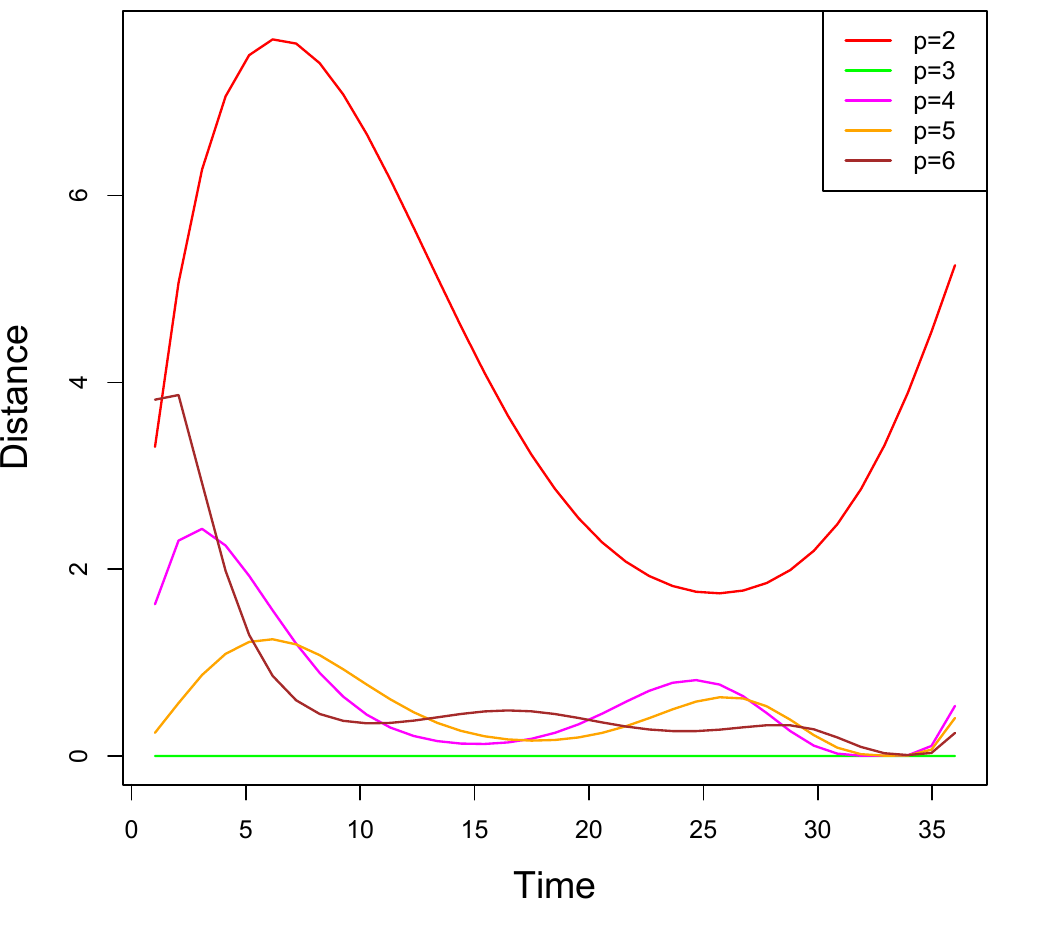}} \quad
	\subfigure[]{\includegraphics[scale=0.32]{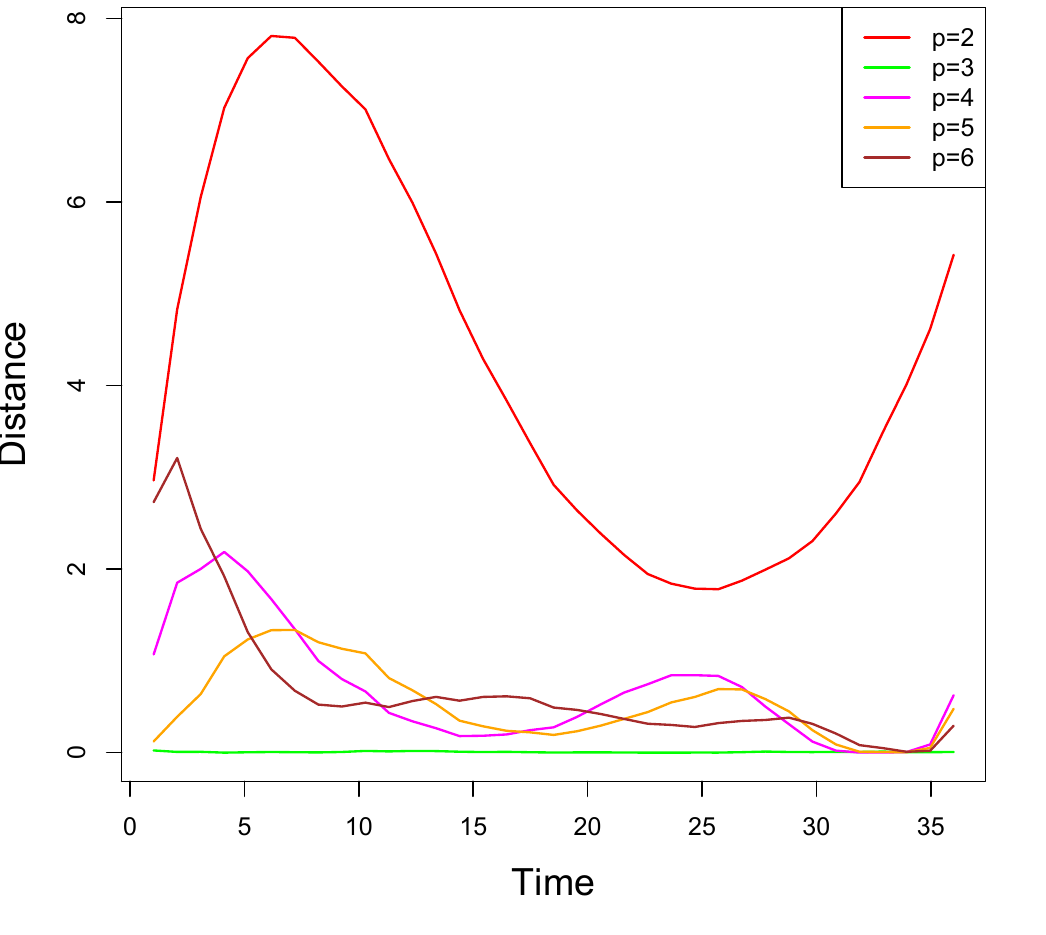}}
	\caption{Resistor-average distance between (a) the theoretical and the estimated distributions and (b) between the sample and the estimated distributions for the FPT density approximation (simulation study - FPT problem).}
	\label{fig:Figure10}
\end{figure}

In this section, the FPT problem is analyzed. With reference to a diffusion process $X(t)$ with a multisigmoidal logistic mean and $Q_\beta(t)= 0.1t-0.009t^2+0.0002t^3$, $\eta=e^{-1}$ and $\sigma=0.01$, we construct $50$ simulated sample paths (see Figure \ref{fig:Figure9}-(a)), each one being formed by $361$ data simulating $X(t_i)$ for $t_i=(i-1)\, 0.1$, $i=1,\dots, 361$. As in Section 5, we first chose the optimal polynomial degree (which corresponds to the best fit), and then we found
the MLEs of the parameters by solving the system \eqref{sist3}.
Further on, the \textbf{\textsf{R}} package \textsf{fptdApprox} is used to approximate the FPT density of the process through a constant threshold $S=15$.
Table \ref{tab:Tabella2003} provides the estimated parameters, whereas Figure \ref{fig:Figure9}-(b) shows the theoretical, the sample and the estimated means, for $p=2,3,4,5,6$.
Table \ref{tab:Tabella2003a} provides the four goodness measures for the considered degrees $p$ of the polynomial $Q_\beta$.   Figure \ref{fig:Figure10} shows the resistor-average distances between the theoretical and the estimated distributions,  and also between the sample and the estimated distributions. From the given results it follows that the best
degree is $p=3$.
\begin{table}
	\caption{The estimated values of the parameters considering $p=2,3,4,5,6$ for the FPT density approximation (simulation study - FPT problem).}
	\label{tab:Tabella2003}
	\tiny
	\centering
	\begin{tabular}{r|r|r|r|r}
		degree 					&$\eta$							&$\beta_1$					&$\beta_2$		&$\beta_3$ \\ \hline
		$p=2$					&$2.5287214$			&$-0.09696011$			&$0.002712063$	&-- \\
		$p=3$					&$0.3425411$			 &$0.09735434$			&$-0.008730983$	&$0.0001936518$\\
		$p=4$					&$0.9005909$			&$0.08465459$			&$-0.003953699$		&$-0.0001821386$\\
		$p=5$					&$0.9005909$			&$0.12376611$			&$-0.014259773$		&$0.0006907176$\\
		$p=6$					&$0.9005909$ 			&$0.04089375$			&$0.016958673$		&$-0.0032783145$\\
	\end{tabular}
	\begin{tabular}{r|r|r|r|r}
		degree 					&$\beta_4$							&$\beta_5$					&$\beta_6$		&$\sigma^2$ \\ \hline
		$p=2$					&--										 &--							   &--					&$0.0017740001$ \\
		$p=3$					&--										&--									&--					&$1.025400e$-$04$\\
		$p=4$					&$7.788128e$-$06$				&--									&--					&$8.511030e$-$05$\\
		$p=5$					&$-2.178008e$-$05$			&$3.477652e$-$07$				&--					&$8.511030e$-$05$\\
		$p=6$					&$2.024839e$-$04$			&$-5.459791e$-$06$		&$5.624755e$-$08$	&$8.511030e$-$055$\\
	\end{tabular}
\end{table}
\begin{table}
	\caption{The $RAE$, the $AIC$, the $BIC$ the median and the mean of the resistor-average distance $D_{RA}$ of the parameters for $p=2,3,4,5,6$. For the resistor-average distance, the estimated and the theoretical distributions are considered (simulation study - FPT problem).}
	\label{tab:Tabella2003a}
	\tiny
	\centering
	\begin{tabular}{r|r|r|r|r|r}
		measure of goodness 					&$p=2$							&$p=3$					&$p=4$		&$p=5$		&$p=6$ \\ \hline
		$RAE$					&$0.313009817$			&$0.001985486$ 		&$0.046745173$		&$0.044842133$		&$0.046064145$\\
		$AIC$					&$-9288.882$			&$-14314.158$				&$-10687.536$		&$-11569.711$		&$-12001.874$ \\
		$BIC$					&$-9267.012$			&$-14286.821$				&$-10654.731$		&$-11531.439$		&$-11958.135$\\
		median of $D_{RA}$	&$3.767780356$	&$0.003568808$	&$0.40969122$	&$0.347181473$	&$0.404933011$\\
		mean of $D_{RA}$ 	&$4.276620795$		&$0.009886785$	&$0.617753008$		&$0.442544796$	&$0.582836790$
	\end{tabular}
\end{table}
%
\par
\begin{figure}[t]
	\centering
	\subfigure[]{\includegraphics[scale=0.33]{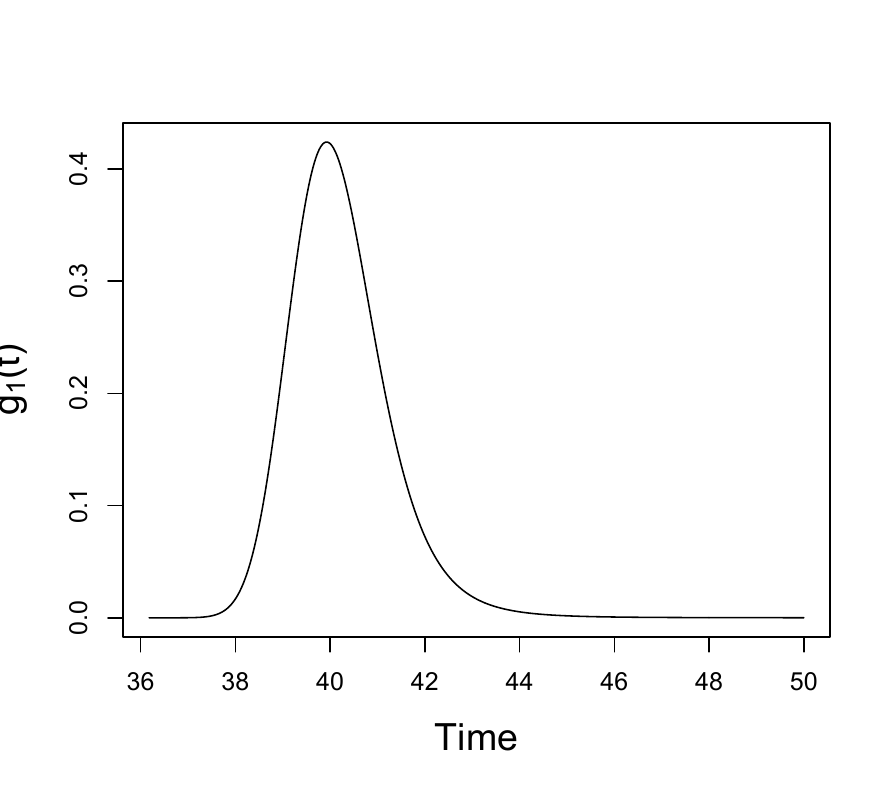}} \quad
	\subfigure[]{\includegraphics[scale=0.33]{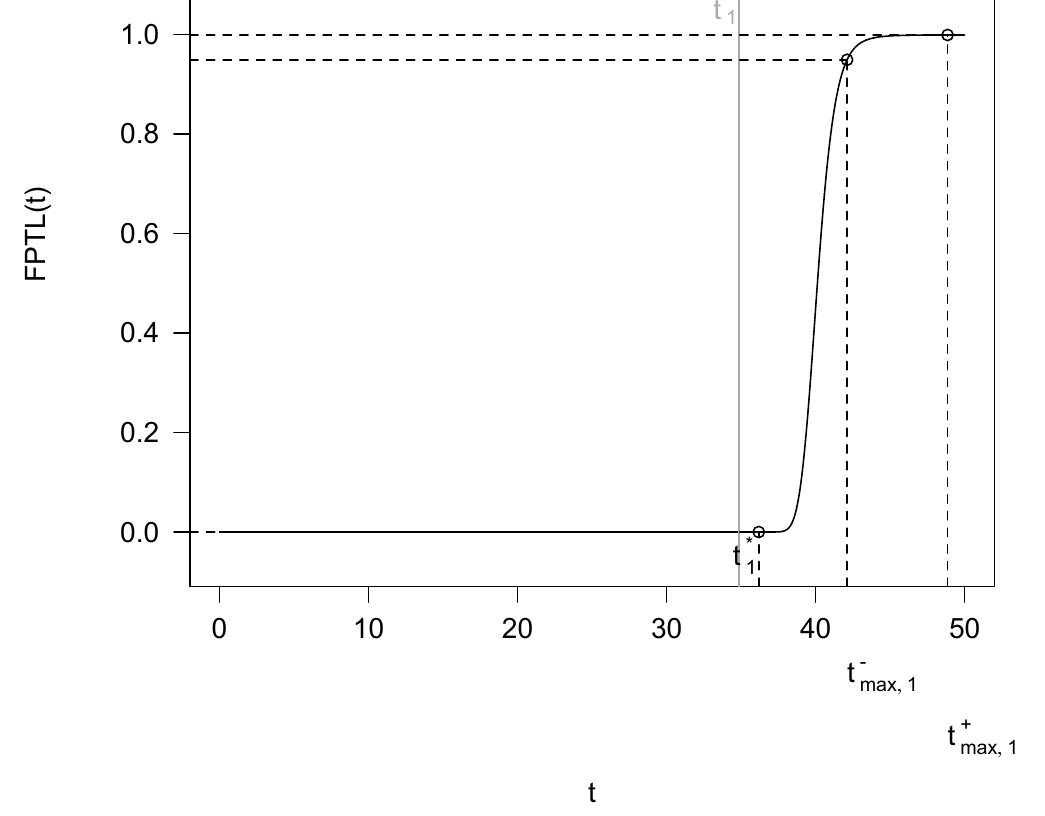}}
	\caption{The approximated FPT density and the FPTL function of the process $X(t)$ through the boundary $S=15$ (simulation study - FPT problem).}
	\label{fig:Figure11}
\end{figure}

Using the estimated model obtained so far, we now focus on the approximation of the FPT density through the boundary $S=15$. Figure \ref{fig:Figure11} shows the approximated FPT density and the FPTL function realized with the package \textsf{fptdApprox}. Finally, in Table \ref{tab:Tabella2003b} other useful quantities related to the FPT density are provided. It is worth noting that the results obtained in this section are in agreement  with those given in Example 4.1.
%
\begin{table}[t]
	\caption{The mean, the standard deviation, the mode, the first, the fifth and the ninth decile of the FPT of the process $X(t)$ through the boundary $S=15$ (simulation study - FPT problem).}
	\label{tab:Tabella2003b}
	\centering
	\begin{tabular}{l|l|l|l|l|l}
		mean 			& st. dev.                	& mode                	& $1^{st}$ decile                   & $5^{th}$ decile  	& $9^{th}$ decile \\ \hline
		$39.88883$ 		&$1.034443$ 	 &$39.70804$ 			&$38.83282$ 						&$39.83612$			& $41.0529$
	\end{tabular}
\end{table}
\section{Application to real data}
\begin{figure}[t]
	\centering
	\subfigure[]{\includegraphics[scale=0.32]{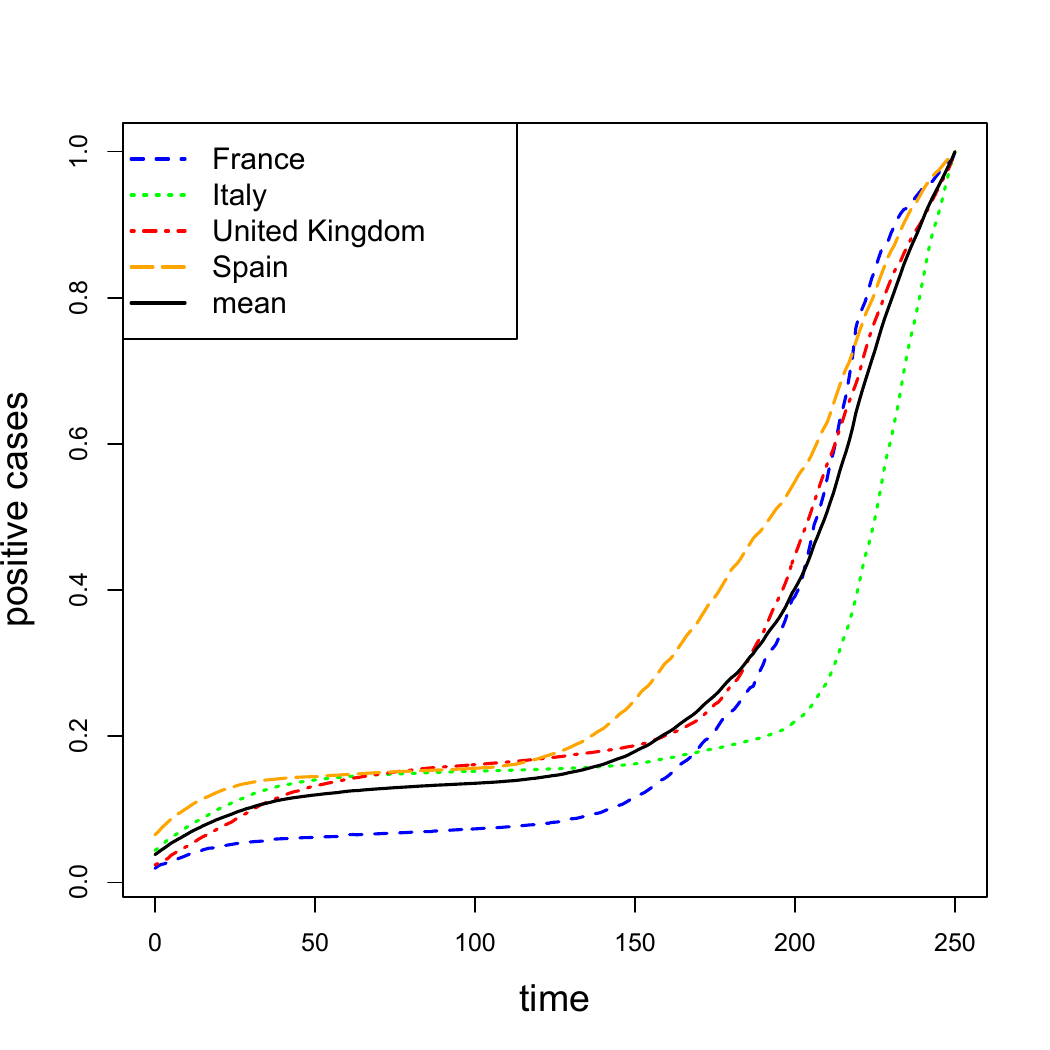}} \quad
	\subfigure[]{\includegraphics[scale=0.32]{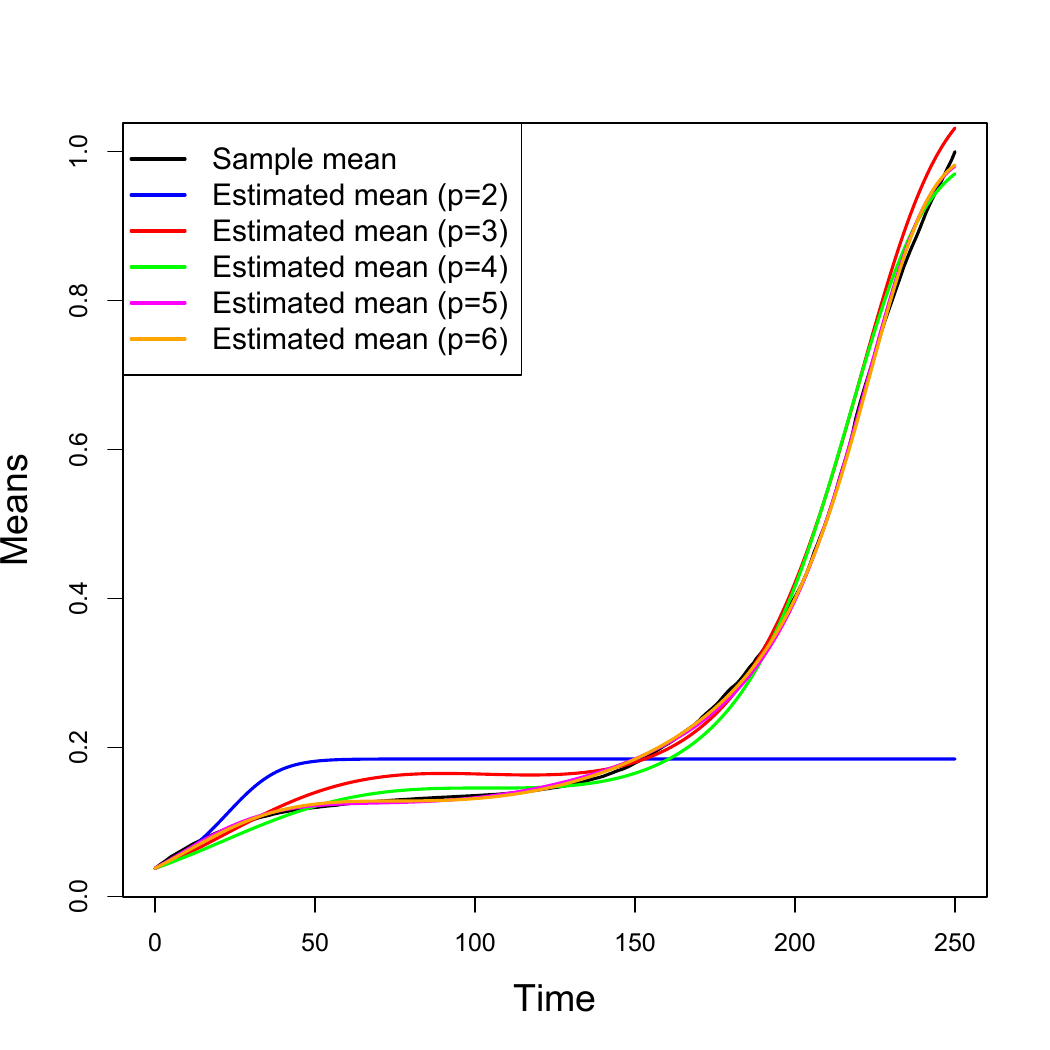}}
	\caption{(a) Number of infections in France, Italy, Spain and United Kingdom, the black line represents the sample mean. (b) Sample and the estimated means obtained by solving the system \eqref{sist3} (real application).}
	\label{fig:Figure12}
\end{figure}

\begin{figure}[t]
	\centering
	\subfigure[]{\includegraphics[scale=0.32]{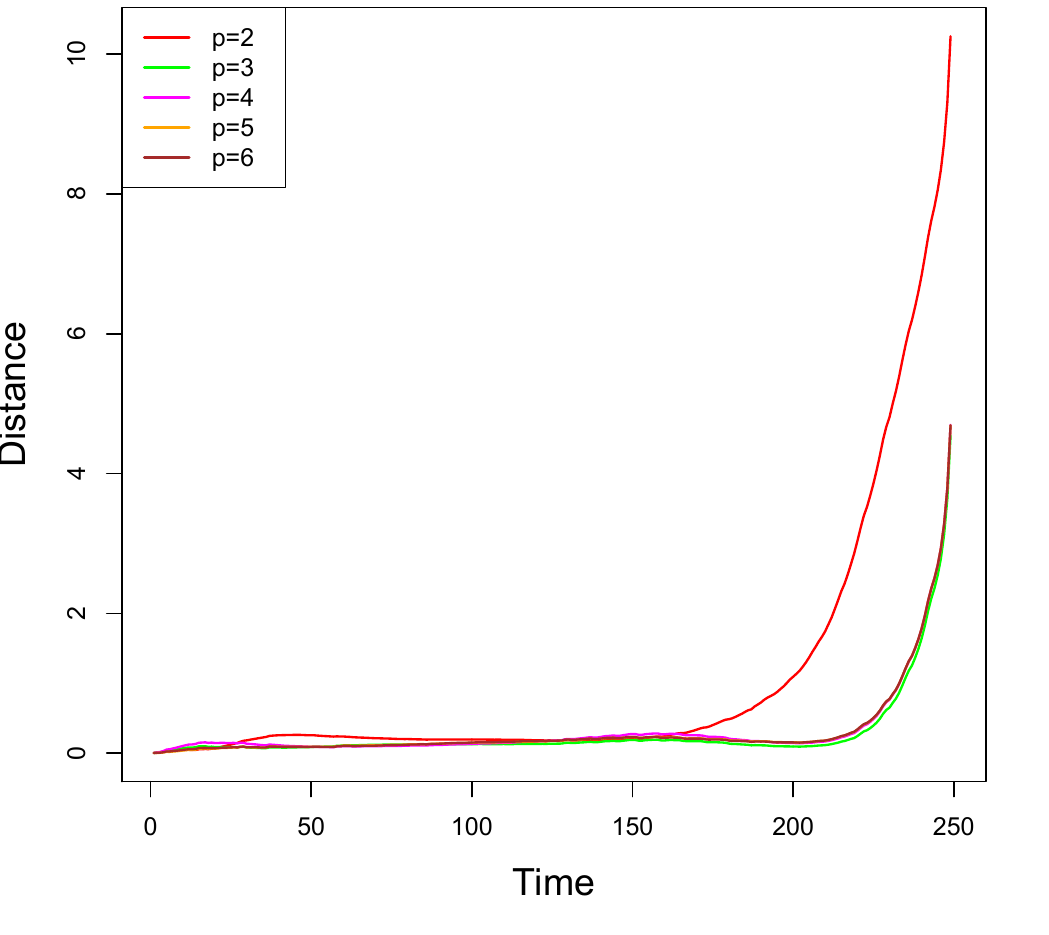}}\quad
	\subfigure[]{\includegraphics[scale=0.32]{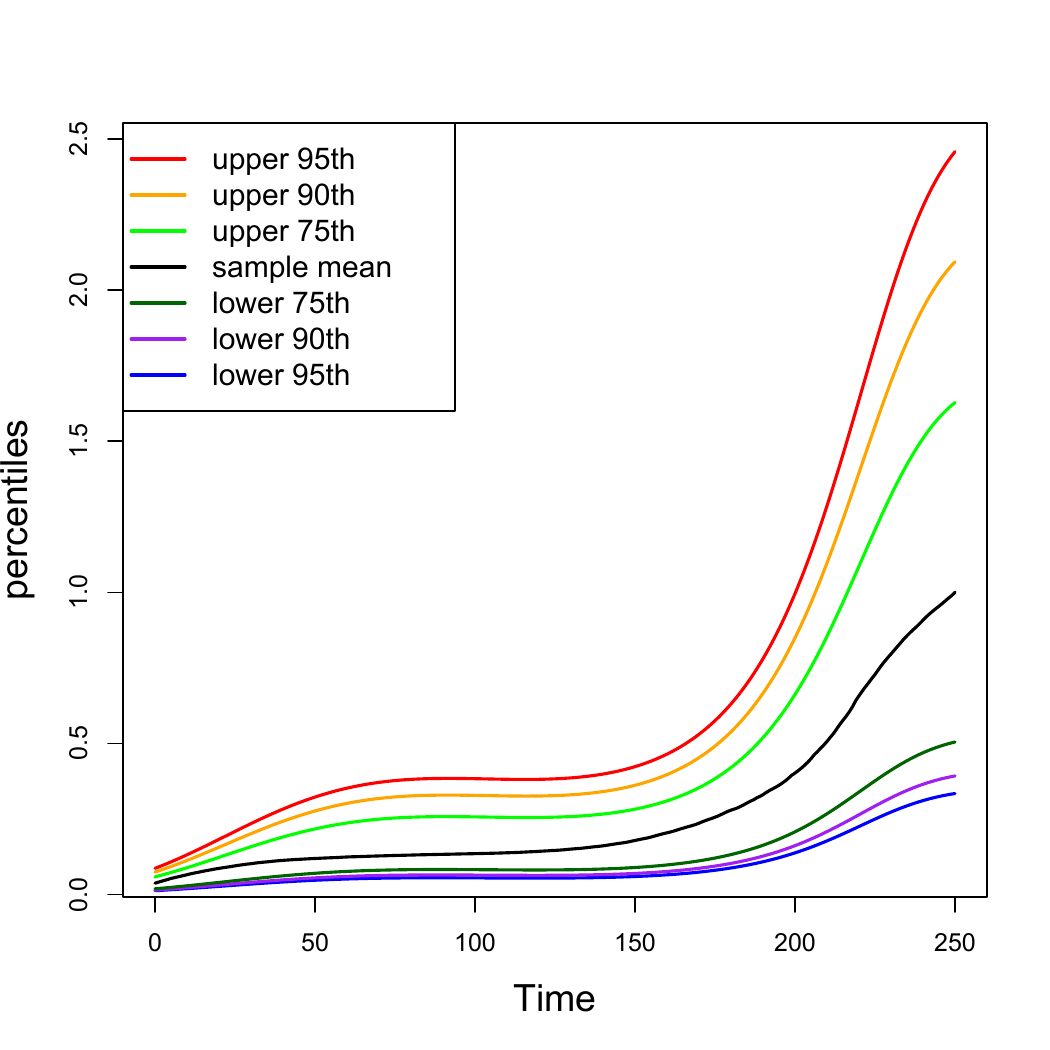}}
	\caption{(a) The resistor-average distances between the sample and the estimated distributions considering different degrees of the polynomial. (b) The $\alpha$-percentiles of the estimated diffusion process $X(t)$ obtained for a degree $p=3$ and for $\alpha=95,90,75$ (real application).}
	\label{fig:Figure13}
\end{figure}

\begin{figure}[t]
	\centering
	\subfigure[]{\includegraphics[scale=0.32]{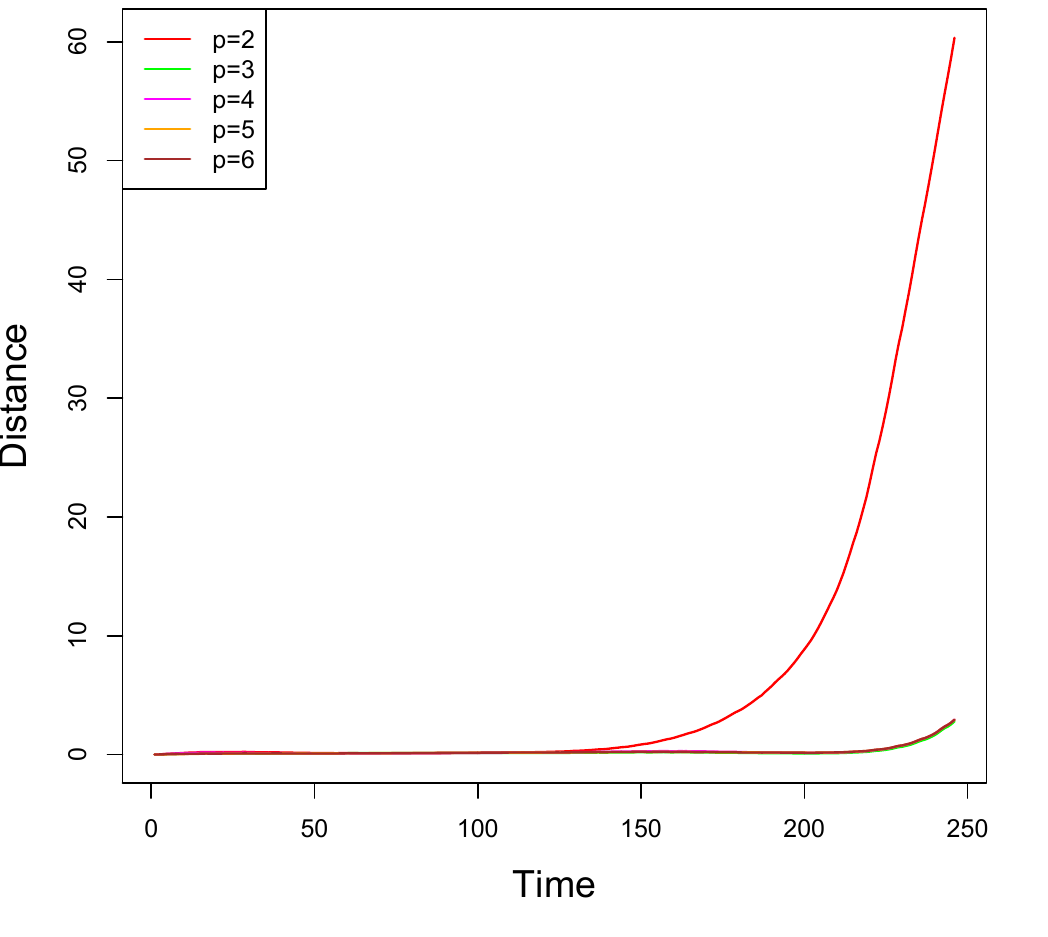}} \quad
	\subfigure[]{\includegraphics[scale=0.32]{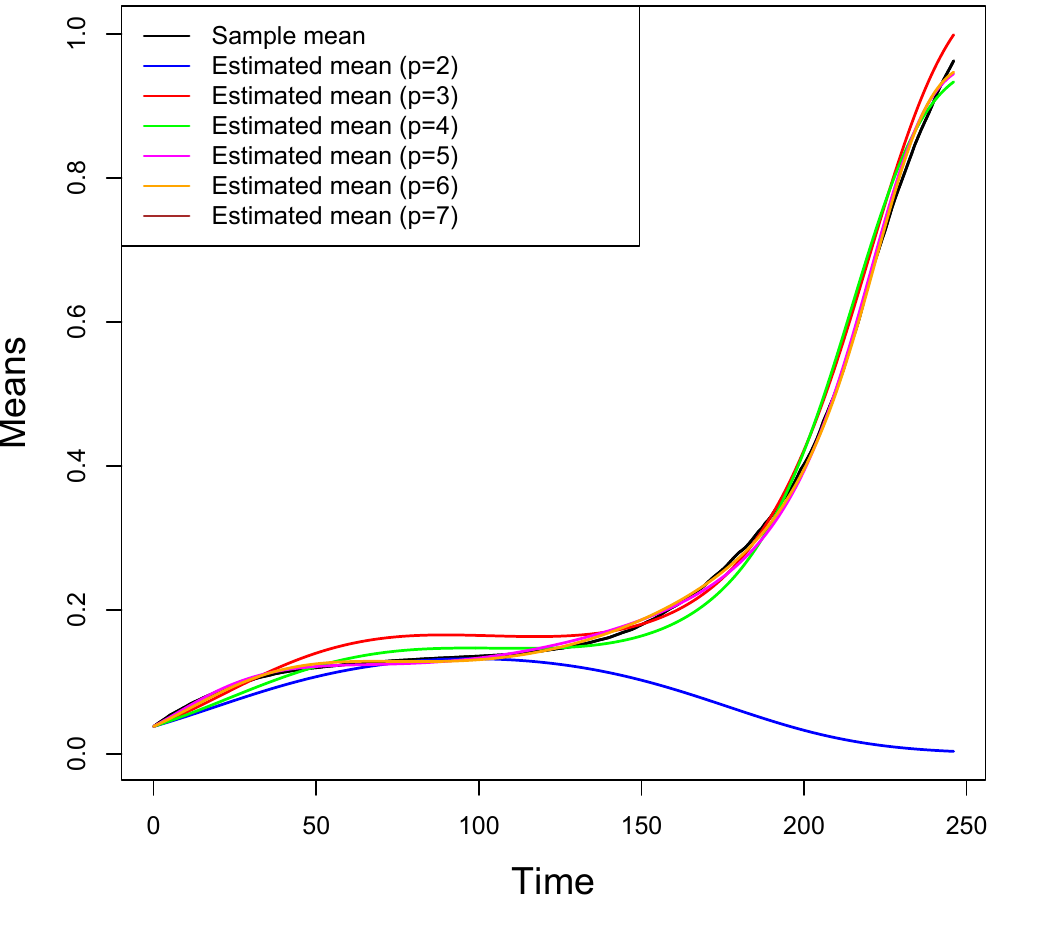}}
	\caption{(a) Resistor-average distances between the sample and the estimated distributions in the restricted time range. (b) Sample and the estimated means obtained by solving the system \eqref{sist3} in the restricted time range (real application with $t_f=246$).}
	\label{fig:Figure14}
\end{figure}

Multisigmoidal functions are suitable to model several special growth phenomena in which the carrying capacity is reached after various stages. In any of these stages a linear growth trend is followed by an explosion of exponential type which finally flattens to a specific value. A growth of this kind is typical of some fruit species, such as peaches or coffee berries (see, for instance the application given in Section 3 of Di Crescenzo {\it et al.}\ (2020) \cite{DiCrescenzoetal2020}). But also some population diseases follow an expansion  with a multisigmoidal trend.
In this section we apply the considered stochastic model to data concerning the COVID-19 infections in four different European countries, taken from \cite{worldometer}.
This is just an example finalized to show an application of the multisigmoidal logistic model, without taking into account specific more sophisticated models that describe epidemiological phenomena with greater precision.
First of all, we note that the trend of infections in France, Italy, Spain and United Kingdom is similar (see Figure \ref{fig:Figure12}-(a)).
This suggests to view these data as different trajectories of the diffusion process $X(t)$ defined on $I=[t_0,t_f]$, having a multisigmoidal logistic mean (cf.\ Section \ref{2.2}). Hence, in order to find the MLEs of the parameters, we apply the procedure described in Section \ref{section3.1}. {For each country, the initial time $t_0=0$ corresponds to the $30$-th day after the one in which the number of infections exceeded $100$ (March 30th for France, March 24th for Italy,  March 21st for Spain, April 5th for UK),} and the final time is chosen as $t_f=250$. For any path, the data are scaled as divided by their maximum value, so to be interpreted as a percentage of the last and therefore the maximum value of the growth curve.
The estimated means obtained for different degrees are plotted in Figure \ref{fig:Figure12}-(b). Table \ref{Tab:table11} provides the initial and the estimated values of the parameters, whereas Table \ref{Tab:table11b} shows the four measures of goodness, for different degrees of the polynomial. Regarding the $RAE$, every time the degree increases, the approximation improves, whereas the $AIC$, the $BIC$ and resistor-average distance show that the best choice is $p=3$.
\begin{table}[t]
	\caption{The initial and the estimated values of the parameters considering different degrees. The results have been obtained by solving the system (real application).}
	\label{Tab:table11}
	\centering
	\tiny
	\begin{tabular}{l|l|l|l|l|l}
		& value     										&$\beta_1$ 						&$\beta_2$ 						&$\beta_3$ 							&$\beta_4$ \\ \hline
		\multirow{2}{*}{$p=2$} & initial    	&$0.00644346$               &$5.375995e$-$05$            &--                         				&--                         \\
		& estimated &$0.05356653$               &$0.0011042774$             &--                                      &--                         \\ \hline
		\multirow{2}{*}{$p=3$} & initial   		&$0.04480694$               &$-4.587843e$-$04$           &$1.540731e$-$06$                &--                         \\
		& estimated &$0.04774851$                &$-0.0004685118$           &$1.506227e$-$06$                &--                         \\ \hline
		\multirow{2}{*}{$p=4$} & initial        &$0.04090832$               &$-3.650280e$-$4$             &$8.831071e$-$07$                &$1.405770e$-$09$                         \\
		& estimated &$0.04090832$               &$-0.0003650280$          &$8.831071e$-$07$                 &$1.405770e$-$09$                         \\ \hline
		\multirow{2}{*}{$p=5$} & initial   		&$0.06509648$                 &$-1.269895e$-$03$           &$1.176365e$-$05$                 &$-5.092724e$-$08$                         \\
		& estimated &$0.06509648$                 &$-0.0012698948$           &$1.176365e$-$05$                 &$-5.092724e$-$08$                         \\ \hline
		\multirow{2}{*}{$p=6$} & initial   		&$0.05891632$                 &$-9.396086e$-$04$            &$5.806280e$-$06$               &$-3.170420e$-$09$                         \\
		& estimated &$0.05891632$                 &$-0.0009396086$            &$5.806280e$-$06$               &$-3.170420e$-$09$
	\end{tabular}
	\begin{tabular}{l|l|l|l|l|l}
		& value     										&$\beta_5$ 						&$\beta_6$ 						&$\eta$ 							&$\sigma^2$ \\ \hline
		\multirow{2}{*}{$p=2$} & initial    	&--               					  &--           						&$0.03942583$                 &$1.931962e$-$02$                         \\
		& estimated &--               					  &--             						&$0.25862727$                  &$2.454394e$-$04$                         \\ \hline		
		\multirow{2}{*}{$p=3$} & initial   		&--               					  &--           						&$0.03942583$                &$1.931962e$-$02$                          \\
		& estimated &--             					&--           							&$0.03605835$                &$1.024422e$-$04$                         \\ \hline
		\multirow{2}{*}{$p=4$} & initial        &--           						&--          							 &$0.03942583$                &$1.931962e$-$02$                             \\
		& estimated &--       							&--         							 &$0.03942583$                 &$3.729000e$-$04$                         \\ \hline
		\multirow{2}{*}{$p=5$} & initial   		&$8.739928e$-$11$                 &--       							&$0.03942583$                 &$1.931962e$-$02$                      \\
		& estimated &$8.739928e$-$11$                &--          							&$0.03942583$                &$3.72900e$-$04$                         \\ \hline
		\multirow{2}{*}{$p=6$} & initial   		&$-8.806843e$-$11$                 &$2.411350e$-$13$            &$0.03942583$               &$1.931962e$-$02$                           \\
		& estimated &$-8.806843e$-$11$              &$2.411350e$-$13$            &$0.03942583$                 &$3.72900e$-$04$
	\end{tabular}
\end{table}
\begin{table}
	\caption{$RAE$, $AIC$, $BIC$, the median and the mean of the resistor-average distance $D_{RA}$ considering different degrees. For the resistor-average distance, the estimated and the sample distributions are considered (real application).}
	\label{Tab:table11b}
	\centering
	\tiny
	\begin{tabular}{r|r|r|r|r|r}
		$p$ 		&$2$          					&$3$        					 &$4$       						&$5$             					&$6$ 		\\ \hline
		$RAE$ 	  &$0.39085660$ 	            &$0.10404805$ 	           &$0.06853298$  	          &$0.02087067$  	           &$0.02030226$		  \\
		$AIC$ 	   &$-7225.192$ 		         &$-8175.819$               &$-7582.180$  	    		&$-7677.528$  	                &$-7664.849$	\\
		$BIC$	   &$	-7205.561$                 &$-8151.281 $                &$-7552.734 $                 &$-7643.173$        &$-7625.587$ \\
		median of $D_{RA}$	&$0.2250753$	&$0.1301707$		&$0.1548300$		&$0.1615877$		&$0.1642668$	\\
		mean of $D_{RA}$		&$1.0586904$	&$0.2741844$	&$0.3218632$	&$0.3103206$	&$0.3106608$
	\end{tabular}
\end{table}
Regarding the last measure of goodness, see also Figure \ref{fig:Figure13}-(a), in which the resistor-average distances between the sample and the estimated distributions are provided. Hence, in view of the results obtained for the measures of goodness, the degree $p=3$ is considered.
Table \ref{tab:Tabella112} shows the estimated values of the parameters, the estimation of their standard error and the $95\%$, $90\%$ and $75\%$ percentiles.
\begin{table}
	\caption{The estimated values, the standard error and the $95\%$, $90\%$ and $75\%$ confidence intervals for the parameters, considering $p=3$ (real application).}
	\label{tab:Tabella112}
	\tiny
	\centering
	\begin{tabular}{r|r|r|r}
		Parametric function 		&$\eta$          					&$\beta_1$        					 &$\beta_2$       							\\ \hline
		Estimated value 	 		 &$0.03605835$ 	               &$0.04774851$ 	           		&$-0.0004685118$  	              \\
		Standard error 	   			  &$6.086478e$-$03$ 		       &$2.342236e$-$03$               &$2.825993e$-$05$  	    		  \\
		$95\%$ confidence interval	   &$(0.0241291,0.0479876)$                 &$(0.0431578,0.0523392 )$                &$(-0.0005239,-0.0004131)$                  \\
		$90\%$ confidence interval	   &$(0.0260470,0.0460697)$                 &$(0.0438959,0.0516011)$                &$(-0.0005150,-0.0004220)$                 \\
		$75\%$ confidence interval	   &$(0.0290568,0.0430599)$                 &$(0.0450541,0.0504429)$                &$(-0.0005010,-0.0004360)$                 \\
	\end{tabular}
	\begin{tabular}{r|r|r}
		Parametric function 		      						&$\beta_3$             					&$\sigma^2$ 		\\ \hline
		Estimated value 	 		   	             &$1.506227e$-$06$  	                &$1.024422e$-$04$		  \\
		Standard error 	   			    	    		  &$9.15127e$-$08$  	                &$4.581353e$-$06$	\\
		$95\%$ confidence interval	                   &$(0.0000013,0.0000017)$        &$(0.0000934,0.0001114)$ \\
		$90\%$ confidence interval	   &$(0.0000014,0.0000017)$        &$(0.0000949,0.0001010)$ \\
		$75\%$ confidence interval	  &$(0.0000014,0.0000016)$        &$(0.0000971,0.0001077)$ \\
	\end{tabular}
\end{table}
Moreover, the $\alpha$-percentiles \eqref{alphapercentile} of the estimated diffusion process with $p=3$ are provided in Figure \ref{fig:Figure13}-(b) with $\alpha=95,90,75$.
Let us now consider a restricted time range from $t_0=0$ to $t_f=246$, in order to predict the trend of the growth curve in a short-term prediction analysis. Indeed, forecasting the number of infections during a disease in progress is interesting also in the case of short terms, especially for the goodness of estimation (better in this case than in the long term analysis) and for the timeliness of the results. The considered procedure is the same of the one used above, so (i) the best degree $p$ for the polynomial $Q_\beta$ is chosen by considering various measures of goodness, and (ii) the estimated values of the parameters are used to construct a diffusion process $X(t)$ defined on $I=[0,250]$. The estimated values of the parameters are given in Table \ref{tab:Tabella1703a}, the values of the four measures of goodness are given in Table \ref{tab:Tabella1703b}, and finally in Figure \ref{fig:Figure14}-(a) we provide the resistor-average distances between the restricted sample and the estimated distributions.  See also Figure \ref{fig:Figure14}-(b) for the plots of the estimated means for different degrees of the polynomial.
\begin{table}
	\caption{The estimated values of the parameters considering $p=2,3,4,5,6$ (real application with $t_f=246$).}
	\label{tab:Tabella1703a}
	\tiny
	\centering
	\begin{tabular}{r|r|r|r|r}
		degree 					&$\eta$							&$\beta_1$					&$\beta_2$		&$\beta_3$ \\ \hline
		$p=2$					&$0.22079965$			&$0.04165889$			&$-0.0002129137$	&-- \\
		$p=3$					&$0.03650929$			&$0.04792200$			&$-0.0004714958$	&$1.518106e$-$06$\\
		$p=4$					&$0.04101207$			 &$0.03985192$			&$-0.0003277795$		&$5.314482e$-$07$\\
		$p=5$					&$0.04101207$			&$0.06778144$			&$-0.0013896339$		&$1.350773e$-$05$\\
		$p=6$					&$0.04101207$ 			&$0.05704750$			&$-0.0008066323$		&$2.820787e$-$06$\\
	\end{tabular}
	\begin{tabular}{r|r|r|r|r}
		degree 					&$\beta_4$							&$\beta_5$					&$\beta_6$		&$\sigma^2$ \\ \hline
		$p=2$					&--										 &--							   &--					&$2.536393e$-$04$ \\
		$p=3$					&--										&--									&--					&$ 1.034147e$-$04$\\
		$p=4$					&$2.392264e$-$09$				&--									&--					&$3.550000e$-$04$\\
		$p=5$					&$-6.103779e$-$08$			&$1.076581e$-$10$				&--					&$3.550000e$-$04$\\
		$p=6$					&$ 2.602935e$-$08$			&$-2.174560e$-$10$		&$4.540655e$-$13$	&$3.550000e$-$04$\\
	\end{tabular}
\end{table}
\begin{table}
	\caption{The $RAE$, the $AIC$, the $BIC$, the median and the mean of the resistor-average distance $D_{RA}$ of the parameters for $p=2,3,4,5,6$. For the resistor-average distance, the estimanted and the sample distributions are considered (real application with $t_f=246$).}
	\label{tab:Tabella1703b}
	\tiny
	\centering
	\begin{tabular}{r|r|r|r|r|r}
		measure of goodness 					&$p=2$							&$p=3$					&$p=4$		&$p=5$		&$p=6$ \\ \hline
		$RAE$					&$0.41031587$			&$0.10421156$ 		&$0.07598928$		&$0.02665995$		&$0.02500666$\\
		$AIC$					&$-2722.134$			&$-8035.566$				&$-7476.246$		&$-7591.347$		&$-7571.929$ \\
		$BIC$					&$-2702.568$			&$-8011.108$				&$-7446.896$		&$-7557.105$		&$-7532.796$\\
		median of $D_{RA}$		&$0.2275183$		&$0.1294749$		&$0.1557215$		&$0.1593042$		&$0.1626228$\\
		mean of $D_{RA}$		&$6.2370324$		&$0.2212618$		&$0.2666561$		&$0.2532703$		&$0.2536295$
	\end{tabular}
\end{table}
%
Also in this case, the $RAE$ does not provide a good measure of goodness, since every time the degree increases, the approximation improves.
From the remaining results, the best choice is $p=3$, which corresponds to the lowest value of the $BIC$ and of the $AIC$, and to the lowest resistor-average distance.
Hence, considering $p=3$ and the corresponding estimated values of the parameters, Figure \ref{fig:Figure15}-(a) provides the sample and the estimated means in the complete time range, i.e.\ in $I_C=[0,250]$. The relative errors between the values of the sample and the estimated means are given in Table \ref{tab:Tabella1703c}; note that in all cases they are less than $3\%$.

\begin{table}
	\caption{The sample, the forecasted mean and relative error for the last $4$ time instants of the complete range $I_C=[0,250]$, considering a degree $p=3$ (real application).}
	\label{tab:Tabella1703c}
	\small
	\centering
	\begin{tabular}{r|rrrr}
		time 					&$247$							&$248$					&$249$		&$250$	 \\ \hline
		sample mean		&$0.97179233$				&$0.98096101$		&$0.98949697$	&$1.00000000$\\ \hline
		forecasted mean	&$1.00554923$			&$1.01157710$		&$1.01720499$	&$1.02244759$\\ \hline
		relative error		&$0.03473674$			&$0.0312103$ 		&$0.02800212$		&$0.02244759$	\\
	\end{tabular}
\end{table}
%
\subsection{Approximation of FPT density}		
This section is devoted to the FPT problem. Considering an initial portion of the available data and setting a constant boundary, an estimate of the FPT density is constructed using the numerical procedures recalled in Section \ref{sec:FPT}. The resulting FPT density is then compared to the approximated FPT density obtained by using the whole data set.
\par
\begin{figure}[t]
	\centering
	\subfigure[]{\includegraphics[scale=0.32]{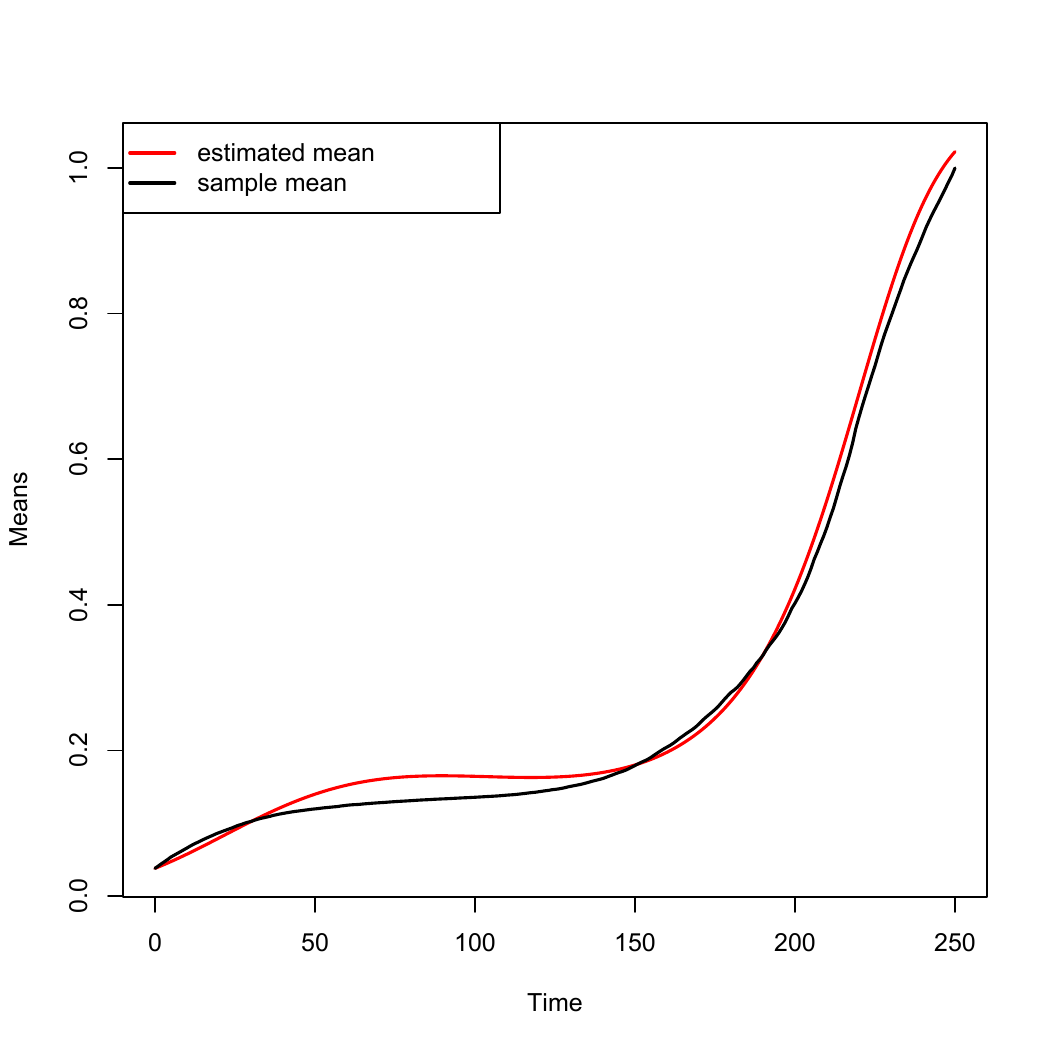}}
	\subfigure[]{\includegraphics[scale=0.32]{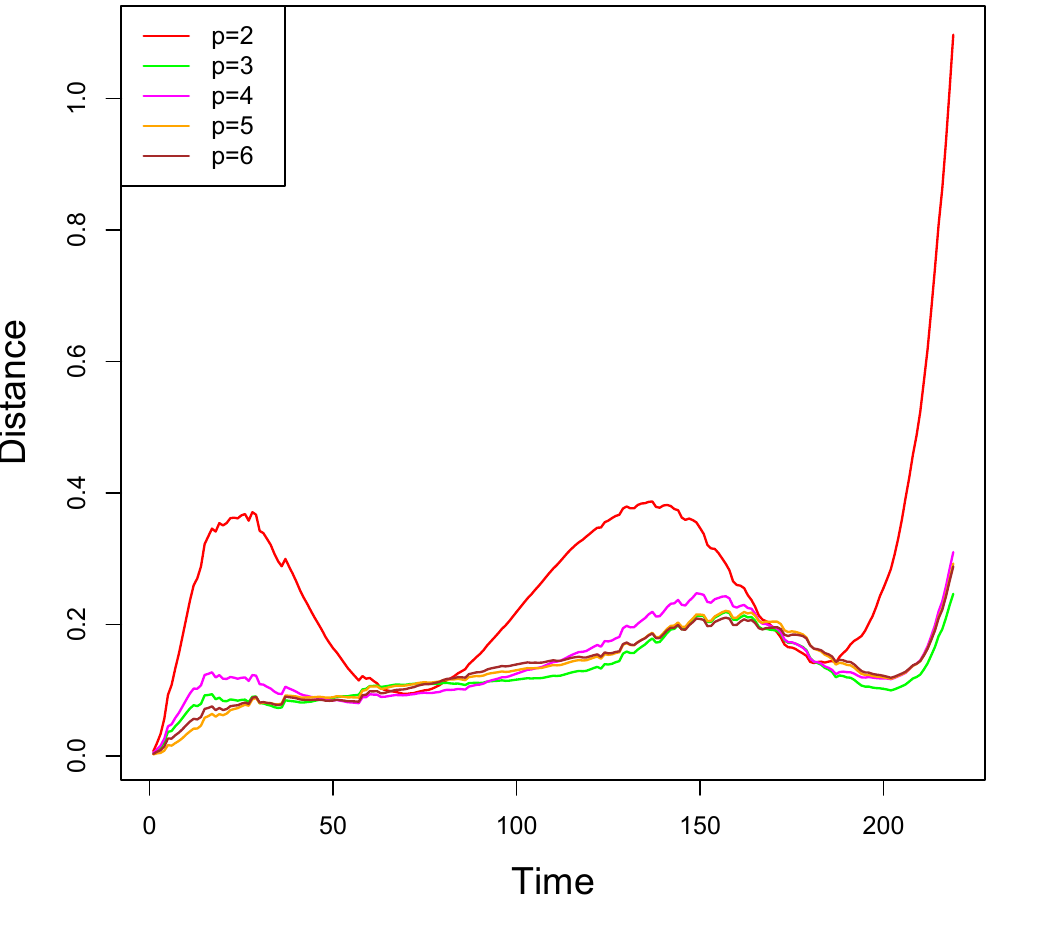}}
	\subfigure[]{\includegraphics[scale=0.32]{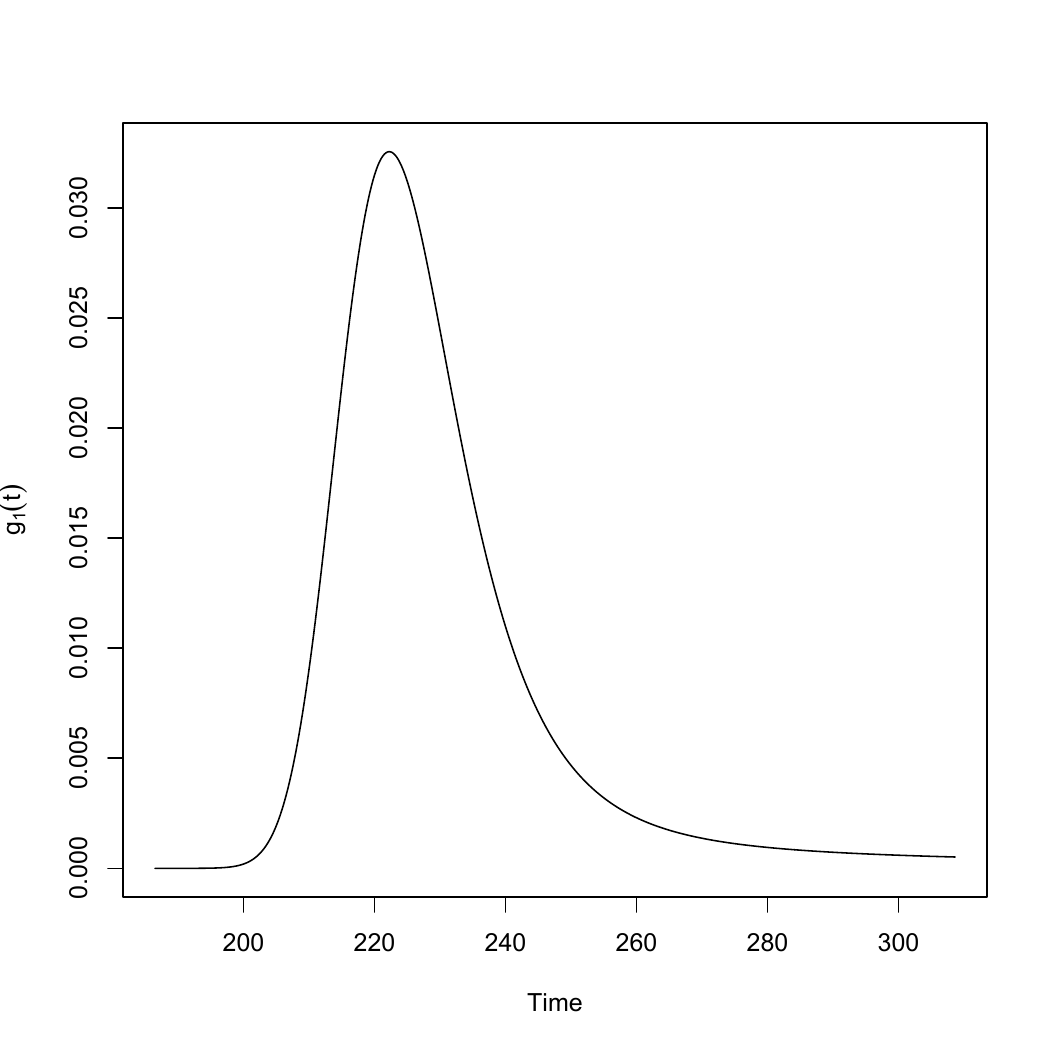}}
	\subfigure[]{\includegraphics[scale=0.32]{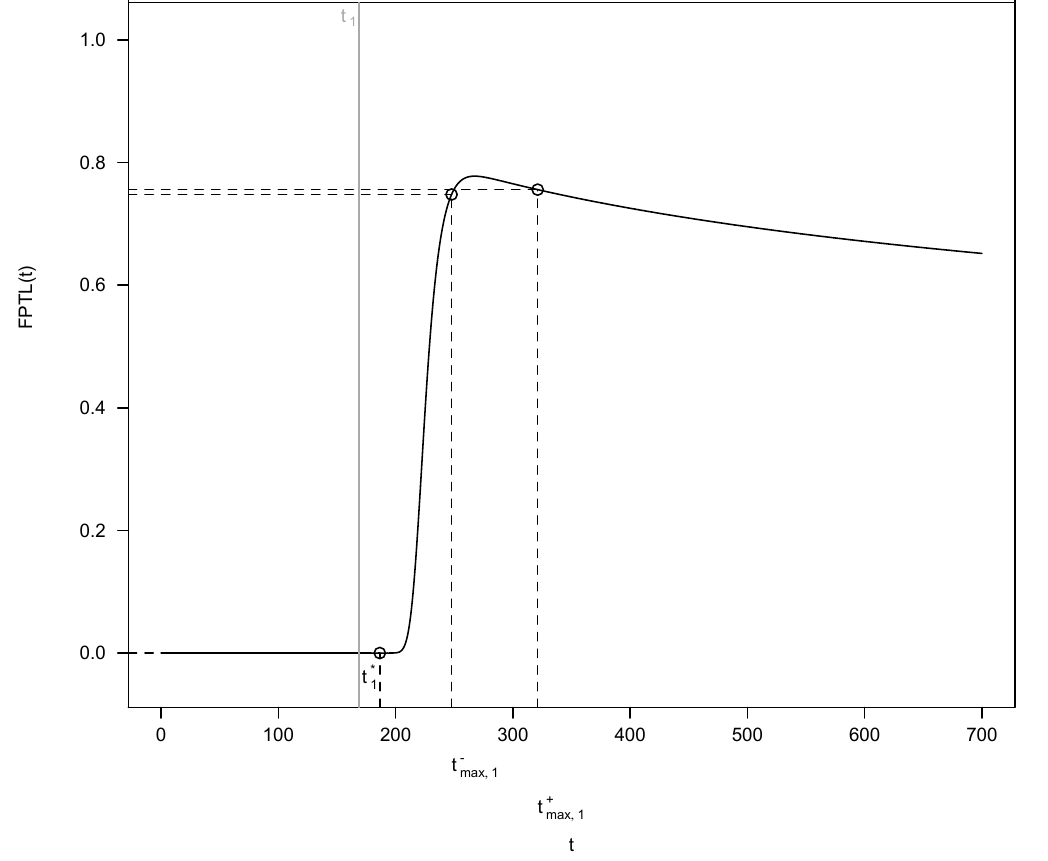}}
	\caption{(a) The sample and the forecasted means in the complete time range $I_C=[0,250]$ considering a degree $p=3$. (b) Resistor-average distances between the sample and the estimated distributions in the restricted time range $I_R$. (c) Approximated FPT density  and (d) FPTL function in the restricted time range $I_R$ through the boundary $S=0.7$ (real application - FPT problem).}
	\label{fig:Figure15}
\end{figure}

More in detail, we consider only the first $220$ data of COVID-19 infections in the restricted time range $I_R=[0, 219]$ and we investigate the best model to fit them. The choice of the optimal degree $p$ of the polynomial $Q_\beta$ is based on  the measures of goodness (i)--(iv) described in Section \ref{sec:simulation}. The estimated parameters (given in Table \ref{tab:Tabella2403}) are obtained by solving the system \eqref{sist3}.
\begin{table}
	\caption{The estimated values of the parameters considering $p=2,3,4,5,6$ in the restricted time range $I_C$ (real application - FPT problem).}
	\label{tab:Tabella2403}
	\tiny
	\centering
	\begin{tabular}{r|r|r|r|r}
		degree 					&$\eta$							&$\beta_1$					&$\beta_2$		&$\beta_3$ \\ \hline
		$p=2$					&$0.09416137$			&$0.02514812$			&$-4.247417e$-$05$	&-- \\
		$p=3$					&$0.04858678$			&$0.05004022$			&$-5.125970e$-$04$	&$1.693253e$-$06$\\
		$p=4$					&$0.06278231$			 &$0.04592707$			&$-4.462889e$-$04$		&$1.183743e$-$06$\\
		$p=5$					&$0.06278231$			&$0.07198803$			&$-1.559548e$-$03$		&$1.646946e$-$05$\\
		$p=6$					&$0.06278231$ 			&$0.05899141$			&$-7.664128e$-$04$		&$1.337821e$-$07$\\
	\end{tabular}
	\begin{tabular}{r|r|r|r|r}
		degree 					&$\beta_4$							&$\beta_5$					&$\beta_6$		&$\sigma^2$ \\ \hline
		$p=2$					&--										 &--							   &--					&$2.63974e$-$04$ \\
		$p=3$					&--										&--									&--					&$1.226112e$-$04$\\
		$p=4$					&$1.721755e$-$09$				&--									&--					&$2.359000e$-$04$\\
		$p=5$					&$-8.223093e$-$08$			&$1.600995e$-$10$				&--					&$2.359000e$-$04$\\
		$p=6$					&$6.730438e$-$08$			&$-4.672810e$-$10$		&$9.845081e$-$13$	&$2.359000e$-$04$\\
	\end{tabular}
\end{table}
By comparing the results given in Table \ref{tab:Tabella2403a} and in Figure \ref{fig:Figure15}-(b), we choose $p=3$ as the optimal degree.
\begin{table}
	\caption{The $RAE$, the $AIC$, the $BIC$, the median and the mean of the resistor-average distance $D_{RA}$ of the parameters for $p=2,3,4,5,6$. For the resistor-average distance, the estimated and the sample distributions are considered (real application - FPT problem).}
	\label{tab:Tabella2403a}
	\tiny
	\centering
	\begin{tabular}{r|r|r|r|r|r}
		measure of goodness 					&$p=2$							&$p=3$					&$p=4$		&$p=5$		&$p=6$ \\ \hline
		$RAE$					&$0.28460545$			&$0.08631377$ 		&$0.06598901$		&$0.02791701$		&$0.03225994$\\
		$AIC$					&$-6330.849$			&$-7150.963$				&$-6876.159$		&$-6999.085$		&$-6971.142$ \\
		$BIC$					&$-6311.748$			&$-7127.086$				&$-6847.507$		&$-6965.657$		&$-6932.939$\\
		median of $D_{RA}$ 	&$0.2448219$		&$0.1144287$		&$0.1238069$		&$0.1284333$		&$0.1355944$\\
		mean of $D_{RA}$		&$0.2635491$	&$0.1260033$		&$0.1405463$		&$0.1325032$		&$0.1328700$
	\end{tabular}
\end{table}
Then, we fix a constant threshold $S=0.7$ which corresponds to the $70\%$ of the last and maximum data in the complete time range $I_C=[0,250]$ and we use the \textbf{\textsf{R}} package \textsf{fptdApprox} to obtain an estimation of the FPT density.
The choice of the constant boundary $S=0.7$ is not random. Indeed, it is worth observing that the descendent inflection points correspond to the peaks of the function representing the daily increments of the infections. More in detail, by means of Eq.\ \eqref{eqinflec}, the function representing the sample mean of the infections shows two descendent inflection points, one at the time $t_{F1}=21.65$ and the other at the time $t_{F2}=220.66$. The population sizes corresponding to the inflection time instants are $S_{F1}=0.08$ and $S_{F2}=0.7$.  In the time interval $[0,t_{F1}]$, the mean function has a logistic trend, hence the FPT problem through the boundary ${S_{F1}}$ is beyond the scope of the present work. Instead, since the mean in the time interval $[0, t_{F2}]$ has a multisigmoidal logistic profile, we focus our attention to the FPT problem through the threshold $S=S_{F2}=0.7$.
\par
The approximated FPT density and the FPTL function of the estimated process $X(t)$ through the boundary $S=0.7$ are plotted in Figure \ref{fig:Figure15}(c)-(d). In order to validate the predicted results concerning the FPT, we consider also the same problem in the complete time range $I_C$. The forecasted results for the restricted time range $I_R$ and the approximated results for the complete time range $I_C$ are given in Table \ref{tab:Tabella2403b}. We note that the most meaningful index is the mode, since it corresponds to the peak of the FPT density and the two modes (namely, the mode in the restricted and in the complete time ranges) are quite close to each other (the relative error between the two values is about $1\%$).
\begin{table}
	\caption{The mean, the mode, the $1$st and the $5$th deciles and the standard deviation of the FPT in the complete time range $I_C$ and in the restricted time range $I_R$ (real application - FPT problem).}
	\label{tab:Tabella2403b}
	\tiny
	\centering
	\begin{tabular}{r|r|r|r|r|r}
		time range 					&mean							&mode					&$1$st decile		&$5$th decile		&st. deviation \\ \hline
		complete $I_C$			&$221.7308$			  &$219.8435$ 		      &$212.9876$		&$221.1406$		&$12.21116$\\
		restricted $I_R$		  &$197.6721$			&$222.2305$				&$214.8875$		&$228.3496$		&$80.0505$ \\
	\end{tabular}
\end{table}
%
\section{Conclusions}
During the recent years, many sigmoidal stochastic models have been introduced to study phenomena of interest in various different scientific areas. In order to model more complex population dynamics in which the maximum level of the growth is reached after many stages, we referred to the multisigmoidal logistic stochastic growth model. More in detail, the present work has been devoted to the analysis of the corresponding statistical inference and of the FPT problem. Two procedures useful to find the MLEs of the parameters have been described, one based on the resolution of the system of the critical points of the likelihood function, and the other one based on the maximization of the likelihood function by means of the S.A.\  algorithm. Then, the described strategies have been validated with a simulation study. The last section of the paper has been devoted to a real application concerning COVID-19 infections in four different European countries (France, Italy, Spain and United Kingdom). The data have been fitted using a suitable multisigmoidal logistic stochastic model. Finally, a study regarding the FPT through a fixed boundary has been also performed.
\par
Future developments can be oriented to find the MLEs of the parameters with other meta-heuristic optimization procedures (such as Variable Neighborhood Search or other swarm-based algorithms) in order to obtain nice estimates in a short computational time. We aim also to introduce a more sophisticated model suitable to describe better epidemiological dynamics with multiple waves, starting from the multisigmoidal logistic equation.
Moreover, aiming at a thorough analysis of the convergence speed
for parameter estimation in stochastic differential equations,
these approaches will be compared with applications of the recent method called
`covariance matrix adaptation evolution strategy'. Indeed, the latter is used often in the presence of several parameters
(cf., for instance, Ghosh {\em et al.}\ \cite{Ghosh2012} and Willjuice and Baskar \cite{Willjuice}).
%
\begin{acknowledgements}
Antonio Di Crescenzo and Paola Paraggio are members of the research group GNCS of INdAM (Istituto Nazionale di
Alta Matematica).
\par
This work was supported in part by the
Ministerio de Ciencia e Innovaci\'on, Spain, under Grant PID2020-1187879GB-100
by FEDER/Junta de Andaluc\'ia-Consejer\'ia de Econom\'ia y Conocimiento,
under Grant A-FQM-456-UGR18, by the ``Mar\'ia de Maeztu'' Excellence Unit IMAG, reference CEX2020-001105-M,
funded by MCIN/AEI/10.13039/501100011033/, and by Italian MIUR-PRIN 2017, project ``Stochastic Models for
Complex Systems'', No. 2017JFFHSH.
\par
Paola Paraggio thanks the Department of Statistics and Operations Research, Faculty of Sciences of the
University of Granada and the Institute of Mathematics of the University of Granada (IMAG) for
the hospitality during the one-month visit carried out in 2019.

\end{acknowledgements}

%
 \section*{Conflict of interest}
 The authors declare that they have no conflict of interest.



\end{document}